\documentclass[preprint]{elsarticle}

\usepackage{lineno,hyperref}
\modulolinenumbers[5]
\usepackage{tabularx}

\newcommand\clearrow{\global\let\rowmac\relax}
\clearrow

\usepackage[T1]{fontenc}
\usepackage[utf8]{inputenc}
\usepackage{amsmath}
\usepackage{natbib}
\usepackage{graphicx}
\usepackage{gensymb}
\usepackage{url}
\usepackage{mathtools}
\usepackage[varg]{txfonts}
\usepackage{textcomp}
\usepackage[super]{nth}
\usepackage{caption}
\usepackage{subcaption}

\usepackage{wasysym}
\usepackage{esvect}

\usepackage{siunitx}
\usepackage{longtable}
\usepackage{lscape}
\usepackage{threeparttable}

\biboptions{authoryear}

\begin{document}

\begin{frontmatter}


\title{First holistic modelling of meteoroid ablation and fragmentation: A case study of the Orionids recorded by the Canadian Automated Meteor Observatory}

\author[uwopa,wiese]{Denis Vida}
\ead{dvida@uwo.ca}

\author[uwopa,wiese]{Peter G. Brown}
\author[uwopa,wiese]{Margaret Campbell-Brown}
\author[montreal_planetarium,uwopa,wiese,imcee]{Auriane Egal}

\address[uwopa]{Department of Physics and Astronomy, University of Western Ontario, London, Ontario, N6A 3K7, Canada}
\address[wiese]{Western Institute for Earth and Space Exploration, University of Western Ontario, London, Ontario, N6A 5B7, Canada}
\address[montreal_planetarium]{Planétarium Rio Tinto Alcan, Espace pour la Vie, 
4801 av. Pierre-de Coubertin, Montréal, Québec, Canada}
\address[imcee]{IMCCE, Observatoire de Paris, PSL Research University, CNRS, Sorbonne Universités,\\ UPMC Univ. Paris 06, Univ. Lille, France}

\begin{abstract}
18 mm-sized Orionid meteoroids were captured in 2019 and 2020 by the Canadian Automated Observatory's mirror tracking system. Meteor position measurements were made to an accuracy of $\sim\SI{1}{\metre}$ and the meteors were tracked to a limiting magnitude of about $+7.5$ at the faintest point.
The trajectory estimation shows the intrinsic physical dispersion of the Orionid radiant is $0.400^{\circ} \pm 0.062^{\circ}$. 
An erosion-based entry model was fit to the observations to reproduce ablation and fragmentation for each meteor, simultaneously reproducing the light curve, the dynamics, and the wake. Wake observations were found to directly inform the grain mass distribution released in the modelled erosion. 
A new luminous efficiency model was derived from simultaneous radar and optical observations and applied in the modelling to improve its accuracy.
The results show that the apparent strength of Orionids varies with radiant location and time of appearance during the period of shower activity. The average differential grain mass distribution index was 2.15, higher than found from in-situ estimates, possibly due to the evolution of the physical properties of meteoroids since ejection.
All Orionids showed leading fragment morphology which was best explained by stopping the erosion at the peak of the light curve, leaving a non-fragmenting meteoroid with $\sim10\%$ of the original mass.
The inverted Orionid meteoroid average bulk density of $\sim\SI{300}{\kilo \gram \per \cubic \metre}$, corresponding to porosities of $\sim90\%$, is consistent with in-situ measurements of larger dust particles by Vega-2 at 1P/Halley and Rosetta at 67P. 
\end{abstract}

\end{frontmatter}


\section{Introduction}

The Orionids are among the strongest annual meteor showers visible at Earth. The shower, which is debris encountered by the Earth near the ascending node of parent comet 1P/Halley, peaks around October 22 each year. The shower flux is moderately high with a mean Zenithal Hourly Rate (ZHR) of around 30 \citep{Egal2020observations}, and a geocentric velocity of \SI{\sim 66.5}{\kilo \metre \per \second}. Meteoroids from 1P/Halley impacting Earth near its descending node produce the $\eta$-Aquariids, the twin shower to the Orionids, in early May each year. 

These twin Halleyid streams represent a unique opportunity to sample dust from a comet for which in situ measurements also exist \citep{Schulze1997}. Moreover, as each stream's evolution is distinct, the particles have different dynamical histories \citep{yeomans1981long, mcintosh1983comet, egal2020modeling}. While both showers contain mm-sized particles, the Orionids show occasional fireball outbursts (cm-sized objects) while the $\eta$-Aquariids are particularly rich in small (sub-mm) sized meteoroids \citep{Galligan2000, Chau2008, Schult2018}. The streams also differ in age - the Orionids represent significantly older ejecta than the $\eta$-Aquariids \citep{egal2020modeling} presenting a rare situation where it becomes possible to examine the evolution of meteoroid physical characteristics and orbital properties as a function of age.

In this work, we use high-precision optical observations of the 2019 and 2020 Orionids collected by the Canadian Automated Meteor Observatory's (CAMO) mirror tracking system \citep[for hardware details see][]{vida2021high} to measure the physical properties of Orionid meteoroids. In addition, the radiant dispersion of mm-sized Orionids is determined with a measurement accuracy an order of magnitude better than previous works at these particle sizes. We estimate trajectory uncertainties on a per-event basis using a Monte Carlo approach \citep{vida2019meteortheory}, with estimated measurement errors extracted from the variance in the observed meteor picks from a straight-line trajectory. This technique allows us to sub-select events having the highest measurement precision.  

The theoretical work by \cite{vida2019meteorresults} shows that in an ideal case, the CAMO mirror tracking system has a radiant precision of \ang{\sim 0.01}, although this is predicated on favourable meteor morphology. As shown in \cite{vida2021high}, radiant measurements might be an order of magnitude less precise for highly fragmenting meteors. Nevertheless, the CAMO system provides the most precise optical measurements available for faint meteor trajectories \citep{koten2019meteors}.

In this work, we apply the meteoroid erosion model by \cite{borovivcka2007atmospheric} to our observations of the Orionids to invert their physical properties and grain size distributions. For the first time, we fit the directly observed wake, a feature that was previously not reproduced by ablation models \citep{campbell2013high} but is critical to constrain the grain distribution. A novel luminous efficiency model was derived on the basis of the deceleration of the leading fragments. With the model fits simultaneously reproducing the light curve, deceleration, wake and the leading fragments (observed to $\mathrm{+7.5^M}$), these are the first holistic ablation model fits which explain the full optically observed meteor phenomenon.

\subsection{Previous observations of the Orionids}

The Orionids show a generally consistent level of annual activity, with some years showing outbursts rich in large meteoroids \citep{egal2020modeling}. Observations of the 2006 Orionid fireball outburst, the strongest in modern times, have shown that large meteoroids are in a 1:6 resonance with Jupiter \citep{sato2007origin, spurny2008exceptional}. Despite the estimated age of the Orionid shower of around 20,000 years \citep{jones1989age}, the 1:6 resonance appears to be impeding the dispersion process, causing stream meteoroids to concentrate and leading to possible future outbursts \citep{sekhar2014resonant}. \citet{egal2020modeling} predict no significant Orionid outbursts comparable to 2006 until after 2050 but shows that the 1:6 MMR is a significant driver of the stream's evolution. 

Many previous studies investigated the structure of Orionid radiants \citep[e.g.][]{znojil1968frequency, porubcan1973telescopic, hajduk1970structure}, but the precision of their measurements was not well defined for individual observations and was likely on the order of the expected physical dispersion itself. \cite{kresak1970dispersion} measured an Orionid radiant dispersion of \ang{0.84} which they defined as the median offset from the mean radiant. They used high-fidelity multi-station Super-Schmidt data reduced by \cite{jacchia1961precision}, although the data set does not contain any individually determined uncertainties. Recently, \cite{moorhead2021meteor} measured the radiant dispersions of major meteor showers using the high-quality Global Meteor Network data set which has robust uncertainty estimates of individual radiants \citep{vida2021global}. They found a median radiant offset of \ang{0.55} for the Orionids, a significantly smaller value than derived by \cite{kresak1970dispersion} but larger than the formal measurement errors of individual radiants, indicating that the physical radiant dispersion at mm sizes was resolved for the first time.

\cite{spurny2008exceptional} measured the median offset from the mean radiant of the resonant Orionid branch in 2006 to be only \ang{0.12}. Although they also do not report uncertainties, their trajectories are likely more precise than the data used in \cite{kresak1970dispersion} as fireballs usually have more data points for the trajectory fit. For larger meteoroids which are less affected by non-gravitational forces and ejected at lower speeds from the nucleus, tighter radiants are naturally expected, particularly if the meteoroids were locked in a resonance for a significant fraction of their lifetime.

Measuring the exact size and location of shower radiants from observations provides a key constraint for dynamical models of meteoroid streams \citep{egal2020forecasting}. This is especially important for younger streams whose activity and radiant distribution are driven by filaments ejected during individual perihelion passages \citep{egal2018draconid}. A recent example is the series of meteor shower outbursts caused by comet 15P/Finlay \citep{ye2015bangs, vaubaillon2020meteor}, the first time that an outburst was predicted before a meteor shower was known. The shower showed two outbursts caused by the 1995 and 2014 trails which happened days apart and had radiants separated by over \ang{10} degrees in the sky \citep{jenniskens2021first}.

Another key parameter in meteoroid stream dynamical models is the meteoroid bulk density which directly influences the terminal velocity of meteoroids ejected from the parent comet \citep{brown1998simulation, vaubaillon2005new, egal2020forecasting} and the magnitude of the radiation pressure, which alters the subsequent orbit \citep{Burns1979}. Roughly, for a meteoroid of the same mass, the difference between a cometary (\SI{300}{\kilo \gram \per \cubic \metre}) and an asteroidal bulk density (\SI{3500}{\kilo \gram \per \cubic \metre}) produces a factor of two difference in the ejection velocity. Less dense meteoroids are larger and are ejected faster as they experience more gas drag. This difference in the ejection velocity can be critical to reproduce observations and make accurate predictions of future shower activity, as was the case with the unexpected outburst of the Draconids in 2012 \citep{ye2014unexpected} and the predicted 2022 $\tau$-Herculid outburst \citep{egal2023modeling}.

In addition to its importance for dynamical models, bulk density is a key parameter in models of meteoroid impact hazard for spacecraft \citep{moorhead2019meteor}. If meteoroids from a particular shower are not monolithic bodies of high strength as traditionally assumed but weak agglomerates of $\micro$m-sized grains that crumble on impact, the overall impact risk is reduced \citep{christiansen1993design}. Nevertheless, very little is known about the actual influence of meteoroid structure on penetration depths as laboratory cratering experiments are limited in the composition of projectiles \citep[usually aluminium and its alloys;][]{ryan2008ballistic} which can be accelerated to meteoric speeds in practice. 

The only accurate in-situ measurements of the physical properties of dust from 1P/Halley were done by the Vega-2 spacecraft. From these data, a bulk density of $\SI{300}{\kilo \gram \per \cubic \metre}$ with an uncertainty of a factor of two \citep{krasnopolsky1988properties} was determined. 

\citet{Verniani1967} provides a value of $\SI{250}{\kilo \gram \per \cubic \metre}$ for the average estimated bulk density of four Orionids recorded by Super-Schmidt cameras with mean peak magnitudes near -1. \citet{Babadzhanov2009} found a range from \SIrange{400}{1400}{\kilo \gram \per \cubic \metre} based on two Orionid fireballs.

In this work, we apply an ablation and fragmentation model to the high-accuracy spatial and temporal measurements of the Orionids. A major goal is to validate our modelling/analysis approach by using the in-situ measurements of dust properties for 1P/Halley as ground truth. This in turn will help to create a robust modelling workflow which we will use to survey the physical properties of mm-sized shower meteoroids and by extension their parent bodies.

\section{Observations and data reduction}

From October 23 - 29, 2019 and October 12 - 17, 2020 the CAMO mirror tracking system observed a total of 18 Orionids that were well-tracked from both CAMO sites and therefore had precisely measurable trajectories. These Orionids were between \SIrange{1}{2}{\milli \metre} in diameter, based on the modeled mass and density derived in Section \ref{subsec:modelling_results}. Five additional Orionids were observed but had either large measurement uncertainties due to unfavourable geometry (low convergence angle hampering trajectory estimation) or inaccurate tracking. For these reasons, they were rejected from further analysis. We also manually rejected 8 meteors that the automated shower association algorithm with very liberal thresholds misclassified as Orionids (all meteors within \ang{5} of the mean Orionid radiant and with geocentric velocities within 30\% of the nominal value). The rejected events had radiants significantly more scattered than the 18 used in this analysis, and when modelled showed distinctly different physical properties, suggesting they were interlopers. We cannot rule out that these are much older (and hence more dispersed) Orionids.


The records were manually reduced following the procedure described in \cite{vida2021high}. This process involved astrometric picks being done by manually centroiding meteor positions using narrow-field data. The photometry was done both using wide-field (for the beginning of the meteor) and narrow-field data (the end of the meteor observed to magnitude $+7.5^{\mathrm{M}}$). The initial velocities were taken from the ablation model which accounts for any pre-observation deceleration \citep[see ][ and Section \ref{sec:ablation_model}]{vida2018modelling}. The velocities, radiants and orbits of all 18 Orionids are given in Table \ref{tab:orionids_orbits}.

\begin{figure}
  \includegraphics[width=\linewidth]{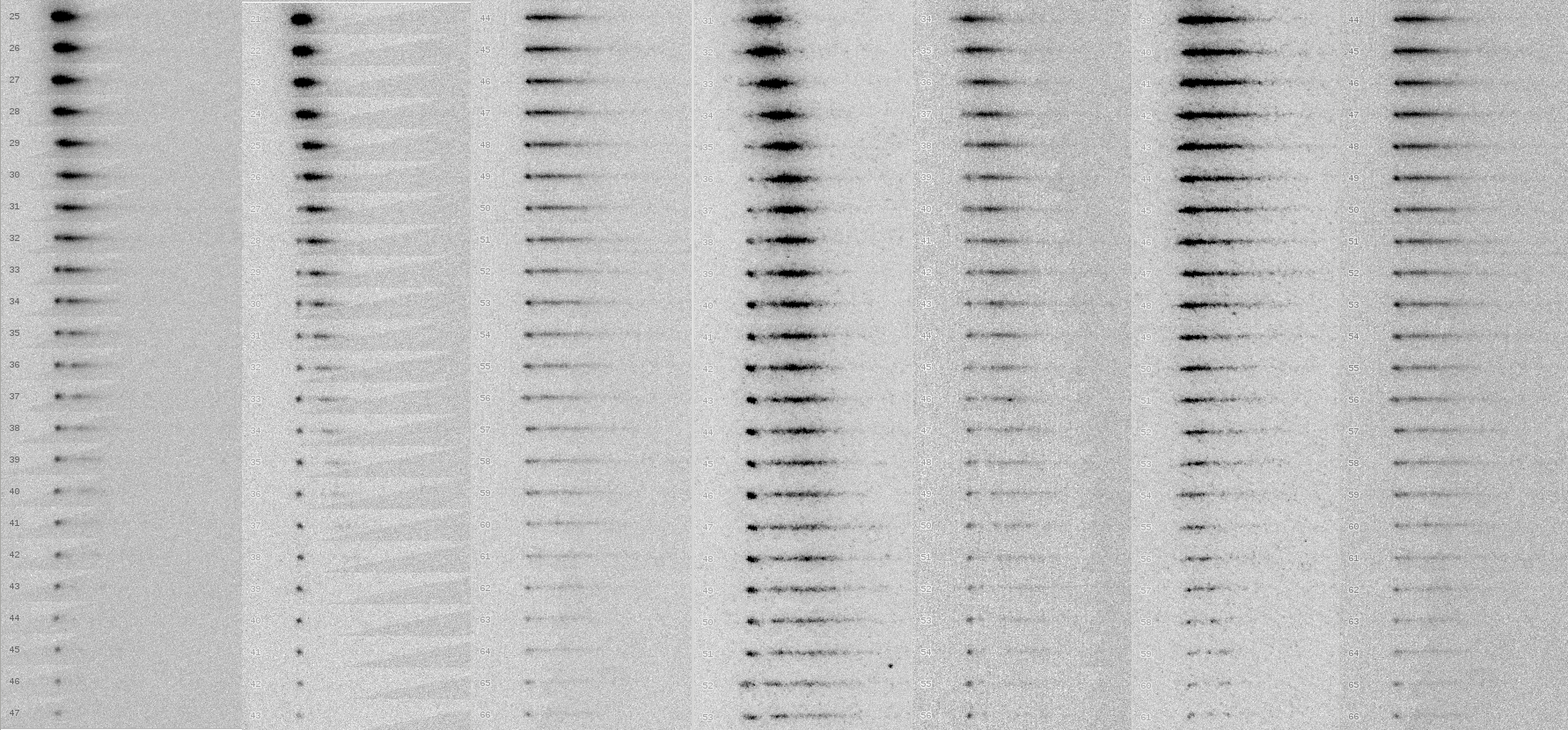}
  \caption[Composite narrow-field image of CAMO Orionids.]{Composite gray-inverted image of seven Orionids recorded using the CAMO narrow-field camera. The frame number increases from top to bottom - time therefore runs down the page. All meteors show the leading fragment morphology type as shown by the remnant point-like objects near the bottom of each sequence.}
  \label{fig:orionid_morphology}
\end{figure}

A particularly unusual qualitative observation we noted is that all 18 Orionids had the same ``leading fragment" fragmentation morphology \citep{subasinghe2018luminous}. Mosaics of video frames showing leading fragments for a selection of seven Orionids are shown in Figure \ref{fig:orionid_morphology}. It is notable that examination of more than 1000 CAMO tracked events showed leading fragments present in only about 3\% of cases \citep{subasinghe2018luminous}. 

At the beginning of their visible trajectory, leading fragment meteors usually exhibit a wake caused by grain erosion (i.e. continuous fragmentation). After some time, one apparently single-body fragment emerges at the front of the meteor, showing no further fragmentation nor wake and ultimately dimming as a star-like point-spread function below the noise floor. 

The leading fragment morphology is favourable for achieving a high trajectory accuracy because there is a consistent leading reference point with a high signal-to-noise ratio. As shown previously \citep{vida2020new}, the main limitation on the precision of CAMO metric measurements is the extended meteor morphology.

\begin{figure}
    \includegraphics[width=.5\linewidth]{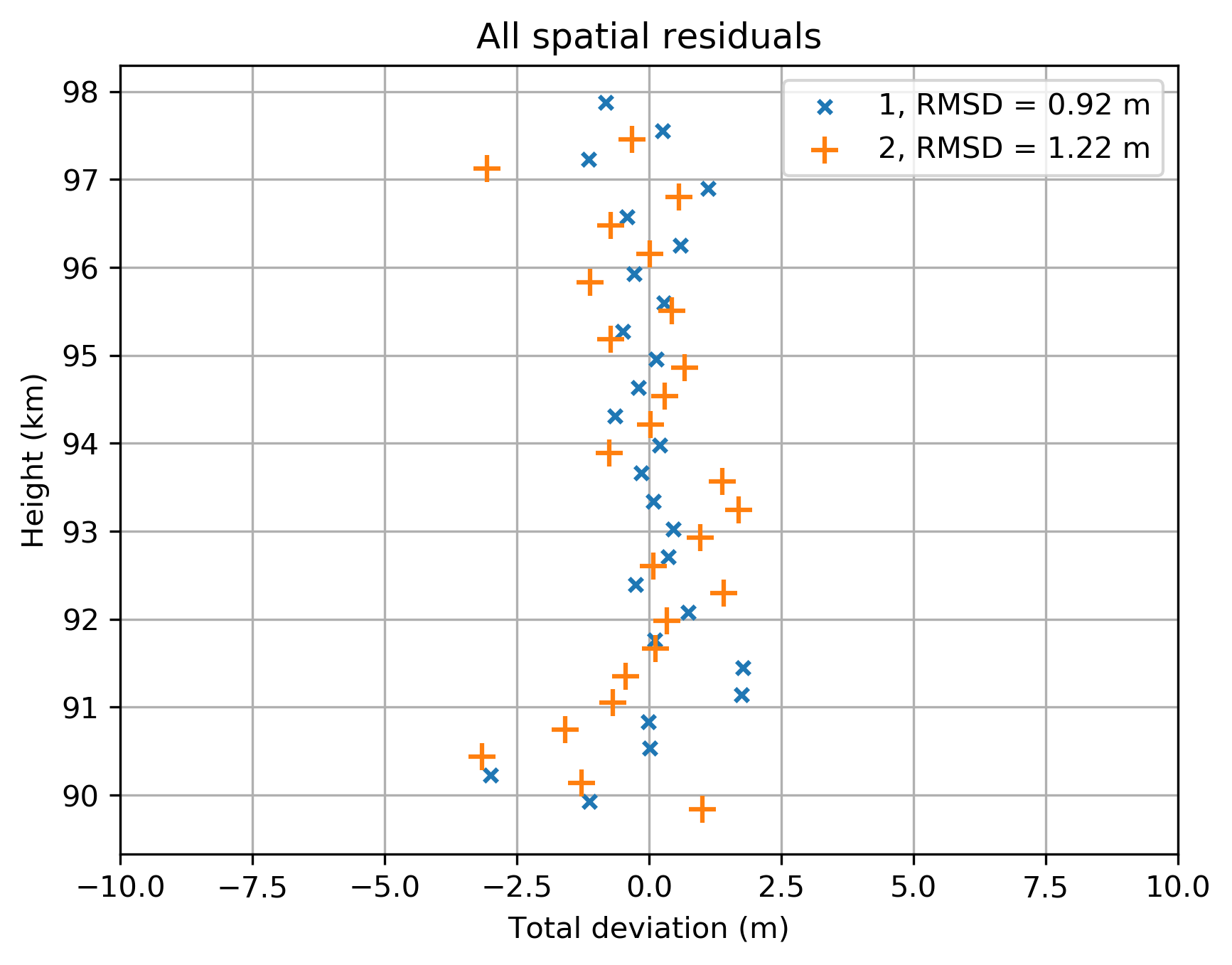}\hfill
    \includegraphics[width=.5\linewidth]{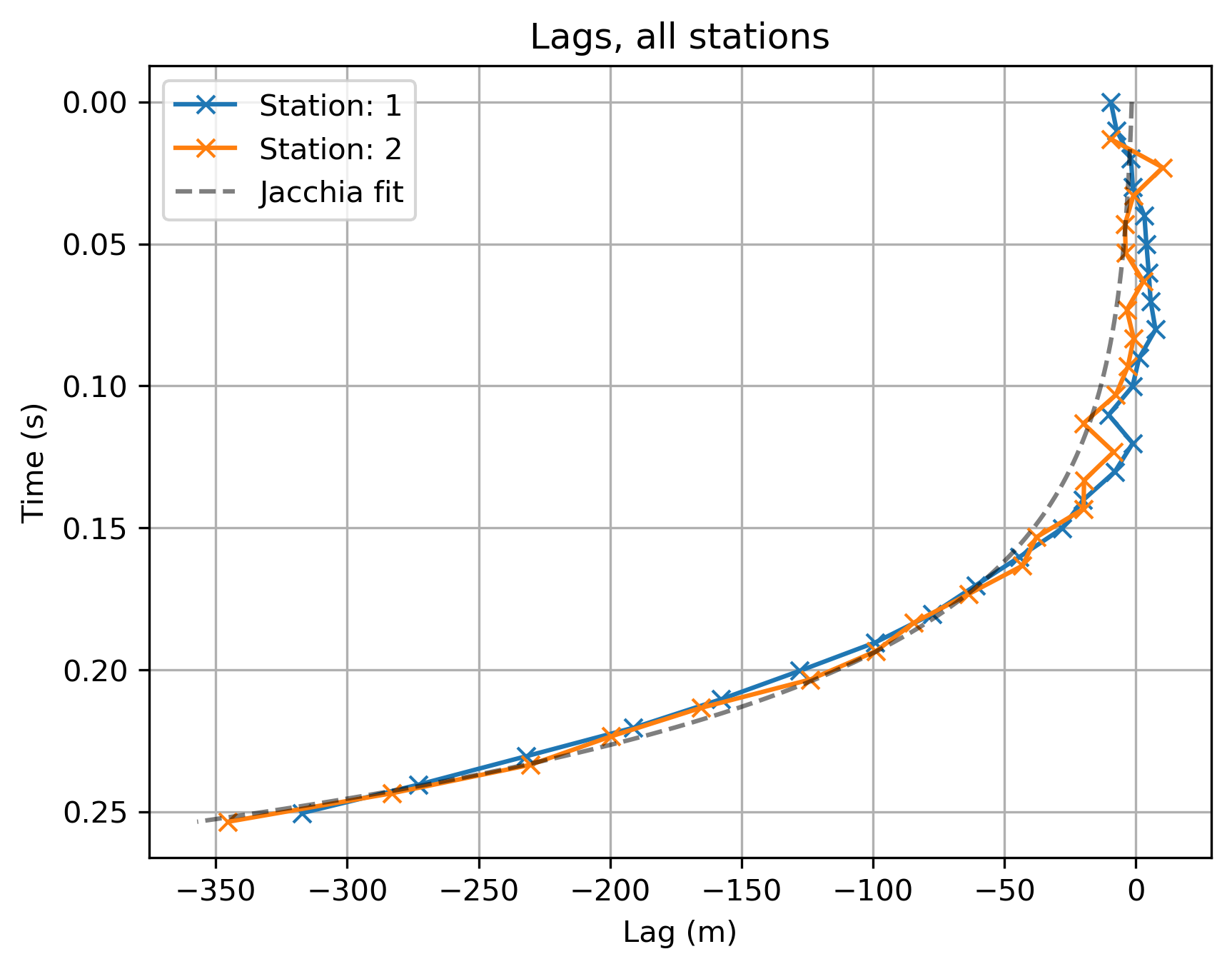}
    \includegraphics[width=.5\linewidth]{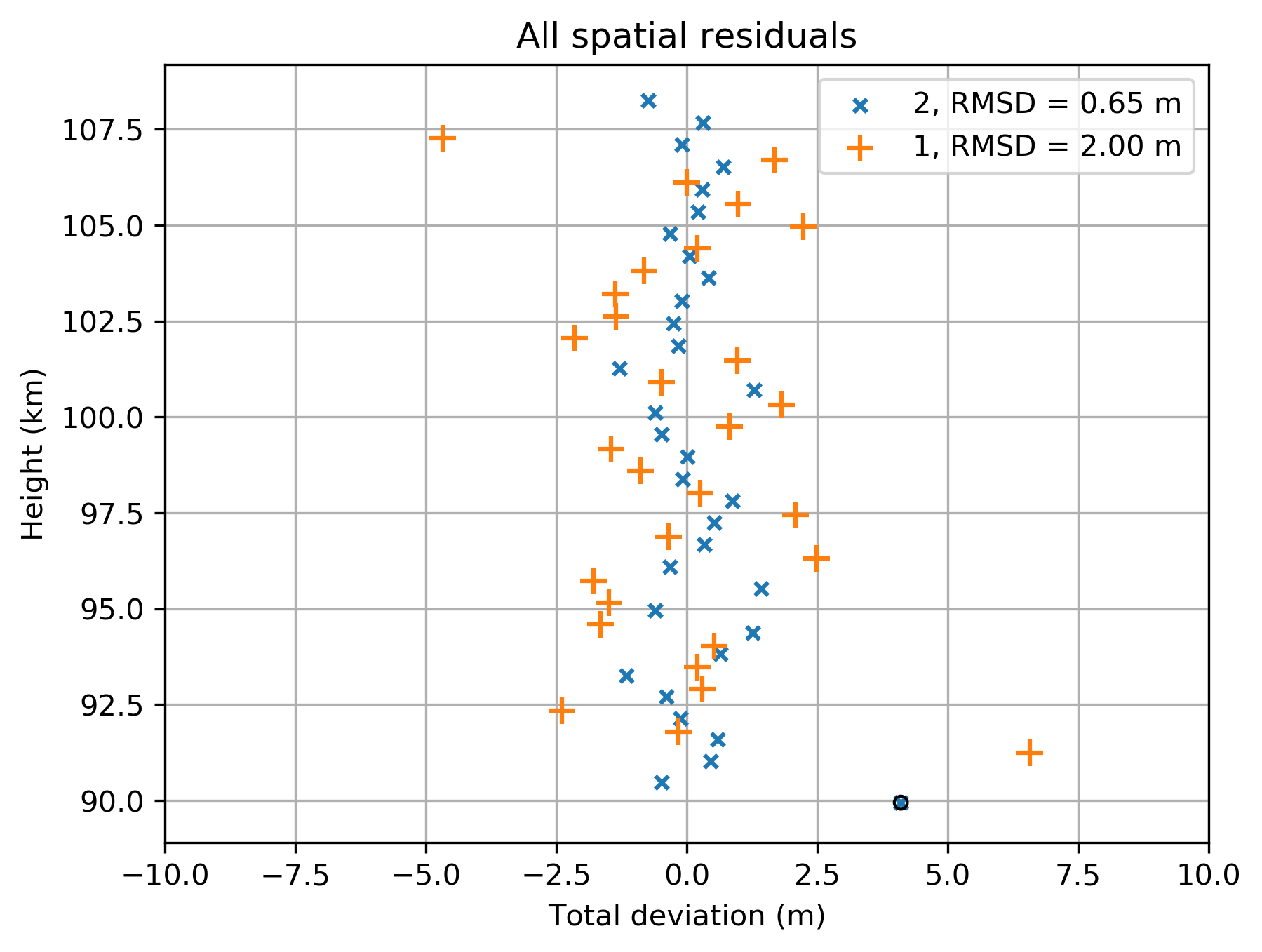}\hfill
    \includegraphics[width=.5\linewidth]{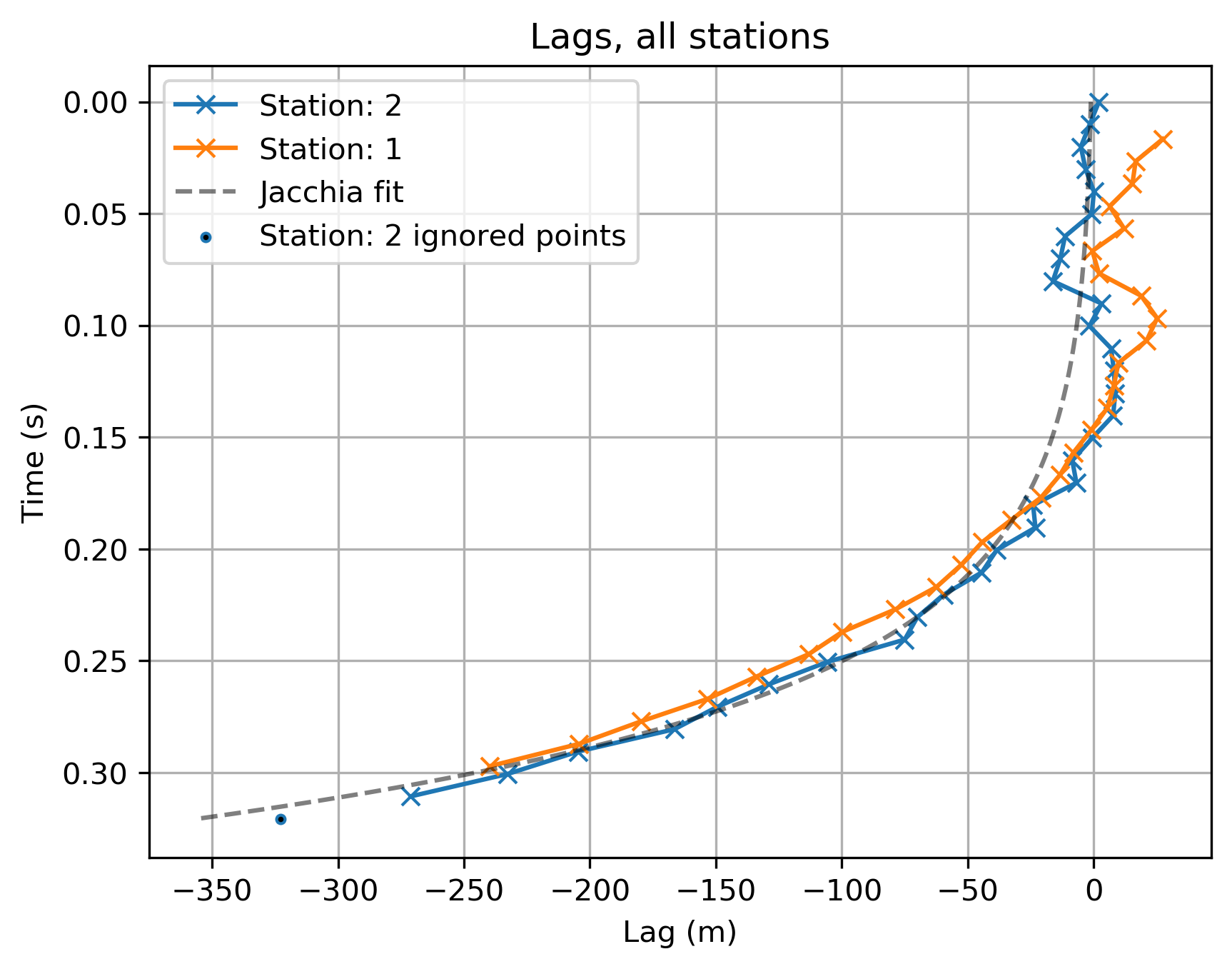}
    \includegraphics[width=.5\linewidth]{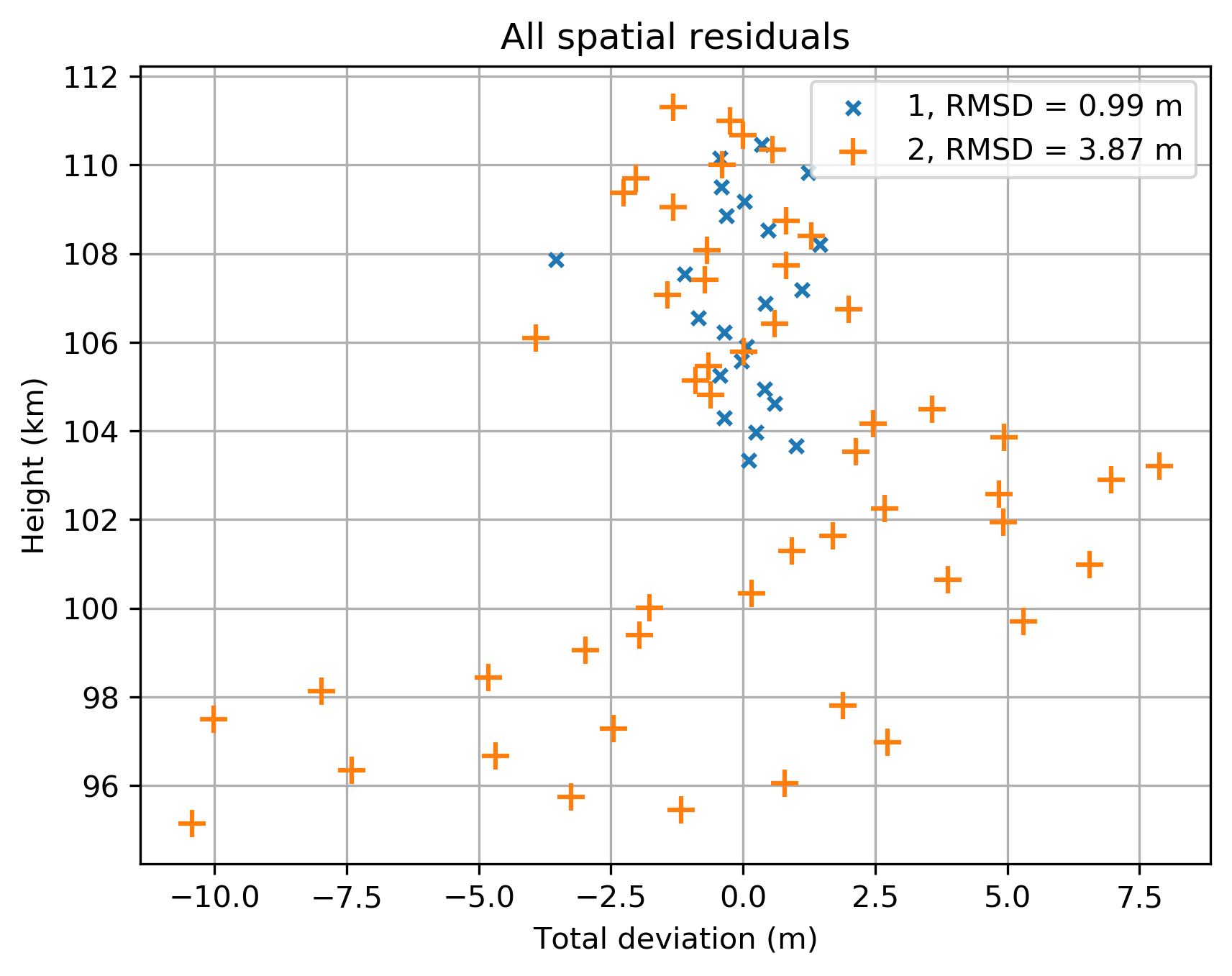}\hfill
    \includegraphics[width=.5\linewidth]{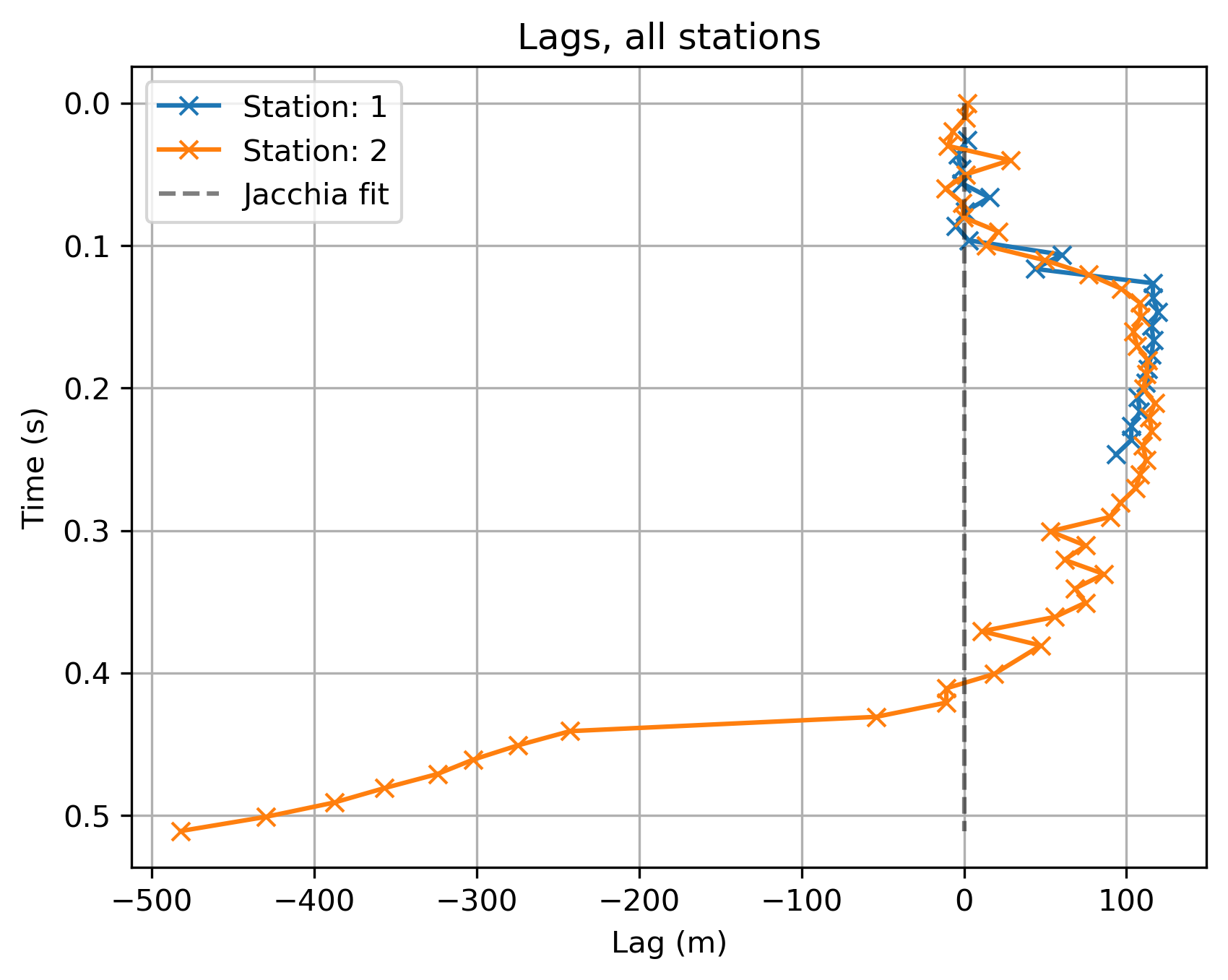}
    \caption[Spatial fit residuals and lag of selected Orionids]{Orionids observed on: Top: October 28, 2019, at 05:16:22 UTC. Middle: October 23, 2019, at 08:45:02 UTC. Bottom: October 28, 2019, at 05:18:34 UTC. Left column: Spatial fit residuals from a straight line. Right column: Lag (distance along the trajectory relative to an assumed constant velocity model). The light curves of these events are given in \ref{app:obs_sim_comp}.}
  \label{fig:orionids_fit_lag}
\end{figure}

As an example of the generally favorable morphology available for the Orionids, the top inset of Figure \ref{fig:orionids_fit_lag} shows a CAMO reduction of an Orionid demonstrating the high precision achievable. However, due to a different perspective from the two sites, it was sometimes difficult to pick the same features in both videos during manual data reduction. Hence, the observed lags from both sites differ in some cases. We define the lag as the distance that a meteoroid falls behind a hypothetical non-decelerating meteoroid \citep{subasinghe2017luminous}.

The middle inset of Figure \ref{fig:orionids_fit_lag} shows one such case where the spatial fit residuals are good, reflecting consistent picks along the meteor straight line trajectory. However, as the same meteor features were not able to be located from both sites, there is an offset in the lag of several to tens of meters (1 to 10 pixels in the image), reflecting poor along-track pick consistency.

The bottom inset of Figure \ref{fig:orionids_fit_lag} shows an Orionid which did not have a constant point of reference. In this case, the transition in pick location happened at the moment of separation between the dust and the leading fragment (at $t = \SI{0.1}{\second}$) where the lags show a discontinuous break. The initial velocity could still be computed with reasonable precision (note the near vertical lag before the break). 

As the lags between stations were sometimes inconsistent, the nominal meteor trajectory was computed using our implementation of the lines of sight method \citep{borovicka1990comparison} in these cases and uncertainties were computed using the Monte Carlo method of \cite{vida2019meteortheory}.

Finally, the brightness and linear extent of the wake (the instantaneous luminous trail behind the meteor head) were measured. This is accomplished on a per-frame basis by rotating the video frame so that the direction of meteor travel is towards the left. A horizontal strip is cropped from the image, encompassing the meteor and its wake. The background is assumed to be uniform and is then subtracted from the image. The wake as a function of distance from the meteor head is estimated as the sum of pixel intensities in every image column.

\section{Ablation and fragmentation model} \label{sec:ablation_model}

The CAMO mirror tracking system observes meteors with visible wakes in more than 90\% of cases \citep{subasinghe2016physical}. As a result, we need to apply an ablation model which explicitly allows for fragmentation which produces particle release - a single-body (non-fragmenting) model does not suffice to reproduce our measurements. To meet this requirement, we employ the quasi-continuous fragmentation model proposed by \cite{borovivcka2007atmospheric}. In this model, a meteoroid begins as a single body at the top of the atmosphere. At a defined height, $h_e$, a steady release of \SIrange{10}{300}{\micro \metre} grains begins, a process called erosion. As smaller grains decelerate more rapidly than the larger ones, they get sorted by mass behind the main body, forming the wake as they continue to ablate. The grains do not fragment further and are treated as ablating single bodies. The model is described in detail in \cite{borovivcka2020two} and our implementation in \cite{vida2023direct}.

This model is fundamentally different from the ``dustball'' ablation model proposed by \cite{hawkes1975quantitative}, who assume that meteoroids completely disintegrate into constituent grains before the start of ablation. This mode of fragmentation would result in the removal of smaller grains as they would ablate first and quickly; only the largest grains would survive for most of the duration of the flight. As we will show, this model is incompatible with our observations where wakes with smaller grains are observed well into the luminous flight.

In the erosion model, the main meteoroid experiences two types of mass loss: ablation and erosion. The ablation mass loss, which represents the rate of mass released by vaporizing atoms of the meteoroid, is controlled through the ablation coefficient $\sigma$. The erosion mass loss represents the rate of release of grains from the main body and is controlled by the erosion coefficient $\eta$. Both the $\sigma$ and $\eta$ have the same units of kg/MJ (equivalent to s$^2$ km$^{-2}$ used in the literature) and are applied in the same way within the classical mass loss equation \citep{vida2023direct}. After release, the eroded grains ablate independently with the same $\sigma$ as the main body through vaporization.

The grain masses are assumed to be distributed according to a power law distribution. The distribution is characterized by the minimum $m_l$ and maximum $m_u$ grain mass, and the differential mass index $s$\footnote{There is a misprint in the equation for the number of fragments of a given mass in \cite{borovivcka2007atmospheric}, see Chapter 1.4.4.4 in \cite{borovivcka2019physical} for the correct relation, or see \cite{borovivcka2020two} for more details.}. 


We assume that the released grains are refractory constituent silicate grains, with makeup similar to micrometeorites \citep{kohout2014density} and cometary grains observed by instruments on board ROSETTA \citep{hornung2016first} which have a bulk density in a narrow \SIrange{3200}{3300}{\kilo \gram \per \cubic \metre} range \citep{borovivcka2019physical}. However, as the grain density is not an independent parameter and the same meteoroid behaviour can be achieved by changing the range of grain masses \citep{borovivcka2007atmospheric}, we assume a fixed bulk density of $\rho_g = \SI{3000}{\kilo \gram \per \cubic \metre}$ for consistency with the previous work by \cite{borovivcka2007atmospheric} and \cite{vojavcek2019properties}.

Instead of ejecting and ablating individual grains, which would number in the millions, the grain mass range ($m_l$, $m_u$) is binned into 10 bins per order of magnitude in our numerical implementation of the model \citep{borovivcka2007atmospheric}. The classical single-body equations are then numerically integrated for each mass bin and the luminosity is multiplied by the number of grains in each mass bin to reconstruct the full light curve and wake at a given time. To allow for a variation in the rate of erosion, an option is added in the model to change the erosion coefficient at a height $h_{e2}$ to a secondary erosion coefficient $\eta_2$.

The simulations are started at the height of \SI{180}{\kilo \metre} (taken as the top of the atmosphere) and the ablation equations are integrated for every fragment/grain separately with a time step of \SI{0.005}{\second}. An object is assumed to stop ablating when its mass drops below \SI{e-14}{\kilo \gram}, its velocity below \SI{3}{\kilo \metre \per \second}, or its height reaches \SI{3}{\kilo \metre} lower than the last observed point on the meteor trajectory. The NRLMSISE-00 model \citep{picone2002nrlmsise} is used for the atmospheric mass density at the mean observed location and time. To speed up the air density lookup during integration, a 7\textsuperscript{th} order polynomial is fit to the logarithm of atmosphere mass density from a height of \SI{180}{\kilo \metre} to \SI{5}{\kilo \metre} below the meteor's terminal height.

The model takes the curvature of the Earth into account, which was found to be critical for accurate fits for data of CAMO's precision. As the simulation starts at the height of \SI{180}{\kilo \metre}, which may be hundreds of kilometers from the observed begin point for shallow trajectories, the initial zenith angle has to be corrected for Earth's curvature. The corrected initial zenith angle $z_c$ used in the simulation is:

\begin{align} \label{eq:zc_sim}
    z_c = \arcsin \left(\frac{h_{\text{b}} + R_E}{h_{0} + R_E} \sin{z_0} \right) \,,
\end{align}

\noindent where $h_{0}$ is the simulation start height (in meters), $h_{\text{b}}$ is the observed begin height (in meters), $z_0$ is the observed zenith angle (in radians), and $R_E$ is the Earth's radius at the latitude of the meteor (in meters) \citep[see the Appendix in][]{vida2019meteortheory}.

The curvature of the Earth and the drop due to gravity also need to be accounted for during each step of the simulation. The height of a meteor under which the Earth curves is higher than in a flat Earth model but decreases slightly due to gravity. The curvature-corrected height is computed as:

\begin{align}
    h = \sqrt{h_0^2 - 2l \cos{z_c}(h_0 + R_E) + 2h_0 R_E + l^2 + R_E^2} - R_E \,,
\end{align}

\noindent where $h$ is the height of the meteor at each time step and $l$ is the cumulative length traversed by the meteor since the beginning of the simulation (i.e. the distance from the start of the simulation to the current position). The drop due to gravity is applied separately using the classical equations but accounting for how the gravitational acceleration changes with height \citep[see the appendices in ][]{vida2019meteortheory}.

For all fragments we assume a fixed drag coefficient of $\Gamma = 1$ and a spherical shape ($A = 1.21$) - we note that the classical ablation equations can only measure the shape-density coefficient $K = \Gamma A \rho_m^{-2/3}$ \citep{ceplecha1975ablation}, thus our measurements of the meteoroid bulk density $\rho_m$ are inherently dependent on this spherical shape assumption. The magnitude is computed from the simulated luminosity assuming a power of a zero magnitude meteor of $P_{0M} = \SI{840}{\watt}$, as appropriate for the CAMO spectral bandpass \citep{weryk2013simultaneous}.

During the numerical integration, the model position of the leading fragment and the model position of the brightest mass bin are tracked. These two positions can differ by several tens of meters - we find that both are needed when matching observations. For example, during the phase of intense erosion, the grains cannot be visually separated and we take the centroid of the bulk of the visible meteor as the reference measurement. In this case, the brightest mass bin location best corresponds to the measured location. In contrast, near the end of the meteor's luminous flight, the grains separate sufficiently to directly measure the position of the leading fragment, which is well defined.

The modelled wake is computed by projecting the positions of the ablating individual grains using the actual perspective of the observing station. These are then convolved for all fragments with a Gaussian point spread function (PSF). The PSF is measured on the calibration images with static stars in the narrow-field instrument. As the focus rarely changes we find the best match uses a full width at half maximum between \SI{7}{\metre} and \SI{20}{\metre} (1 - 3 pixels in the narrow-field camera), depending on the observing conditions. A window of \SI{200}{\metre} behind the leading fragment is usually considered, where the reference point at \SI{0}{\metre} is the location of the leading fragment. The brightness of the observed wake is not calibrated but is scaled to the simulated wake by matching the integrated area below the wake curve $\pm20$ meters surrounding the leading fragment. This procedure produces correctly scaled intensity profiles under the assumption that the simulated and observed light curves match. The restricted window around the leading fragment in which the area is computed helps to ignore the noise in the tail of the wake which can become significant near the end of the meteor when it is barely visible above the noise floor. The along-the-trajectory wake alignment can also be manually adjusted if the leading edge is not a good reference point. An option in the software allows alignment of the simulated wake to observations using cross-correlation.


\subsection{Luminous efficiency model}

A major parameter used in the modelling upon which most absolute physical meteoroid values depend is the conversion between kinetic energy and photon production. This ratio, denoted as $\tau$, is termed luminous efficiency. $\tau$ is challenging to constrain as it requires an accurate independent estimate of the meteoroid's instantaneous mass together with its light production and has been the focus of many laboratory, observational, and theoretical studies \citep{Popova2019}. In this section, we aim to synthesize luminous efficiency measurements done previously by other authors, add our own, and derive an analytical model for fainter meteors that can be used in our ablation modelling. 

A complicating factor in measuring $\tau$ is that meteoroids often fragment into smaller grains \citep{borovivcka2007atmospheric, subasinghe2016physical} which have a larger surface-to-mass ratio compared to the main body, so the measured light is not produced by a single meteoroid of one measurable mass. This introduces large uncertainties in the measurements, artificially increasing the luminous efficiency.

A further complicating factor is that spectra of meteors vary due to compositional differences, while the total light is measured in a single instrumental bandpass \citep{brown2020coordinated}. This means that different instruments with different spectral sensitivities measure different luminous efficiencies \citep{ceplecha1998meteor}. The $\tau$ measurements used herein were done using CAMO, the same instrument we use in the modelling. In this way, our modelling is self-consistent and minimizes spectral bandpass variations of $\tau$.

\cite{subasinghe2018luminous} mitigated the influence of fragmentation on $\tau$ measurements by using CAMO observations of rare non-fragmenting and leading fragment meteoroids. Those meteoroids behave similarly to what we observed for the Orionids - after an initial fragmentation, a single fragment emerges and behaves as a single-body meteoroid. They accurately measured the deceleration and derived the fragment's dynamical mass. The downside of this approach is that a bulk density needs to be assumed to derive a mass - a difference in bulk density of 1000 and \SI{3000}{\kilo \gram \per \cubic \metre} translates to a factor of 5-10 in $\tau$. Regardless of the chosen density, they found a correlation with mass, where smaller meteoroids had a larger luminous efficiency. This was also corroborated by radar estimates \citep{brown2020coordinated}. An almost identical $\tau$-mass dependency was found by \cite{capek2019small} when applying an advanced ablation model to iron meteoroids. 

From these observations, it appears that the $\tau$-mass dependence has an inflection point and reverses around one gram - traditionally, $\tau$ models for fireballs assume an increase in $\tau$ with mass \citep{ceplecha2005fragmentation, borovivcka2013kovsice, borovivcka2020two}. However, these models are optimized for deeply penetrating fireballs which are in the continuous flow regime, while our model is appropriate to free molecular flow. 

To develop an empirical model, we use individual $\tau$ measurements from CAMO and Electron Multiplying Charge Coupled Device (EMCCD) camera data \citep{vida2020new, brown2020coordinated}. These are summarized in Table \ref{tab:tau_list}. We exclude all measurements by \cite{subasinghe2018luminous} except for Halley-type comet (HTC) meteoroids. In their work, they assumed a density for Jupiter-Family comet (JFC) meteoroids of \SI{3190}{\kilo \gram \per \cubic \metre}, as given by \cite{kikwaya2011bulk}. We believe this density estimate is too high, likely due to the optical observations used by \cite{kikwaya2011bulk} being not precise enough to measure meteoroid deceleration for JFC meteoroids, thus the algorithm preferred non-decelerating high-density meteoroids over lower-density meteoroids which show deceleration \citep[as corroborated by][]{buccongello2023physical}. We use most of the measurements by \cite{brown2020coordinated}, excluding obvious outliers and including several new unpublished measurements. We note that the optical measurements from \cite{brown2020coordinated} were made using EMCCD cameras so the bandpass is slightly different from the CAMO-tracking system. However, the zero magnitude power for the EMCCDs has been computed to be \SI{945}{\watt}, very similar to the \SI{840}{\watt} appropriate to the image-intensified CAMO cameras, thus combining the two datasets is reasonable.  Finally, we add three Orionid leading fragment measurements derived with an approach similar to \cite{subasinghe2018luminous}, but using a bulk density of \SI{300}{\kilo \gram \per \cubic \metre}.

\begin{figure}
  \includegraphics[width=\linewidth]{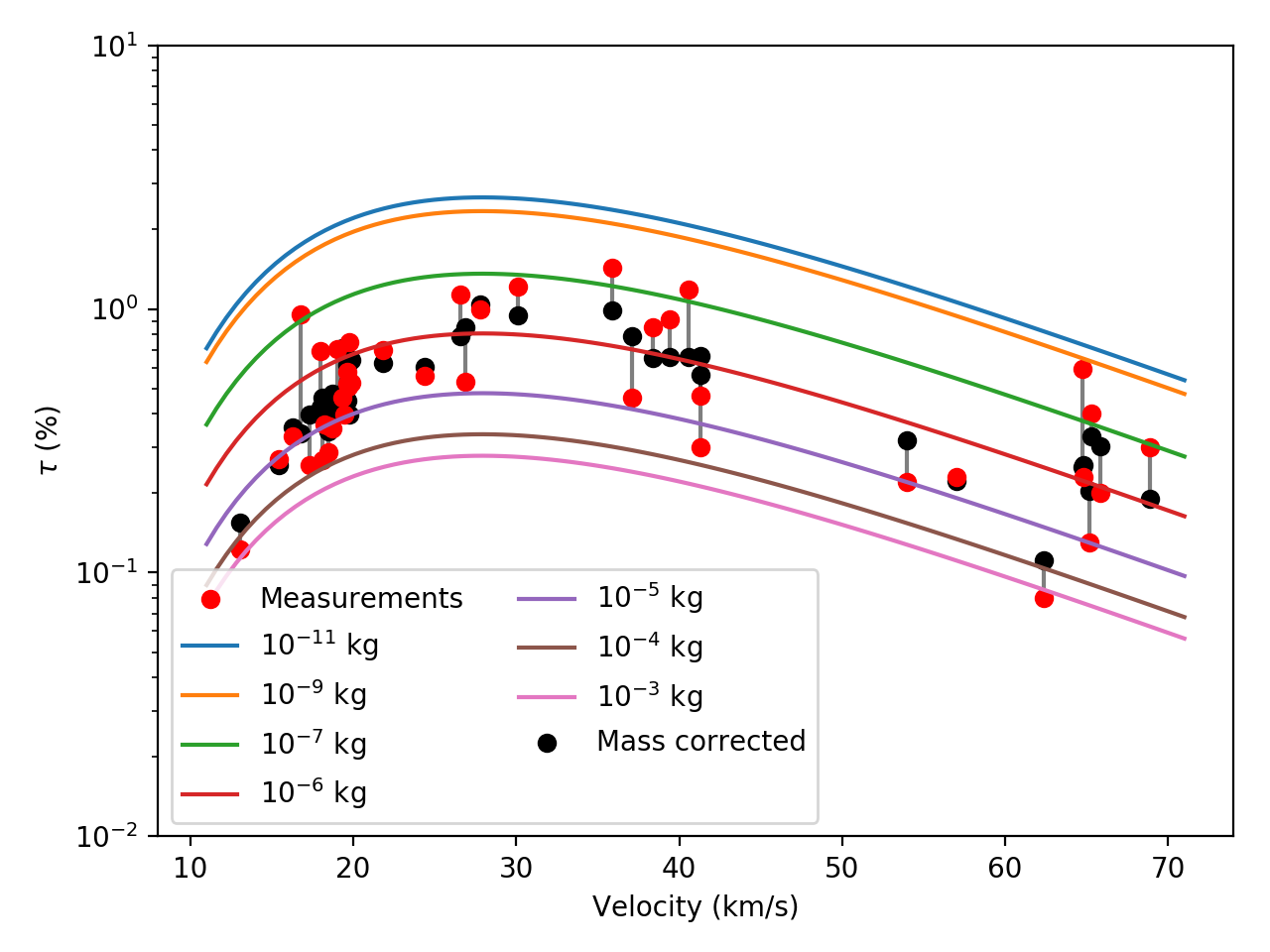}
  \caption{Our empirical model of luminous efficiency dependence on velocity and mass (solid lines per mass). The red dots represent the actual measurements from Table \ref{tab:tau_list}. The black points represent computed $\tau$ values using the new model and the observed velocity and mass, illustrating the relative uncertainty in the fit. Thin black lines connect the observed points with their computed counterparts.}
  \label{fig:tau_vel}
\end{figure}

Motivated by the approach of \citet{pecina1983new} and \citet{ReVelle2001c}, we adopt the following functional form for luminous efficiency:

\begin{equation}
    \ln \tau = A + B\ln{v} + C\ln^3{v} + D\tanh{\left(0.2 \ln(m \times 10^6)\right )} 
\end{equation}

\noindent where $\tau$ is the dimensionless luminous efficiency in percent, $v$ is the meteoroid velocity in $\SI{}{\kilo \metre \per \second}$, and $m$ is the meteoroid mass in $\SI{}{\kilo \gram}$. Using the data in Table \ref{tab:tau_list} we find the best fit parameters $A = -12.59, B = 5.58, C = -0.17, D = -1.21$. The median fit error is $30\%$, and the fit dependence as a function of velocity is shown in Figure \ref{fig:tau_vel}. The velocity dependence is modelled to best suit the trends observed in the data - the luminous efficiency peaks near \SI{25}{\kilo \metre \per \second} and decreases at low and high speeds.

\begin{figure}
  \includegraphics[width=\linewidth]{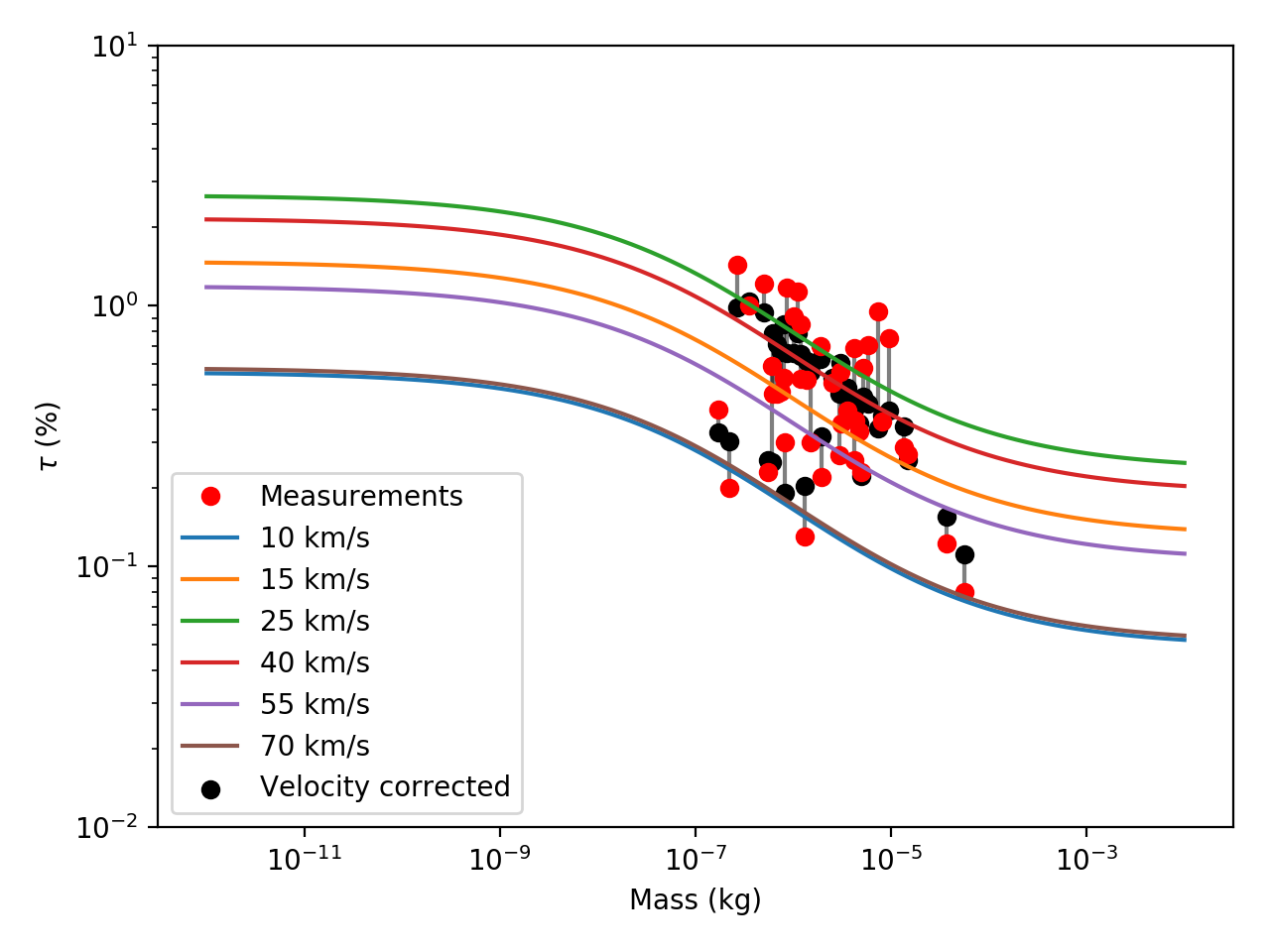}
  \caption{Luminous efficiency dependence on the mass and velocity. The red dots represent the actual measurements, and the black points represent computed $\tau$ values using the new model and the observed velocity and mass. The 10 and \SI{70}{\kilo \metre \per \second} curves overlap and are at the bottom.}
  \label{fig:tau_mass}
\end{figure}

Most models for $\tau$ which include mass dependence use the $\tanh$ function \citep[e.g.][]{ceplecha2005fragmentation, borovivcka2013kovsice}. The measured masses indicate that there is an increase in $\tau$ with smaller masses, but the tapering is not directly observed. As $\tau$ cannot be larger than $100\%$, the $\tanh$ function smoothly levels off at the extremes, as shown in Figure \ref{fig:tau_mass}.

\begin{figure}
  \includegraphics[width=\linewidth]{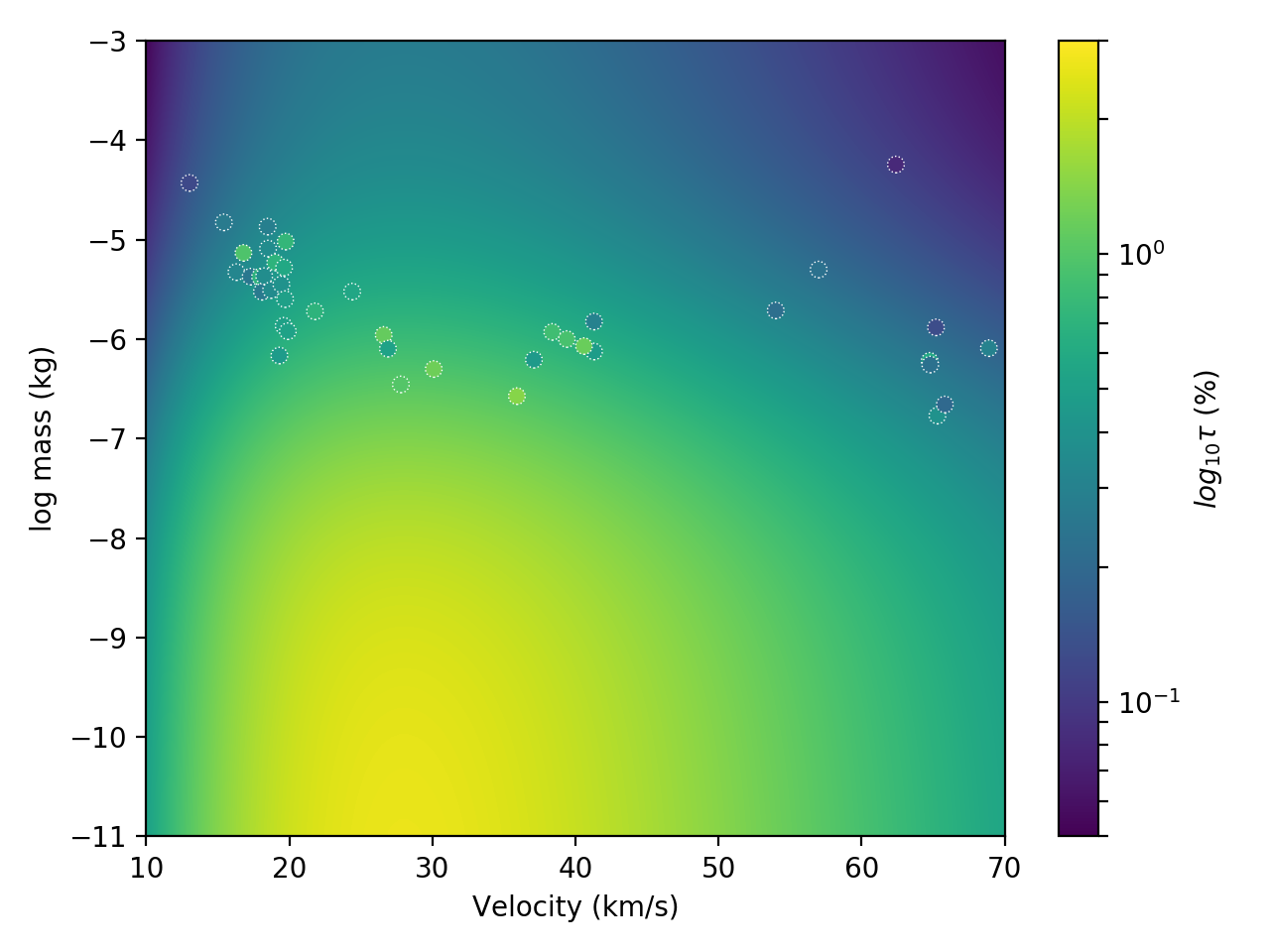} \\
  \includegraphics[width=\linewidth]{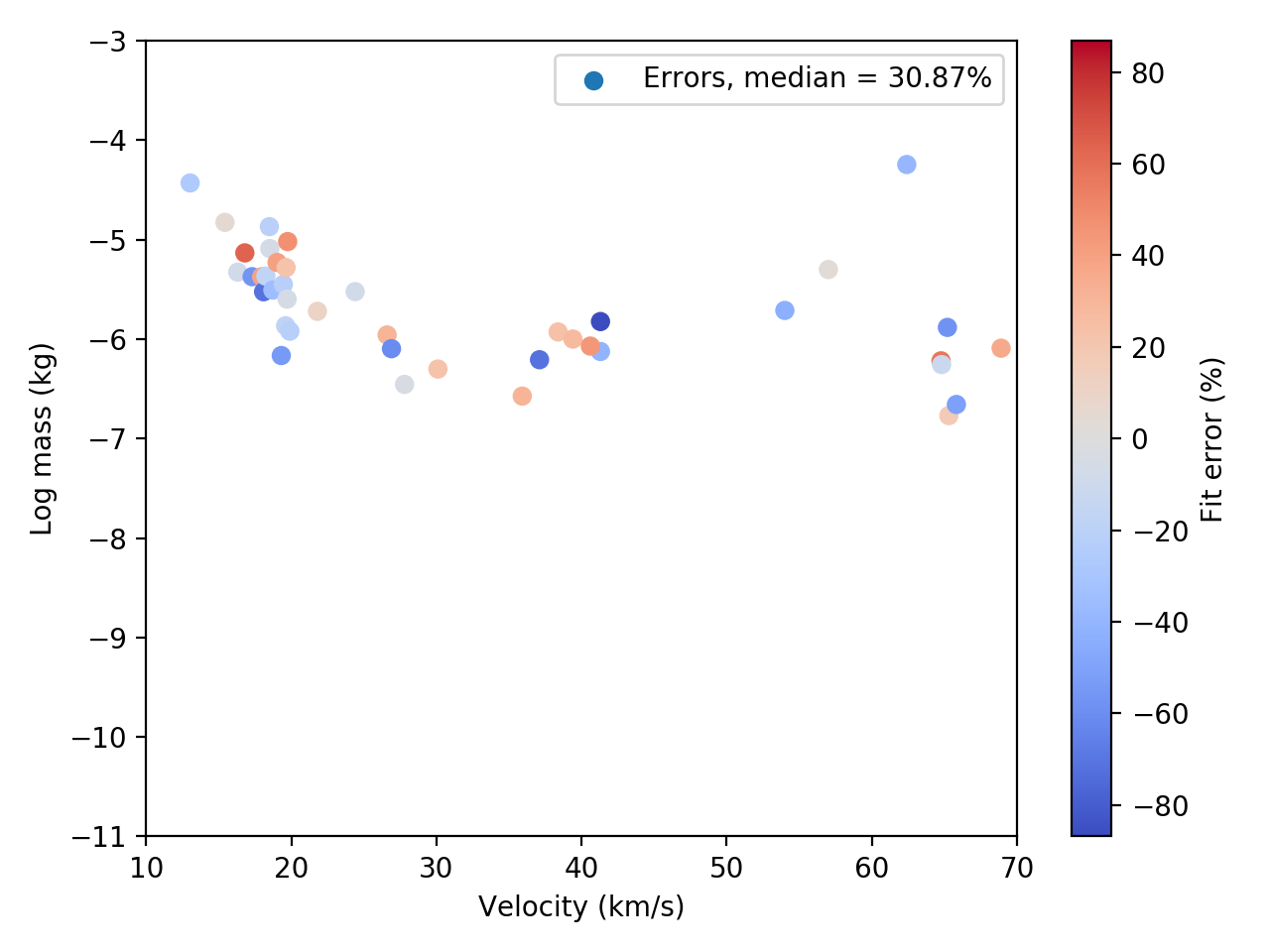}
  \caption{Left: The luminous efficiency dependence on both the mass and velocity. The white circles show the measured values. Right: Fit errors between measurements from Table \ref{tab:tau_list} and our empirical model.}
  \label{fig:tau_vel_mass}
\end{figure}

In general, the average $\tau$ across all speeds and masses is about $0.5\%$. For an average mg meteoroid, it peaks near  $\SI{25}{\kilo \metre \per \second}$ at $0.8\%$, and drops to $0.2\%$ at extreme speeds of 10 and $\SI{65}{\kilo \metre \per \second}$. In terms of the mass, $\tau$ varies an order of magnitude between masses of $10^{-3}$ and $\SI{e-9}{\kilo \gram}$. Figure \ref{fig:tau_vel_mass} is a $\tau$ ``heatmap'' as a function of speed and mass. Note that the $\tau$ values in our model at the very smallest masses shown are extrapolations from the mass trend at larger values - no actual measurements exist below $\SI{e-7}{\kilo \gram}$, although dust as small as $\SI{e-11}{\kilo \gram}$ is used in the erosion model. We empirically estimated the factor of $0.2$ in front of the $\tanh$ mass dependence by forward modelling to achieve an optimal fit to our Orionid data.

Finally, we want to emphasize the importance of using the actual meteoroid mass and speed at the point along the trail where $\tau$ is measured, instead of using the initial values at the beginning of the trajectory \citep[e.g. as in ][]{brown2020coordinated}. The two masses can differ by an order of magnitude and speeds by several kilometres per second, resulting in inaccurate $\tau$ model fits.

\begin{longtable}{l r r r r l}
\caption{List of meteors used for luminous efficiency model fit. The fit errors are given in the last column. 1* - \cite{subasinghe2018luminous}, 2* - \cite{brown2020coordinated}.} \\
    \hline\hline
Event & $v$ (km/s) & $m$ (kg) & $\tau$ (\%) & Fit error (\%) & Dataset \\
    \hline 
 \endfirsthead
 \caption{Continued.} \\
    \hline\hline
Event & $v$ (km/s) & $m$ (kg) & $\tau$ (\%) & Fit error (\%) & Dataset \\
    \hline 
\endhead
    \hline 
\endfoot
\hline 
\endlastfoot
20161022-023854 & 54.00 & $1.94 \times 10^{-06}$ & 0.22 & -43.8 & CAMO, 1* \\
20160906-065905 & 62.40 & $5.69 \times 10^{-05}$ & 0.08 & -39.2 & CAMO, 1* \\
20161105-084501 & 65.20 & $1.31 \times 10^{-06}$ & 0.13 & -57.4 & CAMO, 1* \\
20160902-074008 & 68.90 & $8.10 \times 10^{-07}$ & 0.30 &  36.2 & CAMO, 1* \\
20181015-022016 & 13.03 & $3.71 \times 10^{-05}$ & 0.12 & -26.4 & Radar, 2* \\
20190104-104246 & 15.43 & $1.49 \times 10^{-05}$ & 0.27 &   5.4 & Radar, 2* \\
20180811-035811 & 16.31 & $4.71 \times 10^{-06}$ & 0.33 &  -7.6 & Radar, 2* \\
20170826-042315 & 16.80 & $7.35 \times 10^{-06}$ & 0.95 &  64.4 & Radar, 2* \\
20180718-054557 & 17.30 & $4.24 \times 10^{-06}$ & 0.26 & -56.4 & Radar, 2* \\
20170827-013652 & 17.96 & $4.23 \times 10^{-06}$ & 0.69 &  39.2 & Radar, 2* \\
20190429-022947 & 18.09 & $2.99 \times 10^{-06}$ & 0.27 & -71.6 & Radar, 2* \\
20190524-040736 & 18.25 & $4.29 \times 10^{-06}$ & 0.36 & -17.5 & Radar, 2* \\
20190325-033020 & 18.49 & $1.35 \times 10^{-05}$ & 0.29 & -20.5 & Radar, 2* \\
20180709-052400 & 18.52 & $8.13 \times 10^{-06}$ & 0.36 &  -5.9 & Radar, 2* \\
20180716-030941 & 18.73 & $3.11 \times 10^{-06}$ & 0.35 & -35.1 & Radar, 2* \\
20180707-025102 & 19.02 & $5.89 \times 10^{-06}$ & 0.70 &  40.2 & Radar, 2* \\
20180811-071420 & 19.31 & $6.83 \times 10^{-07}$ & 0.46 & -54.0 & Radar, 2* \\
20190521-023120 & 19.45 & $3.54 \times 10^{-06}$ & 0.40 & -21.7 & Radar, 2* \\
20181231-031707 & 19.61 & $1.36 \times 10^{-06}$ & 0.52 & -17.8 & Radar, 2* \\
20180712-054519 & 19.64 & $5.23 \times 10^{-06}$ & 0.58 &  22.4 & Radar, 2* \\
20180617-025412 & 19.71 & $2.52 \times 10^{-06}$ & 0.51 &  -4.9 & Radar, 2* \\
20180715-035420 & 19.75 & $9.55 \times 10^{-06}$ & 0.75 &  46.9 & Radar, 2* \\
20180719-042518 & 19.90 & $1.20 \times 10^{-06}$ & 0.53 & -22.0 & Radar, 2* \\
20190527-055059 & 26.60 & $1.10 \times 10^{-06}$ & 1.14 &  30.9 & Radar, 2* \\
20190531-075115 & 35.91 & $2.67 \times 10^{-07}$ & 1.44 &  31.1 & Radar, 2* \\
20170703-080601 & 38.38 & $1.18 \times 10^{-06}$ & 0.85 &  23.4 & Radar, 2* \\
20170728-025249 & 27.80 & $3.50 \times 10^{-07}$ & 1.00 &  -3.8 & Radar, new \\
20180709-064514 & 65.30 & $1.70 \times 10^{-07}$ & 0.40 &  17.9 & Radar, new \\
20180711-053305 & 26.90 & $8.00 \times 10^{-07}$ & 0.53 & -60.4 & Radar, new \\
20180718-030228 & 24.40 & $3.00 \times 10^{-06}$ & 0.56 &  -7.7 & Radar, new \\
20180810-042251 & 21.80 & $1.90 \times 10^{-06}$ & 0.70 &  10.6 & Radar, new \\
20180811-031940 & 37.10 & $6.20 \times 10^{-07}$ & 0.46 & -71.4 & Radar, new \\
20180901-015852 & 30.10 & $5.00 \times 10^{-07}$ & 1.22 &  22.5 & Radar, new \\
20190104-052611 & 57.00 & $5.00 \times 10^{-06}$ & 0.23 &   3.5 & Radar, new \\
20190611-074811 & 39.40 & $1.00 \times 10^{-06}$ & 0.91 &  27.8 & Radar, new \\
20190727-061604 & 41.30 & $7.50 \times 10^{-07}$ & 0.47 & -41.0 & Radar, new \\
20190801-050153 & 40.60 & $8.50 \times 10^{-07}$ & 1.18 &  44.2 & Radar, new \\
20190801-054542 & 41.30 & $1.50 \times 10^{-06}$ & 0.30 & -86.8 & Radar, new \\
20191023-084502 & 64.76 & $6.04 \times 10^{-07}$ & 0.59 &  57.3 & Orionids, this work \\
20191028-052806 & 64.80 & $5.54 \times 10^{-07}$ & 0.23 & -11.7 & Orionids, this work \\
20191023-084916 & 65.82 & $2.20 \times 10^{-07}$ & 0.20 & -51.2 & Orionids, this work \\
    \label{tab:tau_list}
\end{longtable}

\subsection{Modelling procedure}

The ablation modelling of observed meteors was done manually, starting with approximate physical values for the meteoroid. The initial mass is first computed from the photometry using a fixed dimensionless luminous efficiency of $0.7\%$ as a starting value \citep{vida2018modelling}. The initial bulk density and the ablation coefficient are taken from the literature for the given orbital type of meteoroid \citep{ceplecha1998meteor} but are adjusted during the modelling procedure. The entry angle is kept fixed at the observed value which is computed using Eq. \ref{eq:zc_sim}.

The nine variables that are estimated during the modelling procedure are:
\begin{itemize}
    \item initial velocity $v_0$,
    \item initial meteoroid mass $m_0$,
    \item meteoroid bulk density $\rho_m$,
    \item ablation coefficient $\sigma$,
    \item the height of the beginning of erosion $h_e$,
    \item erosion coefficient $\eta$,
    \item grain mass distribution index $s$,
    \item smallest grain mass $m_l$,
    \item largest grain mass $m_u$.
\end{itemize}

For the Orionids, additional parameters had to be used to simulate the leading fragment which was a common trait found for all events. At the height $h_{e2}$ the erosion coefficient was changed to $\eta_2 = 0$ for most cases. This makes the meteoroid stop releasing grains and start acting like a single body again, reproducing the leading fragment. In some cases, at the same height $h_{e2}$ the ablation coefficient had to be decreased to $\sigma_2$ and the bulk density increased to $\rho_2$.

As high-resolution observations of the meteor morphology and wake are available to us, we can determine if the initial rise in the meteor's light curve is driven by a high ablation coefficient or the beginning of erosion. As a short wake is always immediately visible at the moment when the tracked cameras lock on the meteor, we model all initial rises in the light curve as caused by erosion. We adjust the ablation coefficient so that the simulated light curve prior to when the meteor is first observed is always below the instrumental sensitivity threshold. The ablation coefficient and the density are later refined to also match the observed deceleration. The initial velocity is also adjusted to match the dynamics; in practice, it is always at least \SI{100}{\metre \per \second} higher than the value measured at the beginning of the trajectory \citep{vida2018modelling}. This means we always detect significant deceleration for all events, as evident by the lag measurements. The begin height of erosion is usually set to around the meteor's observed begin height. The erosion coefficient is set to best match the initial rise in the light curve. 

Table \ref{tab:model_qualitative_change} gives qualitative descriptions of how the simulated light curve and deceleration respond upon changing a given model parameter. For example, a higher value of the erosion coefficient means that the main mass will be exhausted more quickly and thus the meteor will not penetrate as deep in the atmosphere. The sharpness of the rise is also regulated by the amount of smaller grains ejected as these ablate faster than larger grains and thus cause a quicker rise in brightness. This can either be achieved by changing the lower limit of the grain mass or increasing the grain mass index if the range of grain masses spans at least an order of magnitude. 

The total duration of the meteor and the observed deceleration are usually tightly linked to the mass of the largest grains and/or the mass of the leading fragment. If the grain masses are kept fixed and the mass index is adjusted, higher values usually mean quicker brightening and lower values mean a slower rise in the light curve. If the simulated light curve is too bright and the meteoroid is not decelerating enough, this is often an indication that larger grains need to be used and the erosion coefficient should be increased.

Once a satisfactory fit of the observed magnitude and dynamics is achieved, we check if the simulated wake matches the observations. The observed wake profile directly informs the mass distribution of grains at a given point in time. In a majority of cases, some minor tweaking of the model parameters is needed to match the wake at this final stage. This generally included fine-tuning the grain mass size range (most often within factors of 2-3) and the mass index (within 10\%).

For the case of the Orionids, the fitting procedure was simplified as the fitted physical parameters on the first few events were found to be applicable to subsequent events and required only minor adjustments to achieve a close match to observations. In most cases, only the mass, the initial velocity, and the grain distribution had to be adjusted. 

For all Orionids, we assumed that the erosion was severely reduced or had ceased completely when the meteor reached peak brightness, a process which we found accurately reproduced the observed leading fragment and residual wake. If a leading fragment was prominent and observed for a long period of time, matching its exact behaviour was more difficult. The exact moment when the erosion changed or stopped had to be precisely adjusted to achieve a leading fragment mass which resulted in the observed deceleration and brightness.

\begin{landscape}
\begin{table}
    \caption{Qualitative description of the most impactful way the modelled light curve and dynamics change upon a slight change in the given parameter, keeping all others fixed. ``LC'' is shorthand for light curve.}
    {
    \begin{tabular}{l l l l}
    \hline\hline 
    Parameter & Name & Increase & Decrease \\
    \hline 
    $v_0$ [$\mathrm{m~s^{-1}}$] & Initial velocity & Tilts simulated lag right. & Tilts simulated lag left. \\
    $m_0$ [kg] & Initial meteoroid mass & LC brighter, less deceleration, lower end height. & LC fainter, more deceleration, higher end height. \\
    $\rho_m$ [$\mathrm{kg~m^{-3}}$] & Meteoroid bulk density & Heights shift down, less deceleration. & Heights shift up, more deceleration. \\
    $\sigma$ [$\mathrm{kg~{MJ}^{-1}}$] & Ablation coefficient & Heights shift up, more deceleration, shorter wake. & Heights shift down, less deceleration, longer wake. \\
    $h_e$ [m] & Height of erosion start & Erosion starts higher, longer duration, fainter peak. & Erosion starts lower, shorter duration, brighter peak. \\
    $\eta$ [$\mathrm{kg~{MJ}^{-1}}$] & Erosion coefficient & Earlier peak, shorter duration, more deceleration, longer wake. & Later peak, longer duration, less deceleration, shorter wake. \\
    $s$ & Grain mass index & Earlier and sharper peak, longer wake. & Later and flatter peak, shorter wake. \\
    $m_l$ [kg] & Smallest grain mass & Later and flatter peak, longer duration. & Earlier and sharper peak, shorter duration. \\
    $m_u$ [kg] & Largest grain mass & Flatter LC, longer duration. & Earlier and sharper peak, shorter duration. \\
    $m_u - m_l$ & Grain mass range & Longer wake. & Shorter wake. \\
    \hline 
\end{tabular}
    }
    \label{tab:model_qualitative_change}
\end{table}
\end{landscape}

\subsection{Automated parameter refinement}

For this study, all fits were done manually by only one analyst (DV). Ideally, a manual fit would not be necessary at all, we would be able to show that fits are unique and also quantify uncertainty in fits. Automated approaches have found success in modelling fireballs with discrete points of fragmentation and fixed ablation parameters \citep{henych2023semi}; however they still require human supervision and a good starting point for the solution. High-quality observations and accurate dynamics are critical for accurate model inversion, despite the complexity of the fitting algorithm and computational power now available \citep{tarano2019inference}.

For fainter meteors, \cite{kambulow2022inverting} attempted to use machine learning to directly invert model parameters without any forward modelling. However, they found a degeneracy in the single-body ablation equations between the ablation coefficient and the bulk density, where a set of identical light curves and dynamics can be produced using a set of different values of the two parameters. The degeneracy appears to only hold for certain classes of objects (cometary objects appear to be less affected), requiring that at least one of the parameters be independently constrained. A similar finding was noted by \cite{pecina1983new} who developed a semi-analytical method for single-body fits.

\cite{kikwaya2011bulk} successfully applied a fully automated fitting procedure to video meteor observations through a full grid search across the whole parameter space. However, \cite{kikwaya2011bulk} had observations with significantly lesser accuracy than those used in our study, allowing for a greater range of possible model parameters. We attempted a similar approach on CAMO data using a classical grid search, local optimization algorithms, and Particle Swarm Optimization \citep{kennedy1995particle}, but we were unsuccessful in obtaining even an initial estimate of model parameters. Despite running the PSO algorithm for CPU days per individual event, all attempts to produce automated fits failed. We believe that the algorithm gets stuck in local minima as the model is nonlinear, despite using thousands of PSO particles scattered across the parameter space. As a result, while we cannot rigorously prove that our fits are true absolute minima. However, the fact that all events show excellent lightcurve, lag and wake agreement between the model and observations using similar physical model values increases our confidence in the resulting fits.

Another problem with automated fits is constructing a reliable cost function. Including all available observations is not straightforward due to complex meteor morphology. For example, as described in \cite{vida2021high}, some meteors have a break in the dynamics when the reference point that is tracked changes, e.g. when a leading fragment emerges which was not previously visible. Prior to the emergence of the leading fragment, there is no way to make consistent position picks other than computing a centroid of the bright meteor and trying to exclude the wake. But this involves centroiding on a large area many pixels (or often tens of pixels) in size.

Another challenge is how to include the wake in the cost function, which requires significantly more computation time and requires positional alignment within the model to match the changing observed reference point. In addition, it is not clear a priori how to optimally weight the contributions from the light curve, dynamics, and the wake and collapse the dimensionality into a single number that an optimizer can minimize.

With these limitations in mind, we adopted a semi-automated approach whereby the user provides a starting point for the model parameters and the optimizer then iteratively refines the parameters. Once convergence is reached, the user can then inspect the results and make adjustments. In our approach, only a subset of the parameters (5-6 at most) are fit at any time while the rest are kept fixed. A Nelder-Mead optimizer is used \citep{nelder1965simplex} with manually set bounds on each parameter. Different sets of parameters are cycled so that all 11 are eventually fit, but not all at the same time. We found that the optimizer fails to find a good gradient if all parameters are used at once.

For the cost function, we follow the modified approach of \cite{henych2023semi} where we sum up the contribution of the differences in the light curve and the length, scaling them according to their measurement error:

\begin{align}
    d_{len} = \sum \frac{|L_{obs} - L_{model}|}{n_{L} \sigma^2_{L}} \,, \\
    d_{lc}  = \sum \frac{|M_{obs} - M_{model}|}{n_{M} \sigma^2_{M}} \,, \\
    d = d_{len} + d_{lc} \,, \\
\end{align}

\noindent where $L_{obs}$ and $L_{model}$ are the observed and modelled length (measured from the height of the first observed point on the trajectory), and $M_{obs}$ and $M_{model}$ is the absolute magnitude. $\sigma^2_{L}$ and $\sigma^2_{M}$ represent the expected scatter in the observations, and $n_L$ and $n_M$ is the number of length and magnitude observations. We use the sum of absolute differences (also called the Manhattan distance or the $L^1$ norm) to lessen the effect of outliers as compared to the classical $\chi^2$ approach. For a perfect fit and perfectly chosen values of $\sigma^2_{L}$ and $\sigma^2_{M}$, the cost function value would be $d = 2$. The wake is only checked qualitatively after the automated fit but we found it generally matches well if the fit on the light curve and dynamics is good, though in some cases further adjustments were made to better match the wake.

Through trial and error, we found reduced sets of parameters that can be optimized separately and work well in refining a model fit. The first set of parameters, the initial mass and velocity, are the crudest parameters to fit - the initial mass adjusts the overall brightness of the light curve and the velocity of the slope of length vs. time. The second set includes, in addition to the previous two, the bulk density, the height of erosion onset, the erosion coefficient, and the grain mass index. We found that this set of parameters improves the overall quality of the fit on both the light curve and the dynamics. The third set includes the erosion coefficient, the grain mass index and the grain mass range. This set of parameters generally improves the wake, even though the algorithm does not see it directly, but this can be checked by the analyst. The sets of parameters are summarized in Table \ref{tab:auto_fit_sets}. The algorithm runs these sets in sequence, one after the other. The bounds can either be absolute or relative to the initial value, are appropriate to mm-sized meteoroids and are physically informed.

\begin{table}
  \caption{Sets of fit parameters used for automated fitting in the three-step fit procedure (see text for details and variable definitions). The relative boundary allows the parameters to vary within a range defined by the nominal value as [lower $\times$ nominal, upper $\times$ nominal]. The absolute value puts a hard limit on the parameters in the [lower, upper] range.}
  \centering
  {
  \begin{tabular}{l l l l}
  \hline\hline 
  Parameter & Bound & Lower & Upper \\
  \hline 
  \multicolumn{4}{c}{Setting the main components} \\
  $m_0$ & Relative & 0.5 & 2.0 \\
  $v_0$ & Relative & 0.98 & 1.02 \\
  \hline 
  \multicolumn{4}{c}{Overall fit} \\
  $m_0$    & Relative & 0.5 & 2.0 \\
  $v_0$    & Relative & 0.90 & 1.10 \\
  $\rho_m$ & Relative & 0.8 & 1.2 \\
  $h_e$    & Relative & 0.95 & 1.05 \\
  $\eta$   & Relative & 0.5 & 2.0 \\
  $s$      & Absolute & 1.5 & 3.0 \\
  \hline 
  \multicolumn{4}{c}{Fine tuning of erosion} \\
  $\eta$   & Relative & 0.75 & 1.25 \\
  $s$      & Absolute & 1.5 & 3.0 \\
  $m_l$    & Relative & 0.1 & 10.0 \\
  $m_u$    & Relative & 0.1 & 10.0 \\
  \hline 
\end{tabular}
  }
  \label{tab:auto_fit_sets}
\end{table}

\section{Results}

\subsection{Trajectories and radiants}

Table \ref{tab:orionids_orbits} lists the measured orbits of the observed Orionids. The radiants match well to simulated radiant positions produced by \cite{egal2020modeling}, who successfully modelled the past activity of both the Orionids and $\eta$-Aquariids. In 2019, all meteors were observed at solar longitudes $>209^{\circ}$, and in 2020 at solar longitudes $\lesssim 204$, meaning that the observing periods in the two years did not overlap.

\begin{figure}
  \includegraphics[width=\linewidth]{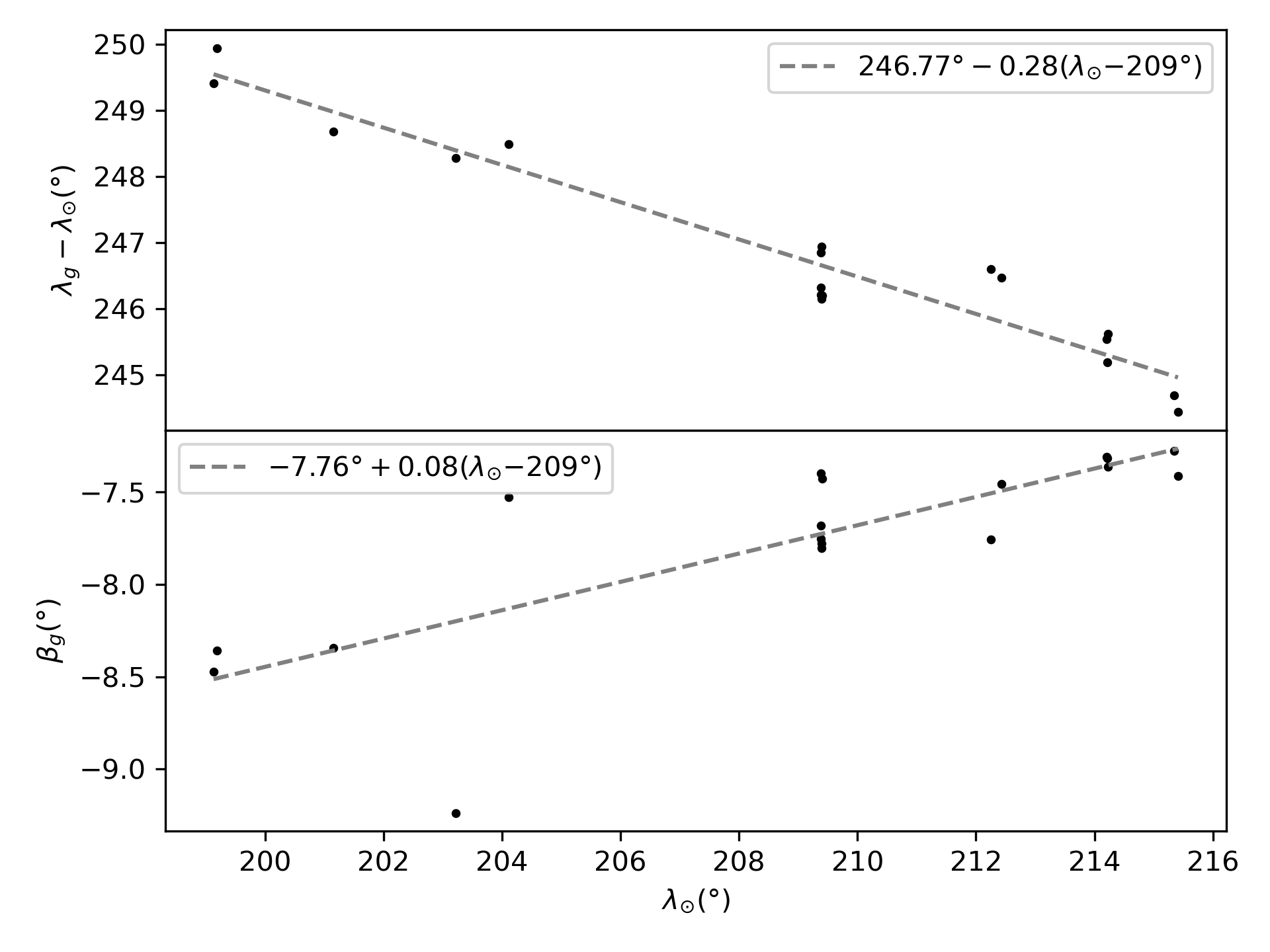}
  \caption{Radiant drift of the CAMO-observed Orionids.}
  \label{fig:radiant_drift_fit}
\end{figure}

Figure \ref{fig:radiant_drift_fit} shows the observed radiant drift for our Orionids in sun-centred ecliptic coordinates. The reference solar longitude was fixed on $209^{\circ}$ which corresponds to the annual peak of activity \citep{vida2022flux, Egal2020observations}. We measured a mean Sun-centered geocentric radiant of $\lambda_G - \lambda_{\astrosun} = 246.765^{\circ} \pm 0.099^{\circ}$, $\beta_G = -7.758^{\circ} \pm 0.078^{\circ}$ at the activity peak and a radiant drift $\frac{d(\lambda_G - \lambda_{\astrosun})}{d\lambda_{\astrosun}} = -0.282 \pm 0.019$, $\frac{d\beta_G}{d\lambda_{\astrosun}} = 0.077 \pm 0.015$. 

\begin{figure}
  \includegraphics[width=\linewidth]{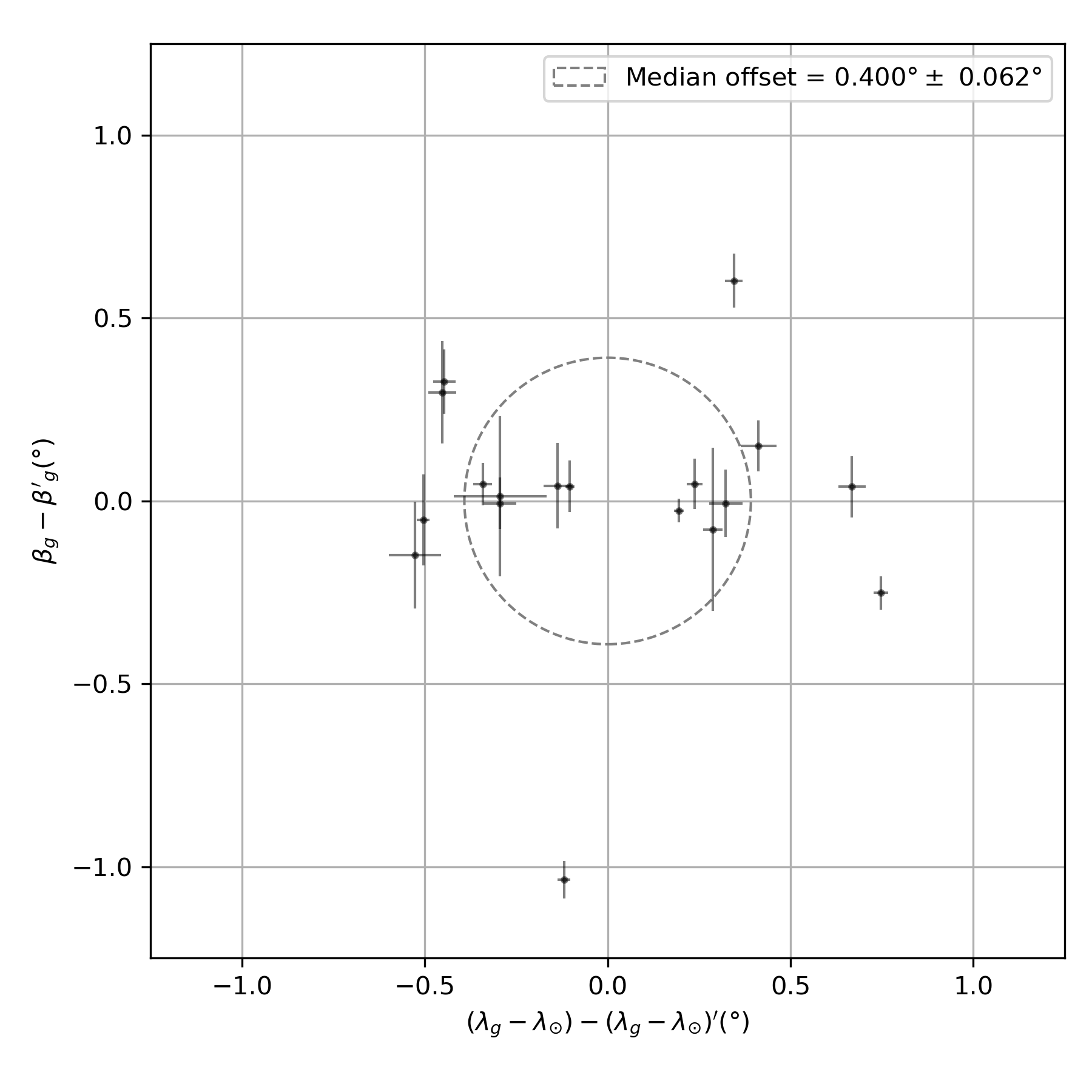}
  \caption{Offsets for each drift-corrected radiant from the mean radiant. The dashed circle shows the radius of the median offset.}
  \label{fig:radiant_offsets}
\end{figure}

The drift-corrected radiants, together with their individually determined uncertainties following the procedure of \citet{vida2019meteortheory} are shown in Figure \ref{fig:radiant_offsets}. Using the method of \cite{moorhead2021meteor}, a median offset from the mean radiant of $0.400^{\circ} \pm 0.062^{\circ}$ was measured using a Rayleigh distribution fit. Only one outlier is present: the Orionid which occurred on 2020-10-16 10:06:10 was over a degree from the mean radiant. This event is also an outlier because it is the only meteor with a second brightness peak during flight at a height of $\sim99.4$~km which had to be modelled using a discrete disruption. Otherwise, it has similar physical properties to other Orionids.

Compared to the radiant dispersion measured in \cite{moorhead2021meteor}, our Orionid dispersion is about $0.12^{\circ}$ smaller. However, due to the small sample size and the fact that the difference is within the $2\sigma$ uncertainty, we do not consider the difference to be statistically significant.


\begin{landscape}
\begin{longtable}{r l r r r r r r r r r r}
\caption{Radiants and orbits of the Orionids observed by CAMO. Rows below every entry list $1\sigma$ uncertainties. The uncertainties only state the measurement precision and not the total accuracy. The geocentric velocity represents the model-corrected speed accounting for early deceleration.} \label{tab:orionids_orbits} \\
	\hline\hline 
	
Num & 	Date and time (UTC) & $\lambda_{\astrosun}$ & $\alpha_g$ & $\delta_g$ & $v_g$ & $a$ & $e$ & $q$ & $\omega$ & $i$ & $\pi$ \\
&	            & (deg) & (deg) & (deg) & (\SI{}{\kilo \metre \per \second}) & (AU) &  & (AU) & (deg) & (deg) & (deg) \\
\hline
\endfirsthead
\caption{continued.} \\
	\hline\hline 
	
Num &	Date and time (UTC) & $\lambda_{\astrosun}$ & $\alpha_g$ & $\delta_g$ & $v_g$ & $a$ & $e$ & $q$ & $\omega$ & $i$ & $\pi$ \\
&	            & (deg) & (deg) & (deg) & (\SI{}{\kilo \metre \per \second}) & (AU) &  & (AU) & (deg) & (deg) & (deg) \\
\hline
\endhead 
\hline
\endfoot
 1 & 2019-10-23 08:45:02 & 209.3781 & 96.410 & 15.544 & 67.020 & 33.629 & 0.9824 & 0.5915 & 79.54 & 163.80 & 108.91 \\
  &   &   & 0.013 & 0.033 & 0.014 & 0.884 & 0.0008 & 0.00023 & 0.01 & 0.07 & 0.01 \\
 2 & 2019-10-23 08:49:16 & 209.3811 & 95.768 & 15.923 & 67.172 & -733.299 & 1.0008 & 0.5784 & 80.62 & 164.41 & 110.00 \\
  &   &   & 0.028 & 0.089 & 0.049 & 8.596 & 0.0026 & 0.00039 & 0.06 & 0.19 & 0.06 \\
 3 & 2019-10-23 08:54:37 & 209.3848 & 95.867 & 15.639 & 66.728 & 25.371 & 0.9774 & 0.5742 & 81.70 & 163.76 & 111.08 \\
  &   &   & 0.024 & 0.059 & 0.027 & 0.335 & 0.0015 & 0.00066 & 0.08 & 0.12 & 0.08 \\
 4 & 2019-10-23 09:12:25 & 209.3971 & 96.517 & 15.489 & 67.114 & 42.345 & 0.9859 & 0.5953 & 79.00 & 163.74 & 108.40 \\
  &   &   & 0.021 & 0.224 & 0.072 & 1.019 & 0.0029 & 0.00049 & 0.06 & 0.47 & 0.06 \\
 5 & 2019-10-23 09:13:10 & 209.3976 & 95.705 & 15.548 & 66.798 & 37.409 & 0.9847 & 0.5724 & 81.73 & 163.54 & 111.12 \\
  &   &   & 0.015 & 0.125 & 0.044 & 0.655 & 0.0018 & 0.00048 & 0.07 & 0.27 & 0.07 \\
 6 & 2019-10-23 09:29:14 & 209.4087 & 95.782 & 15.896 & 66.760 & 28.485 & 0.9800 & 0.5710 & 82.01 & 164.27 & 111.42 \\
  &   &   & 0.034 & 0.142 & 0.046 & 0.894 & 0.0023 & 0.00068 & 0.03 & 0.29 & 0.03 \\
 7 & 2019-10-26 06:06:16 & 212.2575 & 99.100 & 15.398 & 66.671 & 17.721 & 0.9674 & 0.5784 & 81.41 & 163.64 & 113.67 \\
  &   &   & 0.017 & 0.047 & 0.010 & 0.155 & 0.0007 & 0.00050 & 0.04 & 0.09 & 0.04 \\
 8 & 2019-10-26 10:16:08 & 212.4306 & 99.166 & 15.698 & 66.719 & 19.536 & 0.9706 & 0.5748 & 81.75 & 164.24 & 114.18 \\
  &   &   & 0.034 & 0.086 & 0.053 & 1.406 & 0.0029 & 0.00055 & 0.09 & 0.18 & 0.09 \\
 9 & 2019-10-28 05:06:17 & 214.2116 & 100.049 & 15.786 & 66.497 & 22.845 & 0.9760 & 0.5492 & 84.59 & 164.27 & 118.80 \\
  &   &   & 0.018 & 0.070 & 0.031 & 0.472 & 0.0015 & 0.00083 & 0.12 & 0.14 & 0.12 \\
10 & 2019-10-28 05:18:34 & 214.2201 & 99.703 & 15.802 & 66.582 & 40.718 & 0.9867 & 0.5434 & 84.98 & 164.19 & 119.20 \\
  &   &   & 0.009 & 0.071 & 0.000 & 0.176 & 0.0001 & 0.00062 & 0.07 & 0.15 & 0.07 \\
11 & 2019-10-28 05:28:06 & 214.2267 & 100.143 & 15.727 & 66.410 & 18.589 & 0.9704 & 0.5496 & 84.70 & 164.16 & 118.92 \\
  &   &   & 0.040 & 0.095 & 0.033 & 0.506 & 0.0020 & 0.00088 & 0.06 & 0.19 & 0.06 \\
12 & 2019-10-29 08:21:22 & 215.3456 & 100.341 & 15.799 & 66.224 & 24.423 & 0.9785 & 0.5253 & 87.30 & 164.07 & 122.64 \\
  &   &   & 0.041 & 0.074 & 0.039 & 0.944 & 0.0023 & 0.00062 & 0.03 & 0.15 & 0.03 \\
13 & 2019-10-29 09:55:41 & 215.4110 & 100.140 & 15.676 & 65.973 & 18.697 & 0.9724 & 0.5160 & 88.56 & 163.65 & 123.97 \\
  &   &   & 0.063 & 0.151 & 0.066 & 0.682 & 0.0036 & 0.00103 & 0.03 & 0.32 & 0.03 \\
14 & 2020-10-12 06:58:50 & 199.1292 & 88.506 & 14.958 & 67.430 & 20.021 & 0.9669 & 0.6624 & 71.58 & 163.05 & 90.71 \\
  &   &   & 0.041 & 0.116 & 0.023 & 0.503 & 0.0013 & 0.00173 & 0.24 & 0.22 & 0.24 \\
15 & 2020-10-12 08:17:16 & 199.1831 & 89.108 & 15.075 & 67.366 & 13.860 & 0.9515 & 0.6726 & 70.66 & 163.36 & 89.84 \\
  &   &   & 0.050 & 0.070 & 0.029 & 0.235 & 0.0019 & 0.00116 & 0.13 & 0.13 & 0.13 \\
16 & 2020-10-14 07:58:50 & 201.1513 & 89.831 & 15.092 & 66.974 & 13.572 & 0.9531 & 0.6371 & 74.95 & 163.03 & 96.10 \\
  &   &   & 0.130 & 0.219 & 0.057 & 0.343 & 0.0043 & 0.00296 & 0.26 & 0.42 & 0.26 \\
17 & 2020-10-16 10:06:10 & 203.2223 & 91.525 & 14.194 & 67.240 & 33.143 & 0.9808 & 0.6363 & 74.35 & 161.24 & 97.58 \\
  &   &   & 0.018 & 0.051 & 0.060 & 1.067 & 0.0033 & 0.00096 & 0.18 & 0.10 & 0.18 \\
18 & 2020-10-17 07:40:07 & 204.1141 & 92.682 & 15.884 & 67.257 & 18.459 & 0.9657 & 0.6323 & 75.17 & 164.66 & 99.28 \\
  &   &   & 0.024 & 0.074 & 0.026 & 0.290 & 0.0015 & 0.00045 & 0.03 & 0.15 & 0.03 \\
\end{longtable}
\end{landscape}

\begin{figure}
  \includegraphics[width=\linewidth]{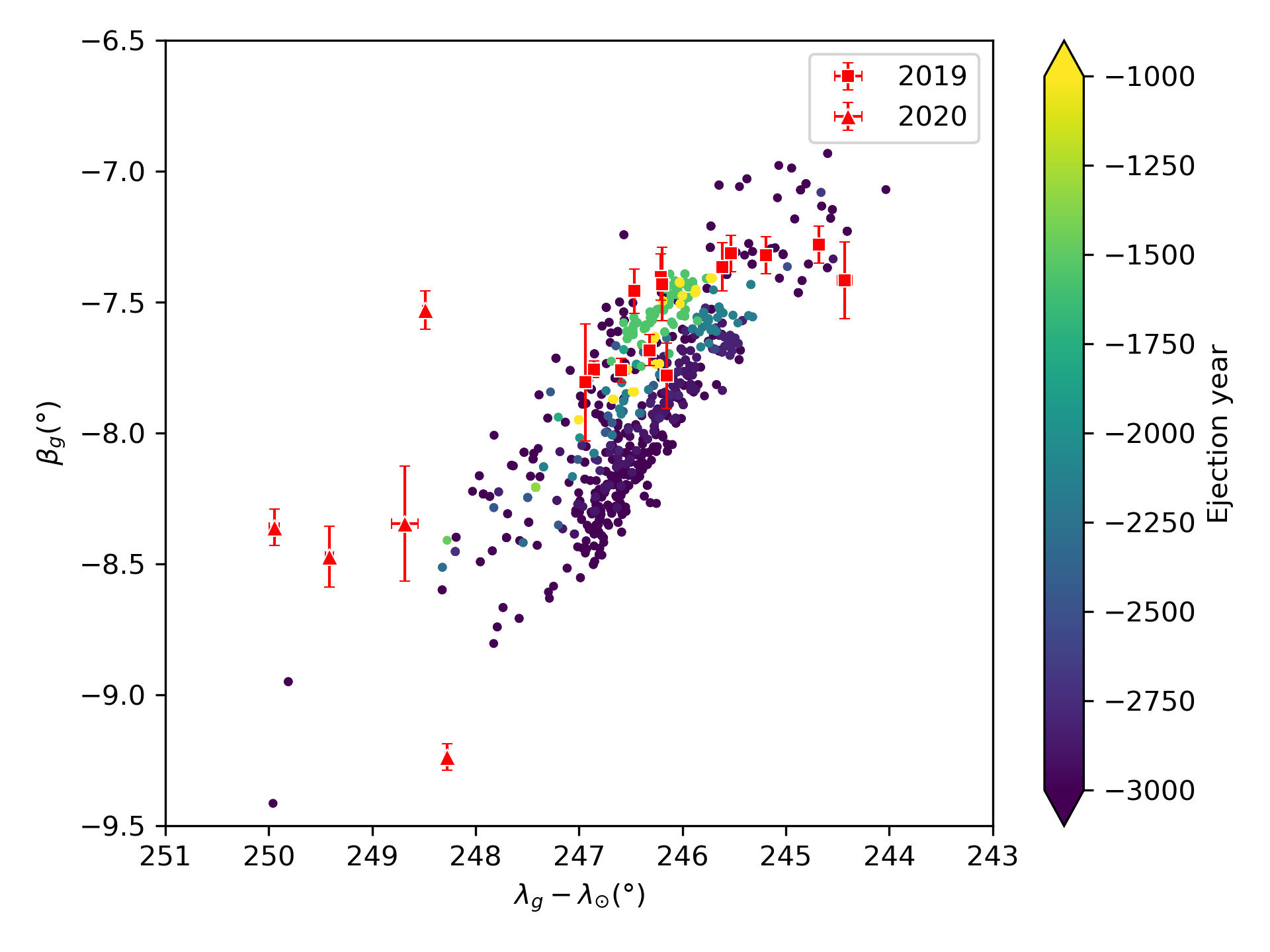}
  \caption{Comparison of observed radiants to simulated radiants by \cite{egal2020modeling}. Red squares mark meteors observed in 2019 and red triangles meteors observed in 2020. The simulated radiants are shown as circles color-coded by ejection year.}
  \label{fig:radiants_obs_vs_sims}
\end{figure}

Figure \ref{fig:radiants_obs_vs_sims} shows a comparison between the observed radiants and the radiant locations simulated by \cite{egal2020modeling}. All meteors in 2020 were observed during the beginning of the shower activity window, some 5 days before the peak, at $\lambda_{\astrosun} \lesssim 204^{\circ}$, and had radiants in locations which were sparsely populated with simulated radiants predicted by \cite{egal2020modeling}. Their locations best correspond to having been ejected from 1P/Halley more than 3000 years in the past. The radiants observed in 2019, near the peak of the shower, better correspond to more recent ejections which occurred within the last 1500 years. The meteors occurring after the peak also have radiants corresponding to modelled ejections more than 3000 years ago.

\subsection{Ablation modelling applied to the Orionids} \label{subsec:modelling_results}

We applied our numerical erosion model, described earlier, to each observed Orionid. A series of plots comparing observations to model fits are given in \ref{app:obs_sim_comp}. We find an excellent match to the light curve, dynamics, and wake for each event. In only a handful of cases, the wake match was not optimal for one part of the trajectory while it was good for another, potentially pointing to a change in the grain mass distribution, which was not modelled. 

\begin{figure}
  \includegraphics[width=\linewidth]{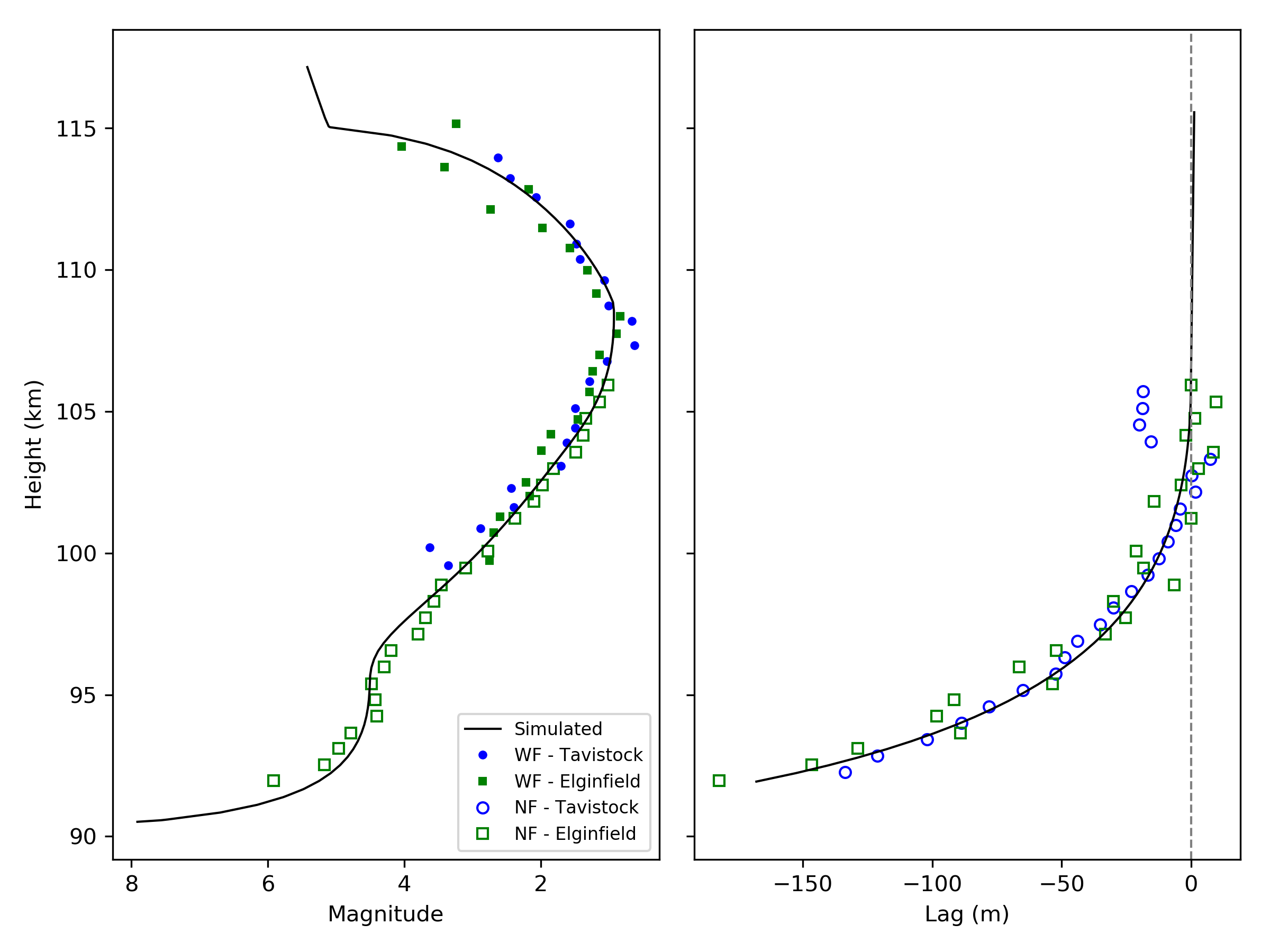}
  \caption{Comparison between the observed and modelled light curve and lag for the 2019-10-23 08:49:16 Orionid. WF are the wide-field and NF are the narrow-field measurements. The solid black line is the simulation.}
  \label{fig:ori2_fit}
\end{figure}

\begin{figure}
  \includegraphics[width=\linewidth]{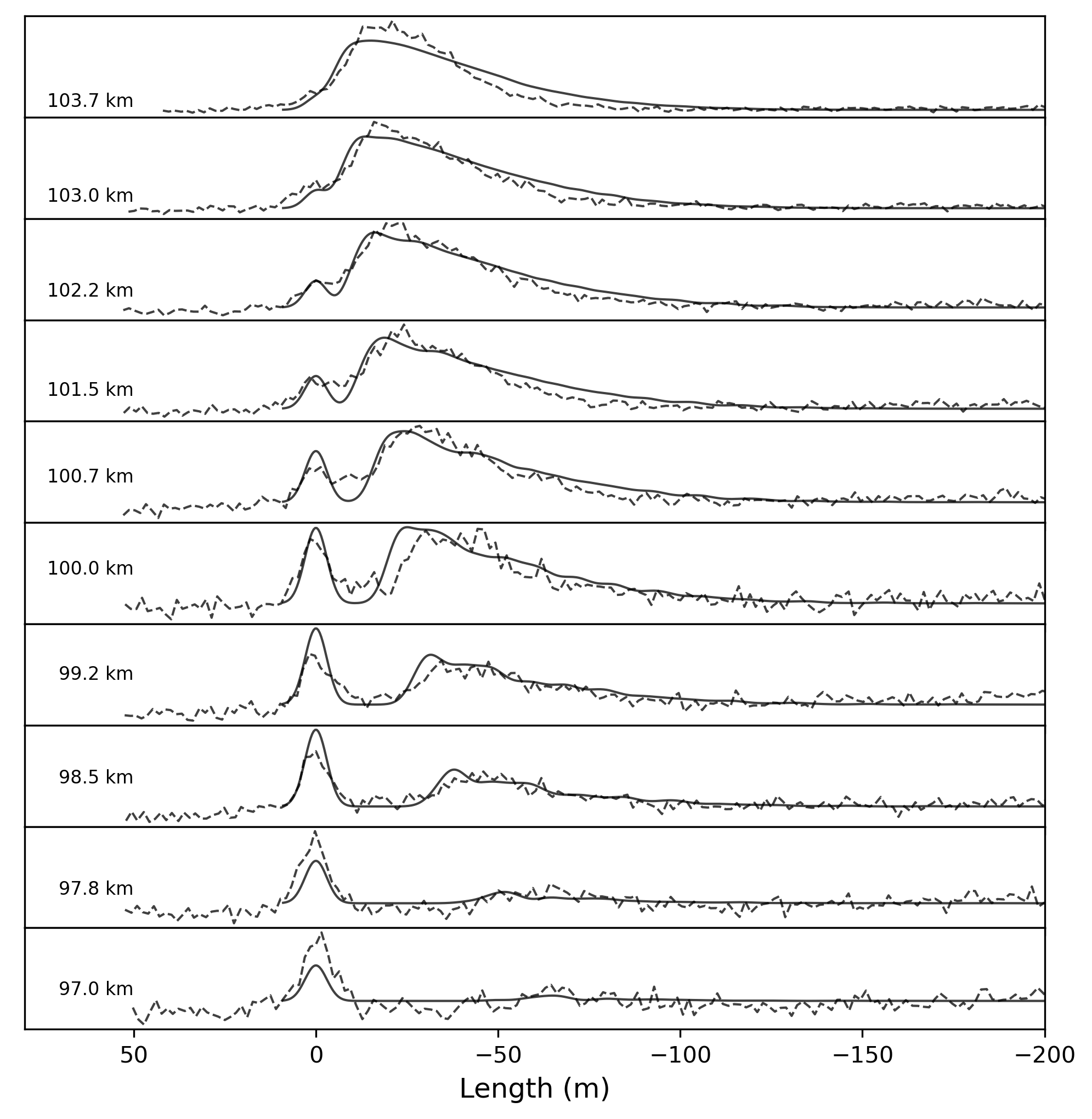}
  \caption{Comparison between the observed (dashed) and modelled (solid) wake for the 2019-10-23 08:49:16 Orionid. Each video frame produces a single measurement of the wake - the inset values per plot show the height at which the frame captured the meteor.  }
  \label{fig:ori2_wake_overview}
\end{figure}

As an example, Figure \ref{fig:ori2_fit} shows a detailed comparison between the observed and modelled light curve and lag and Figure \ref{fig:ori2_wake_overview} shows the wake for the 2019-10-23 08:49:16 Orionid. The model fully explains the light curve, deceleration, and the separation of the leading fragment and its brightness in comparison to the grains forming the wake.

Tables \ref{tab:physical_properties} and \ref{tab:erosion_properties} list the erosion model parameters measured for the Orionids in this work. Table \ref{tab:physical_properties} gives inferred bulk parameters and the observed initial conditions, while Table \ref{tab:erosion_properties} gives the erosion parameters. A bulk density of around \SI{300}{\kilo \gram \per \cubic \metre} was derived for every Orionid. Other bulk densities in increments of \SI{\pm100}{\kilo \gram \per \cubic \metre} were attempted (\SIrange{200}{600}{\kilo \gram \per \cubic \metre}), but they always produced fits of inferior quality. These findings are consistent with the Vega-2 in-situ measurements \citep{krasnopolsky1988properties}.

The uncertainties on individual parameters are not calculated as the model is fit manually and only a subset is used in automated refinement. This makes it difficult to subjectively decide when to stop refining the parameters and call the fit ``optimal''. Nevertheless, we found that the model is very sensitive to changes in the bulk density, initial velocity, initial mass, ablation coefficient, and erosion coefficient. The sensitivity to changes in the grain distribution differs on a case-by-case basis. For example, the upper mass grain limit can be very well constrained for cases when the leading fragment is prominently visible.

\begin{table}
    \caption{The modelled bulk physical properties of the observed Orionids. Here the initial mass is given as $m_0$, the initial velocity $v_0$, the observed zenith angle $Z_c$, the ablation coefficient $\sigma$, the bulk density $\rho$, energy per unit cross section needed to begin erosion $E_S$ and the energy per unit mass section needed to begin erosion $E_V$.}
    {
    \begin{tabular}{r r r r r r r r}
    \hline\hline 
Num & $m_0$ & $v_0$                                   & $Z_c$ & $\sigma$                                 & $\rho$                                       & $E_S$                                            & $E_V$ \\
    & (kg)  & (\SI{}{\kilo \metre \per \second}) & (deg) & (\SI{}{\kilogram \per \mega \joule}) & (\SI{}{\kilo \gram \per \metre \cubed}) & (\SI{}{\mega \joule \per \metre \squared}) & (\SI{}{\mega \joule \per \kilogram}) \\
\hline
 1 & \num{1.3e-05} & 67.932 & 30.604 & 0.030 & 307 & 1.0 & 1.1 \\
 2 & \num{5.3e-06} & 68.082 & 29.657 & 0.030 & 301 & 1.0 & 1.5 \\
 3 & \num{5.7e-06} & 67.644 & 29.786 & 0.025 & 273 & 0.9 & 1.5 \\
 4 & \num{3.4e-06} & 68.025 & 28.859 & 0.030 & 289 & 0.9 & 1.7 \\
 5 & \num{4.5e-06} & 67.714 & 28.579 & 0.030 & 295 & 1.1 & 1.8 \\
 6 & \num{3.6e-06} & 67.676 & 27.725 & 0.030 & 306 & 0.8 & 1.4 \\
 7 & \num{1.0e-05} & 67.588 & 53.154 & 0.030 & 325 & 1.1 & 1.3 \\
 8 & \num{6.8e-06} & 67.635 & 28.448 & 0.030 & 317 & 0.8 & 1.1 \\
 9 & \num{1.2e-05} & 67.416 & 62.343 & 0.030 & 303 & 1.5 & 1.8 \\
10 & \num{9.0e-06} & 67.500 & 60.081 & 0.032 & 300 & 1.4 & 1.9 \\
11 & \num{4.1e-06} & 67.330 & 58.663 & 0.030 & 350 & 1.7 & 2.6 \\
12 & \num{9.2e-06} & 67.147 & 31.649 & 0.032 & 304 & 1.0 & 1.3 \\
13 & \num{8.5e-06} & 66.900 & 27.917 & 0.032 & 300 & 0.9 & 1.2 \\
14 & \num{2.9e-05} & 68.336 & 46.672 & 0.030 & 287 & 1.3 & 1.2 \\
15 & \num{7.3e-06} & 68.274 & 34.845 & 0.032 & 302 & 1.3 & 1.8 \\
16 & \num{4.4e-06} & 67.887 & 36.806 & 0.032 & 298 & 1.3 & 2.2 \\
17 & \num{7.5e-06} & 68.150 & 29.299 & 0.032 & 300 & 1.4 & 1.9 \\
18 & \num{7.1e-06} & 68.166 & 38.897 & 0.032 & 354 & 1.2 & 1.5 \\
    \hline 
\end{tabular}
    }
    \label{tab:physical_properties}
\end{table}

\begin{landscape}
\begin{longtable}{r r r r r r r r r r r r r}
    \caption{The model-inferred erosion properties for all Orionids. The erosion parameters include the height of the erosion beginning $h_e$, the dynamic pressure at erosion beginning $p_{dyn,e}$, the erosion coefficient $\eta$, the upper grain mass limit $m_u$, the lower grain mass limit $m_l$, the grain differential mass index $s$, the height of the erosion change $h_{e2}$, the changed erosion coefficient $\eta_2$, the changed ablation coefficient $\sigma_2$, the changed bulk density $\rho_2$, the total mass loss prior to erosion change $\Delta m_e$ and the mass of the leading fragment $m_2$.}
    \label{tab:erosion_properties} \\
    \hline\hline 
Num & $h_e$ & $p_{dyn,e}$ & $\eta$                                 & $m_u$ & $m_l$ & $s$ & $h_{e2}$ & $\eta_2$                       & $\sigma_2$                                 & $\rho_2$ & $\Delta m_e$ & $m_2$ \\
    & (km)  & (kPa)       & (\SI{}{\kilogram \per \mega \joule}) & (kg)  & (kg)  &     & (km)     & (\SI{}{\kilogram \per \mega \joule}) & (\SI{}{\kilogram \per \mega \joule}) & (\SI{}{\kilo \gram \per \metre \cubed}) & (\%) & (kg) \\
\hline
\endfirsthead
\caption{continued.} \\
	\hline\hline 
	
Num & $h_e$ & $p_{dyn,e}$ & $\eta$                                 & $m_u$ & $m_l$ & $s$ & $h_{e2}$ & $\eta_2$                       & $\sigma_2$                                 & $\rho_2$ & $\Delta m_e$ & $m_2$ \\
    & (km)  & (kPa)       & (\SI{}{\kilogram \per \mega \joule}) & (kg)  & (kg)  &     & (km)     & (\SI{}{\kilogram \per \mega \joule}) & (\SI{}{\kilogram \per \mega \joule}) & (\SI{}{\kilo \gram \per \metre \cubed}) & (\%) & (kg) \\
\endhead 
\hline
\endfoot
 1 & 115.81 & 0.149695 & 0.507 & \num{1e-08}   & \num{3e-11}   & 2.28 & 108.60 & 0.005 & 0.012 &  450 & 93.9 & \num{8e-07} \\
 2 & 115.50 & 0.183304 & 0.445 & \num{9e-09}   & \num{3e-11}   & 2.26 & 108.90 & 0.000 & 0.020 & 1000 & 92.9 & \num{4e-07} \\
 3 & 116.05 & 0.194501 & 0.208 & \num{5e-08}   & \num{1e-11}   & 2.22 & -      & -     & -     & -    & -    & -           \\
 4 & 116.06 & 0.179346 & 0.461 & \num{1e-08}   & \num{1e-11}   & 2.01 & -      & -     & -     & -    & -    & -           \\
 5 & 114.58 & 0.204986 & 0.355 & \num{1e-08}   & \num{1e-11}   & 2.09 & 107.50 & 0.010 & 0.008 &  295 & 95.2 & \num{2e-07} \\
 6 & 117.60 & 0.154318 & 0.241 & \num{5e-08}   & \num{1e-11}   & 2.30 & 106.20 & 0.050 & 0.005 &  306 & 96.9 & \num{1e-07} \\
 7 & 117.95 & 0.143692 & 0.296 & \num{1e-08}   & \num{5e-11}   & 2.15 & 109.00 & 0.015 & 0.012 &  500 & 90.5 & \num{1e-06} \\
 8 & 117.59 & 0.131967 & 1.049 & \num{2e-08}   & \num{2e-11}   & 2.24 & -      & -     & -     & -    & -    & -           \\
 9 & 117.24 & 0.143136 & 0.313 & \num{9e-09}   & \num{2e-11}   & 1.95 & 110.00 & 0.007 & 0.010 &  400 & 91.8 & \num{1e-06} \\
10 & 117.00 & 0.144776 & 0.650 & \num{5e-08}   & \num{1e-11}   & 2.05 & 113.20 & 0.100 & 0.010 &  300 & 85.3 & \num{1e-06} \\
11 & 115.06 & 0.191182 & 1.014 & \num{2e-08}   & \num{1e-11}   & 2.13 & 112.70 & 0.005 & 0.012 & 1200 & 93.6 & \num{3e-07} \\
12 & 115.01 & 0.188881 & 0.373 & \num{1e-08}   & \num{5e-11}   & 2.06 & 107.00 & 0.000 & 0.015 &  304 & 94.2 & \num{5e-07} \\
13 & 116.00 & 0.157103 & 0.250 & \num{8e-09}   & \num{5e-11}   & 2.15 & 104.50 & 0.000 & 0.015 &  400 & 97.8 & \num{2e-07} \\
14 & 115.23 & 0.198162 & 0.428 & \num{8e-08}   & \num{1e-11}   & 2.27 & 108.20 & 0.200 & 0.011 & 1000 & 96.1 & \num{1e-06} \\
15 & 113.59 & 0.246516 & 0.252 & \num{5e-09}   & \num{2e-11}   & 2.20 & 106.20 & 0.015 & 0.015 & 1100 & 95.5 & \num{3e-07} \\
16 & 113.50 & 0.248731 & 0.250 & \num{1e-08}   & \num{5e-11}   & 2.48 & 105.80 & 0.000 & 0.020 & 1000 & 99.1 & \num{4e-08} \\
17 & 112.50 & 0.280846 & 0.100 & \num{1e-08}   & \num{5e-11}   & 2.15 & 103.70 & 0.050 & 0.013 &  300 & 81.6 & \num{1e-06} \\
18 & 114.65 & 0.215728 & 0.206 & \num{1e-08}   & \num{1e-11}   & 2.18 & 105.50 & 0.000 & 0.015 & 1100 & 97.7 & \num{2e-07} \\
\end{longtable}
\end{landscape}

\begin{figure}
  \includegraphics[width=\linewidth]{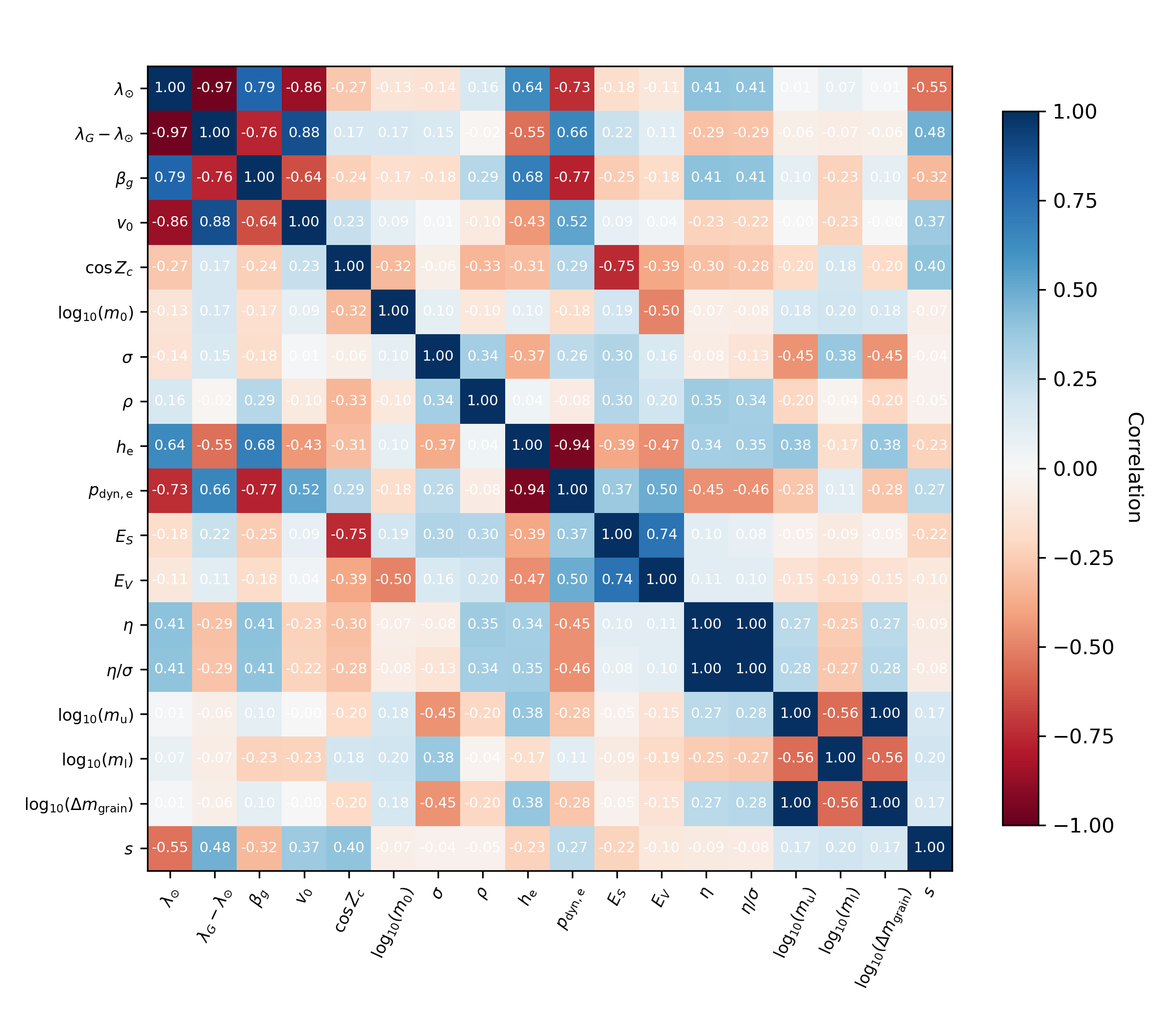}
  \caption{Correlation matrix of all Orionids. Here each value shown on the ordinate is correlated with the variable in the abscissa for all 18 Orionids and the resulting Pearson r-coefficient displayed.}
  \label{fig:corr_matrix_all}
\end{figure}

\begin{figure}
  \includegraphics[width=\linewidth]{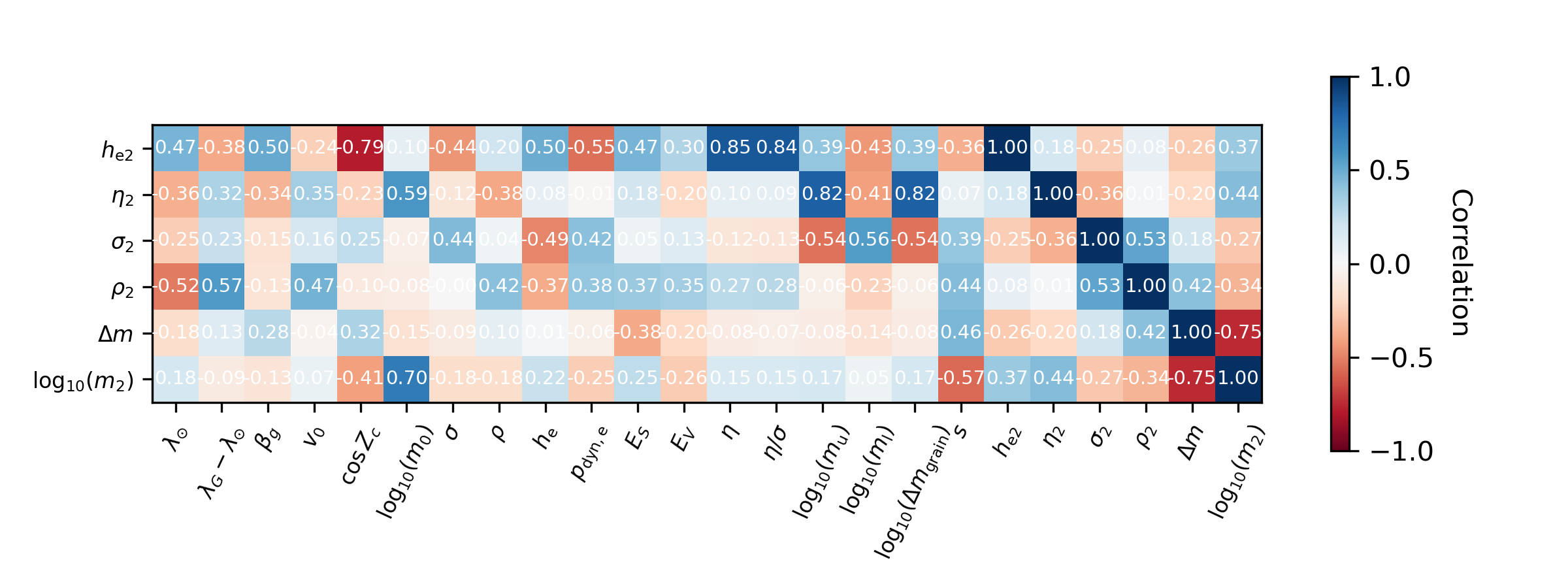}
  \caption{Correlation matrix using only events which had the erosion change.}
  \label{fig:corr_matrix_erosion_change}
\end{figure}

Figures \ref{fig:corr_matrix_all} and \ref{fig:corr_matrix_erosion_change} show correlation matrices of the radiant information, velocity, and inverted physical meteoroid properties for the Orionids. Figure \ref{fig:corr_matrix_all} shows all Orionids, while Figure \ref{fig:corr_matrix_erosion_change} only includes Orionids for which an erosion change was modelled (15/18) and includes additional parameters pertaining to the leading fragment. In addition to the physical parameters listed in Table \ref{tab:erosion_properties}, the correlation matrix also shows the dynamic pressure at the height of erosion ($p_{\mathrm{dyn,e}} = \rho_{\mathrm{air}}(h_e) v^2$), the ratio of the erosion and the ablation coefficient ($\eta/\sigma$), and the logarithm of the difference between the grain masses ($m_u - m_l$).

The observed radiant drift explains the strong correlation between the solar longitude, radiant coordinates, and speed. The height, dynamic pressure at the beginning of erosion, the erosion coefficient and the grain mass index also correlate with the time and radiant location, indicating that the strength of Orionid meteoroids varies with the radiant location. Later Orionids start eroding at higher pressures, with lower erosion coefficients and grain mass indices preferring smaller grains. As these later Orionids appear to be younger based on the simulations from \citet{egal2020modeling} (though the age gradient is not strong) this correlation hints at more evolved (older) Orionids become progressively weaker over time.

The bulk density of the leading fragment is also correlated with the radiant, where leading fragments from later Orionids appear to be denser. The mass of the leading fragment also appears to be correlated to the initial mass and is on average around 10\% of the initial mass.

\subsection{Grain size distribution and porosity}

Given that we have been able to constrain the grain size distribution by using the wake measurements, we can investigate the theoretical porosity of the typical Orionid meteoroid assuming random packing. \cite{brouwers2006particle} gives a relation for the void fraction (i.e. porosity) of an object composed of particles with a power law distribution of sizes:

\begin{equation}
    \varphi = \varphi_1 \left ( \frac{d_{max}}{d_{min}} \right )^{-(1-\varphi_1) \beta / (1 + u^2)}
\end{equation}

\noindent where $\varphi_1$ is the void fraction of a single constituent, $\beta$ is the maximum gradient of the single-sized void fraction on the onset to bimodal packing \citep[][$\varphi_1 \simeq 0.36$, $\beta \simeq 0.2$ for hard spheres]{kansal2002computer, karayiannis2009structure}, and $u$ is the size distribution exponent.

Given a mean differential grain mass distribution index of $s = 2.15$, the differential grain size distribution is $u = 4.45$ \citep{vida2021high}. For most of our modelled meteoroids, the mean ratio between the diameters of the largest and the smallest grains was $\frac{d_{max}}{d_{min}} \sim 10$; thus we can derive a theoretical minimum porosity of $\varphi \sim 0.355$. This value is consistent with the experimental data of \cite{zangmeister2014packing} who explored porosities of various aggregates across $\micro m$ - $mm$ sizes.

On the other hand, for our assumed grain density of $\SI{3000}{\kilo \gram \per \cubic \metre}$ our observed bulk density leads to derived porosities of around 0.9, similar to previous values derived for cometary meteoroids \citep{borovivcka2007atmospheric, vojavcek2019properties} and chondritic porous interplanetary dust particles \citep{Bradley2003}. When compared to a theoretical minimum of 0.36, this indicates that cometary meteoroids are not composed of densely packed grains, but are fluffy aggregates, a picture consistent with that emerging from Rosetta results \citep{hornung2016first}.

We note that the structural strength of meteoroids needs to be sufficient to support such high porosity when exposed to aerodynamic pressures, as the shape-density coefficient is measured during the flight. \cite{hulfeld2021three} construct a 3D model of a meteoroid with an empirical grain size distribution and porosity, but immediately after the meteoroid is subjected to aerodynamic drag it collapses and changes its shape (it flattens) and porosity decreases from 90\% to 80\%. A similar effect is observed for micrometeorites as porosity changes during ablation \citep{kohout2014density}. This makes only a minor change in the shape-density coefficient as the drag coefficient decreases as the bulk density increases. \cite{hulfeld2021three} only assumed that the meteoroid is held together by weak adhesive forces which were able to preserve the high porosity and the structural strength.

Recently, \cite{hornung2023structural} have shown that porous mm-sized dust aggregates of comet 67P that survive a low-velocity impact with the COSIMA sensor can be fragmented if an electrical charge is applied over them that exerts a pressure on the order of kilopascals. They observed that after fragmentation, \SIrange{10}{50}{\micro \metre} grains remain, consistent with our modelling.

Taking all of the above, our interpretation is that mm-sized Orionid/cometary meteoroids must possess sufficient structural strength provided by contact forces which are able to maintain a porosity $\sim2.5\times$ higher than densely packed material of similar composition. This high porosity is maintained despite the meteoroid experiencing dynamic pressures on the order of a kilopascal during its flight.

\section{Conclusions}

Multi-station observations of the Orionid meteor shower were performed using the CAMO mirror tracking system in October 2019 and 2020. The narrow field data was manually reduced and a total of 18 high-precision trajectories were computed. The median radiant error for individual Orionids detected by CAMO was \ang{0.05}. All observed Orionids show a leading fragment morphology, permitting high trajectory precision.

The Orionid radiant spread shows an elongated structure which is also visible in the high-precision Global Meteor Network data set \citep{moorhead2021meteor}. The radiant dispersion was measured as the median offset from the mean drift-corrected radiant. Including all observations, the dispersion is $\ang{0.40} \pm \ang{0.06} $, but almost entirely in the Sun-centred ecliptic longitude direction - the median spread in the ecliptic latitude is only \ang{\sim 0.1}. This dispersion is considerably larger than our median measurement precision, implying that the true physical dispersion is being measured. Compared to previously measured radiant dispersions of the Orionids, it is about half the dispersion measured by \cite{kresak1970dispersion} and \ang{0.12} less than measured by \cite{moorhead2021meteor} but still within $2\sigma$ from their value. Our radiant dispersion value of \ang{0.40} represents our best estimate of the physical radiant dispersion of the Orionid shower near the time of its maximum, appropriate to \SIrange{1}{10}{\milli \gram} masses. 

We developed a data-driven luminous efficiency model with dependence on speed and mass which we apply to our Orionid ablation modelling. The model was developed using a fit to previous luminous efficiency measurements estimated using simultaneous CAMO \citep{subasinghe2018luminous} and CMOR radar \citep{brown2020coordinated} observations, also utilizing the observed Orionid leading fragment data gathered as part of this work. The average model-estimated luminous efficiency across all speeds and masses is about 0.5\%, but it varies an order of magnitude with mass and a factor of two with speed.

The erosion model of \cite{borovivcka2007atmospheric} was applied to all observed Orionids. For the first time, we successfully fit the light curve, the dynamics, and the wake. We find that the accurate reproduction of the wake directly constrains the mass distribution of released grains, confirming the assumption that grains are distributed as a power law. We find a differential mass index of $\sim2.15$ fits the structural grain distribution of most Orionids, appropriate for the \SIrange{1e-11}{1e-8}{\kilo \gram} mass range. This is slightly higher than the mass index found for the distribution of grains from the inner coma of 1P/Halley \citep{McDonnell1987} and Rosetta \citep{hornung2016first} where $s < 2.0$ predominated. This may reflect a time evolution in structure, whereby meteoroids become weaker with time, reflecting a predominance of smaller grain size units compared to fresh cometary dust.

The ablation and fragmentation of the Orionids are characterized by a two-stage erosion process. The modelling is best explained using a meteoroid bulk density of $\sim\SI{300}{\kilogram \per \cubic \metre}$ and the erosion of \SIrange{10}{100}{\micro \metre} refractory grains starting near the beginning of visible trajectory ($\sim\SI{115}{\kilo \metre}$) and ending at the light curve peak. This cessation of erosion is followed by a second stage dominated by a leading single-body fragment which has $\sim10\%$ of the initial meteoroid mass and does not erode or fragment further. \citet{vojavcek2019properties} also noted that about 10\% of their sample of 152 meteoroids showed two stages of erosion, with shower meteors showing a higher fraction. They interpreted this two-stage erosion as evidence of a bimodal structure in the meteoroid made from a common material. However, in our data and model fits, the ablation coefficient is roughly half for the second stage erosion, in contrast to the finding of \citet{vojavcek2019properties} where the ablation coefficient is the same. Together with our high-resolution imagery, it suggests a different material with higher density and more resistance to ablation. The ablation properties of leading fragments vary, but they appear to represent a form of denser cometary material. Another interpretation is that the leading fragment represents the in-flight equivalent of refractory metal nuggets (RMNs) \citep{Rudraswami2014} found in cosmic spherules. This might indicate that the dust of 1P/Halley is rich in refractory inclusions, such as CAIs or Fe/Ni metal which separate in flight as "beads" \citep{Brownlee1984}. However, we cannot conclusively identify leading fragment composition without spectral data. Further confirmation of these observations could be done by analyzing $\eta$-Aquariid meteors with CAMO using the same approach. If these really are refractory inclusions, similar to those found in the dust of comet 81P/Wild 2 \citep{simon2008refractory}, this would further support the notion of large-scale radial mixing between reservoirs of solids in the inner and outer solar nebula during the formation of the Solar System.

\cite{borovivcka2007atmospheric} and \cite{borovivcka2014spectral} derived similar densities and porosities for the Draconids as we have found for the Orionids. These meteoroids are considered to be a more fragile cometary material than the Halleyids \citep[][D group for Draconids, C for Orionids]{ceplecha1988earth}, leading us to conclude that the main difference between the two is in their structural strength and not in material composition. \cite{ceplecha1988earth} suggests bulk densities of \SI{270}{\kilo \gram \per \cubic \metre} for group D which seem to be applicable for the Orionids, despite their higher structural strength \citep{ceplecha1966dynamic}.

\section{Acknowledgements}

This work was supported in part by the NASA Meteoroid Environment Office under cooperative agreement 80NSSC21M0073. PGB also acknowledges funding support from the Natural Sciences and Engineering Research council of Canada and the Canada Research Chairs program. We thank Z. Krzeminski for help in optical data reduction, J. Gill, M. Mazur, and P. Quigley for hardware and software support.

\bibliography{main}

\begin{thebibliography}{87}
\expandafter\ifx\csname natexlab\endcsname\relax\def\natexlab#1{#1}\fi
\providecommand{\bibinfo}[2]{#2}
\ifx\xfnm\relax \def\xfnm[#1]{\unskip,\space#1}\fi
\bibitem[{Babadzhanov \& Kokhirova(2009)}]{Babadzhanov2009}
\bibinfo{author}{Babadzhanov, P.~B.}, \& \bibinfo{author}{Kokhirova, G.~I.}
  (\bibinfo{year}{2009}).
\newblock \bibinfo{title}{{Densities and porosities of meteoroids}}.
\newblock {\it \bibinfo{journal}{Astronomy and Astrophysics}\/},  {\it
  \bibinfo{volume}{495}\/}, \bibinfo{pages}{353}.
\bibitem[{Borovi{\v{c}}ka(1990)}]{borovicka1990comparison}
\bibinfo{author}{Borovi{\v{c}}ka, J.} (\bibinfo{year}{1990}).
\newblock \bibinfo{title}{The comparison of two methods of determining meteor
  trajectories from photographs}.
\newblock {\it \bibinfo{journal}{Bulletin of the Astronomical Institutes of
  Czechoslovakia}\/},  {\it \bibinfo{volume}{41}\/}, \bibinfo{pages}{391--396}.
\bibitem[{Borovi{\v{c}}ka et~al.(2014)Borovi{\v{c}}ka, Koten, Shrben{\`y},
  {\v{S}}tork \& Hornoch}]{borovivcka2014spectral}
\bibinfo{author}{Borovi{\v{c}}ka, J.}, \bibinfo{author}{Koten, P.},
  \bibinfo{author}{Shrben{\`y}, L.}, \bibinfo{author}{{\v{S}}tork, R.}, \&
  \bibinfo{author}{Hornoch, K.} (\bibinfo{year}{2014}).
\newblock \bibinfo{title}{Spectral, photometric, and dynamic analysis of eight
  draconid meteors}.
\newblock {\it \bibinfo{journal}{Earth, Moon, and Planets}\/},  {\it
  \bibinfo{volume}{113}\/}, \bibinfo{pages}{15--31}.
\bibitem[{Borovi{\v{c}}ka et~al.(2019)Borovi{\v{c}}ka, Macke, Campbell-Brown,
  Levasseur-Regourd, Rietmeijer \& Kohout}]{borovivcka2019physical}
\bibinfo{author}{Borovi{\v{c}}ka, J.}, \bibinfo{author}{Macke, R.~J.},
  \bibinfo{author}{Campbell-Brown, M.~D.}, \bibinfo{author}{Levasseur-Regourd,
  A.-C.}, \bibinfo{author}{Rietmeijer, F.~J.}, \& \bibinfo{author}{Kohout, T.}
  (\bibinfo{year}{2019}).
\newblock \bibinfo{title}{Physical and chemical properties of meteoroids}.
\newblock {\it \bibinfo{journal}{Meteoroids: Sources of Meteors on Earth and
  Beyond}\/},  (p.~\bibinfo{pages}{37}).
\bibitem[{Borovi{\v{c}}ka et~al.(2007)Borovi{\v{c}}ka, Spurn{\`y} \&
  Koten}]{borovivcka2007atmospheric}
\bibinfo{author}{Borovi{\v{c}}ka, J.}, \bibinfo{author}{Spurn{\`y}, P.}, \&
  \bibinfo{author}{Koten, P.} (\bibinfo{year}{2007}).
\newblock \bibinfo{title}{Atmospheric deceleration and light curves of draconid
  meteors and implications for the structure of cometary dust}.
\newblock {\it \bibinfo{journal}{Astronomy \& Astrophysics}\/},  {\it
  \bibinfo{volume}{473}\/}, \bibinfo{pages}{661--672}.
\bibitem[{Borovi{\v{c}}ka et~al.(2020)Borovi{\v{c}}ka, Spurn{\`y} \&
  Shrben{\`y}}]{borovivcka2020two}
\bibinfo{author}{Borovi{\v{c}}ka, J.}, \bibinfo{author}{Spurn{\`y}, P.}, \&
  \bibinfo{author}{Shrben{\`y}, L.} (\bibinfo{year}{2020}).
\newblock \bibinfo{title}{Two strengths of ordinary chondritic meteoroids as
  derived from their atmospheric fragmentation modeling}.
\newblock {\it \bibinfo{journal}{The Astronomical Journal}\/},  {\it
  \bibinfo{volume}{160}\/}, \bibinfo{pages}{42}.
\bibitem[{Borovi{\v{c}}ka et~al.(2013)Borovi{\v{c}}ka, T{\'o}th, Igaz,
  Spurn{\`y}, Kalenda, Haloda, Svore{\v{n}}, Korno{\v{s}}, Silber, Brown
  et~al.}]{borovivcka2013kovsice}
\bibinfo{author}{Borovi{\v{c}}ka, J.}, \bibinfo{author}{T{\'o}th, J.},
  \bibinfo{author}{Igaz, A.}, \bibinfo{author}{Spurn{\`y}, P.},
  \bibinfo{author}{Kalenda, P.}, \bibinfo{author}{Haloda, J.},
  \bibinfo{author}{Svore{\v{n}}, J.}, \bibinfo{author}{Korno{\v{s}}, L.},
  \bibinfo{author}{Silber, E.}, \bibinfo{author}{Brown, P.} et~al.
  (\bibinfo{year}{2013}).
\newblock \bibinfo{title}{The ko{\v{s}}ice meteorite fall: Atmospheric
  trajectory, fragmentation, and orbit}.
\newblock {\it \bibinfo{journal}{Meteoritics \& Planetary Science}\/},  {\it
  \bibinfo{volume}{48}\/}, \bibinfo{pages}{1757--1779}.
\bibitem[{Bradley(2007)}]{Bradley2003}
\bibinfo{author}{Bradley, J.} (\bibinfo{year}{2007}).
\newblock {\it \bibinfo{title}{Interplanetary Dust Particles}\/}.
\newblock \bibinfo{publisher}{Elsevier}.
\bibitem[{Brouwers(2006)}]{brouwers2006particle}
\bibinfo{author}{Brouwers, H.} (\bibinfo{year}{2006}).
\newblock \bibinfo{title}{Particle-size distribution and packing fraction of
  geometric random packings}.
\newblock {\it \bibinfo{journal}{Physical review E}\/},  {\it
  \bibinfo{volume}{74}\/}, \bibinfo{pages}{031309}.
\bibitem[{Brown \& Jones(1998)}]{brown1998simulation}
\bibinfo{author}{Brown, P.}, \& \bibinfo{author}{Jones, J.}
  (\bibinfo{year}{1998}).
\newblock \bibinfo{title}{Simulation of the formation and evolution of the
  perseid meteoroid stream}.
\newblock {\it \bibinfo{journal}{Icarus}\/},  {\it \bibinfo{volume}{133}\/},
  \bibinfo{pages}{36--68}.
\bibitem[{Brown \& Weryk(2020)}]{brown2020coordinated}
\bibinfo{author}{Brown, P.}, \& \bibinfo{author}{Weryk, R.~J.}
  (\bibinfo{year}{2020}).
\newblock \bibinfo{title}{Coordinated optical and radar measurements of low
  velocity meteors}.
\newblock {\it \bibinfo{journal}{Icarus}\/},  {\it \bibinfo{volume}{352}\/},
  \bibinfo{pages}{113975}.
\bibitem[{Brownlee et~al.(1984)Brownlee, Bates \& Wheelock}]{Brownlee1984}
\bibinfo{author}{Brownlee, D.~E.}, \bibinfo{author}{Bates, B.~A.}, \&
  \bibinfo{author}{Wheelock, M.~M.} (\bibinfo{year}{1984}).
\newblock \bibinfo{title}{{Extraterrestrial platinum group nuggets in deep-sea
  sediments}}.
\newblock {\it \bibinfo{journal}{Nature}\/},  {\it \bibinfo{volume}{309}\/},
  \bibinfo{pages}{693--695}.
\bibitem[{Buccongello et~al.(2023)Buccongello, Brown, Vida \&
  Pinhas}]{buccongello2023physical}
\bibinfo{author}{Buccongello, N.}, \bibinfo{author}{Brown, P.~G.},
  \bibinfo{author}{Vida, D.}, \& \bibinfo{author}{Pinhas, A.}
  (\bibinfo{year}{2023}).
\newblock \bibinfo{title}{A physical survey of meteoroid streams: Comparing
  cometary reservoirs}.
\newblock {\it \bibinfo{journal}{Icarus}\/},  {\it
  \bibinfo{volume}{submitted}\/}.
\bibitem[{Burns et~al.(1979)Burns, Lamy \& Soter}]{Burns1979}
\bibinfo{author}{Burns, J.~A.}, \bibinfo{author}{Lamy, P.~L.}, \&
  \bibinfo{author}{Soter, S.} (\bibinfo{year}{1979}).
\newblock \bibinfo{title}{{Radiation forces on small particles in the solar
  system}}.
\newblock {\it \bibinfo{journal}{Icarus}\/},  {\it \bibinfo{volume}{40}\/},
  \bibinfo{pages}{1--48}.
\bibitem[{Campbell-Brown et~al.(2013)Campbell-Brown, Borovi{\v{c}}ka, Brown \&
  Stokan}]{campbell2013high}
\bibinfo{author}{Campbell-Brown, M.}, \bibinfo{author}{Borovi{\v{c}}ka, J.},
  \bibinfo{author}{Brown, P.}, \& \bibinfo{author}{Stokan, E.}
  (\bibinfo{year}{2013}).
\newblock \bibinfo{title}{High-resolution modelling of meteoroid ablation}.
\newblock {\it \bibinfo{journal}{Astronomy \& Astrophysics}\/},  {\it
  \bibinfo{volume}{557}\/}, \bibinfo{pages}{A41}.
\bibitem[{{\v{C}}apek et~al.(2019){\v{C}}apek, Koten, Borovi{\v{c}}ka,
  Voj{\'a}{\v{c}}ek, Spurn{\`y} \& {\v{S}}tork}]{capek2019small}
\bibinfo{author}{{\v{C}}apek, D.}, \bibinfo{author}{Koten, P.},
  \bibinfo{author}{Borovi{\v{c}}ka, J.}, \bibinfo{author}{Voj{\'a}{\v{c}}ek,
  V.}, \bibinfo{author}{Spurn{\`y}, P.}, \& \bibinfo{author}{{\v{S}}tork, R.}
  (\bibinfo{year}{2019}).
\newblock \bibinfo{title}{Small iron meteoroids-observation and modeling of
  meteor light curves}.
\newblock {\it \bibinfo{journal}{Astronomy \& Astrophysics}\/},  {\it
  \bibinfo{volume}{625}\/}, \bibinfo{pages}{A106}.
\bibitem[{Ceplecha(1966)}]{ceplecha1966dynamic}
\bibinfo{author}{Ceplecha, Z.} (\bibinfo{year}{1966}).
\newblock \bibinfo{title}{Dynamic and photometric mass of meteors}.
\newblock {\it \bibinfo{journal}{Bulletin of the Astronomical Institute of
  Czechoslovakia, vol. 17, p. 347}\/},  {\it \bibinfo{volume}{17}\/},
  \bibinfo{pages}{347}.
\bibitem[{Ceplecha(1975)}]{ceplecha1975ablation}
\bibinfo{author}{Ceplecha, Z.} (\bibinfo{year}{1975}).
\newblock \bibinfo{title}{Ablation and shape-density coefficients in meteors}.
\newblock {\it \bibinfo{journal}{Bulletin of the Astronomical Institutes of
  Czechoslovakia}\/},  {\it \bibinfo{volume}{26}\/}, \bibinfo{pages}{242--248}.
\bibitem[{Ceplecha(1988)}]{ceplecha1988earth}
\bibinfo{author}{Ceplecha, Z.} (\bibinfo{year}{1988}).
\newblock \bibinfo{title}{Earth's influx of different populations of sporadic
  meteoroids from photographic and television data}.
\newblock {\it \bibinfo{journal}{Astronomical Institutes of Czechoslovakia,
  Bulletin (ISSN 0004-6248), vol. 39, July 1988, p. 221-236.}\/},  {\it
  \bibinfo{volume}{39}\/}, \bibinfo{pages}{221--236}.
\bibitem[{Ceplecha et~al.(1998)Ceplecha, Borovi{\v{c}}ka, Elford, ReVelle,
  Hawkes, Porub{\v{c}}an \& {\v{S}}imek}]{ceplecha1998meteor}
\bibinfo{author}{Ceplecha, Z.}, \bibinfo{author}{Borovi{\v{c}}ka, J.},
  \bibinfo{author}{Elford, W.~G.}, \bibinfo{author}{ReVelle, D.~O.},
  \bibinfo{author}{Hawkes, R.~L.}, \bibinfo{author}{Porub{\v{c}}an, V.}, \&
  \bibinfo{author}{{\v{S}}imek, M.} (\bibinfo{year}{1998}).
\newblock \bibinfo{title}{Meteor phenomena and bodies}.
\newblock {\it \bibinfo{journal}{Space Science Reviews}\/},  {\it
  \bibinfo{volume}{84}\/}, \bibinfo{pages}{327--471}.
\bibitem[{Ceplecha \& ReVelle(2005)}]{ceplecha2005fragmentation}
\bibinfo{author}{Ceplecha, Z.}, \& \bibinfo{author}{ReVelle, D.~O.}
  (\bibinfo{year}{2005}).
\newblock \bibinfo{title}{Fragmentation model of meteoroid motion, mass loss,
  and radiation in the atmosphere}.
\newblock {\it \bibinfo{journal}{Meteoritics \& Planetary Science}\/},  {\it
  \bibinfo{volume}{40}\/}, \bibinfo{pages}{35--54}.
\bibitem[{Chau \& Galindo(2008)}]{Chau2008}
\bibinfo{author}{Chau, J.}, \& \bibinfo{author}{Galindo, F.}
  (\bibinfo{year}{2008}).
\newblock \bibinfo{title}{{First definitive observations of meteor shower
  particles using a high-power large-aperture radar}}.
\newblock {\it \bibinfo{journal}{Icarus}\/},  {\it \bibinfo{volume}{194}\/},
  \bibinfo{pages}{23--29}.
\bibitem[{Christiansen(1993)}]{christiansen1993design}
\bibinfo{author}{Christiansen, E.~L.} (\bibinfo{year}{1993}).
\newblock \bibinfo{title}{Design and performance equations for advanced
  meteoroid and debris shields}.
\newblock {\it \bibinfo{journal}{International Journal of Impact
  Engineering}\/},  {\it \bibinfo{volume}{14}\/}, \bibinfo{pages}{145--156}.
\bibitem[{Egal(2020)}]{egal2020forecasting}
\bibinfo{author}{Egal, A.} (\bibinfo{year}{2020}).
\newblock \bibinfo{title}{Forecasting meteor showers: A review}.
\newblock {\it \bibinfo{journal}{Planetary and Space Science}\/},  {\it
  \bibinfo{volume}{185}\/}, \bibinfo{pages}{104895}.
\bibitem[{Egal et~al.(2020{\natexlab{a}})Egal, Brown, Rendtel, Campbell-Brown
  \& Wiegert}]{Egal2020observations}
\bibinfo{author}{Egal, A.}, \bibinfo{author}{Brown, P.~G.},
  \bibinfo{author}{Rendtel, J.}, \bibinfo{author}{Campbell-Brown, M.}, \&
  \bibinfo{author}{Wiegert, P.} (\bibinfo{year}{2020}{\natexlab{a}}).
\newblock \bibinfo{title}{{Activity of the Eta-Aquariid and Orionid meteor
  showers}}.
\newblock {\it \bibinfo{journal}{Astronomy \& Astrophysics}\/},  {\it
  \bibinfo{volume}{640}\/}, \bibinfo{pages}{A58}.
\bibitem[{Egal et~al.(2020{\natexlab{b}})Egal, Wiegert, Brown, Campbell-Brown
  \& Vida}]{egal2020modeling}
\bibinfo{author}{Egal, A.}, \bibinfo{author}{Wiegert, P.},
  \bibinfo{author}{Brown, P.~G.}, \bibinfo{author}{Campbell-Brown, M.}, \&
  \bibinfo{author}{Vida, D.} (\bibinfo{year}{2020}{\natexlab{b}}).
\newblock \bibinfo{title}{Modeling the past and future activity of the halleyid
  meteor showers}.
\newblock {\it \bibinfo{journal}{Astronomy \& Astrophysics}\/},  {\it
  \bibinfo{volume}{642}\/}, \bibinfo{pages}{A120}.
\bibitem[{Egal et~al.(2018)Egal, Wiegert, Brown, Moser, Moorhead \&
  Cooke}]{egal2018draconid}
\bibinfo{author}{Egal, A.}, \bibinfo{author}{Wiegert, P.},
  \bibinfo{author}{Brown, P.~G.}, \bibinfo{author}{Moser, D.~E.},
  \bibinfo{author}{Moorhead, A.~V.}, \& \bibinfo{author}{Cooke, W.~J.}
  (\bibinfo{year}{2018}).
\newblock \bibinfo{title}{The draconid meteoroid stream 2018: prospects for
  satellite impact detection}.
\newblock {\it \bibinfo{journal}{The Astrophysical journal letters}\/},  {\it
  \bibinfo{volume}{866}\/}, \bibinfo{pages}{L8}.
\bibitem[{Egal et~al.(2023)Egal, Wiegert, Brown \& Vida}]{egal2023modeling}
\bibinfo{author}{Egal, A.}, \bibinfo{author}{Wiegert, P.~A.},
  \bibinfo{author}{Brown, P.~G.}, \& \bibinfo{author}{Vida, D.}
  (\bibinfo{year}{2023}).
\newblock \bibinfo{title}{Modeling the 2022 $\tau$-herculid outburst}.
\newblock {\it \bibinfo{journal}{The Astrophysical Journal}\/},  {\it
  \bibinfo{volume}{949}\/}, \bibinfo{pages}{96}.
\bibitem[{Galligan(2000)}]{Galligan2000}
\bibinfo{author}{Galligan, D.~P.} (\bibinfo{year}{2000}).
\newblock {\it \bibinfo{title}{{Structural Analysis of Radar Meteoroid
  Data}}\/}.
\newblock \bibinfo{type}{Doctoral} Canterbury.
\bibitem[{Hajduk(1970)}]{hajduk1970structure}
\bibinfo{author}{Hajduk, A.} (\bibinfo{year}{1970}).
\newblock \bibinfo{title}{Structure of the meteor stream associated with comet
  halley}.
\newblock {\it \bibinfo{journal}{Bulletin of the Astronomical Institutes of
  Czechoslovakia}\/},  {\it \bibinfo{volume}{21}\/}, \bibinfo{pages}{37}.
\bibitem[{Hawkes \& Jones(1975)}]{hawkes1975quantitative}
\bibinfo{author}{Hawkes, R.}, \& \bibinfo{author}{Jones, J.}
  (\bibinfo{year}{1975}).
\newblock \bibinfo{title}{A quantitative model for the ablation of dustball
  meteors}.
\newblock {\it \bibinfo{journal}{Monthly Notices of the Royal Astronomical
  Society}\/},  {\it \bibinfo{volume}{173}\/}, \bibinfo{pages}{339--356}.
\bibitem[{Henych et~al.(2023)Henych, Borovi{\v{c}}ka \&
  Spurn{\`y}}]{henych2023semi}
\bibinfo{author}{Henych, T.}, \bibinfo{author}{Borovi{\v{c}}ka, J.}, \&
  \bibinfo{author}{Spurn{\`y}, P.} (\bibinfo{year}{2023}).
\newblock \bibinfo{title}{Semi-automatic meteoroid fragmentation modeling using
  genetic algorithms}.
\newblock {\it \bibinfo{journal}{Astronomy and Astrophysics}\/},  {\it
  \bibinfo{volume}{671}\/}, \bibinfo{pages}{A23}.
\bibitem[{Hornung et~al.(2023)Hornung, Mellado, Stenzel, Langevin, Merouane,
  Fray, Fischer, Paquette, Baklouti, Bardyn et~al.}]{hornung2023structural}
\bibinfo{author}{Hornung, K.}, \bibinfo{author}{Mellado, E.~M.},
  \bibinfo{author}{Stenzel, O.~J.}, \bibinfo{author}{Langevin, Y.},
  \bibinfo{author}{Merouane, S.}, \bibinfo{author}{Fray, N.},
  \bibinfo{author}{Fischer, H.}, \bibinfo{author}{Paquette, J.},
  \bibinfo{author}{Baklouti, D.}, \bibinfo{author}{Bardyn, A.} et~al.
  (\bibinfo{year}{2023}).
\newblock \bibinfo{title}{On structural properties of comet 67/p dust particles
  collected in situ by rosetta/cosima from observations of electrical
  fragmentation}.
\newblock {\it \bibinfo{journal}{Planetary and Space Science}\/},  (p.
  \bibinfo{pages}{105747}).
\bibitem[{Hornung et~al.(2016)Hornung, Merouane, Hilchenbach, Langevin,
  Mellado, Della~Corte, Kissel, Engrand, Schulz, Ryno
  et~al.}]{hornung2016first}
\bibinfo{author}{Hornung, K.}, \bibinfo{author}{Merouane, S.},
  \bibinfo{author}{Hilchenbach, M.}, \bibinfo{author}{Langevin, Y.},
  \bibinfo{author}{Mellado, E.~M.}, \bibinfo{author}{Della~Corte, V.},
  \bibinfo{author}{Kissel, J.}, \bibinfo{author}{Engrand, C.},
  \bibinfo{author}{Schulz, R.}, \bibinfo{author}{Ryno, J.} et~al.
  (\bibinfo{year}{2016}).
\newblock \bibinfo{title}{A first assessment of the strength of cometary
  particles collected in-situ by the cosima instrument onboard rosetta}.
\newblock {\it \bibinfo{journal}{Planetary and Space Science}\/},  {\it
  \bibinfo{volume}{133}\/}, \bibinfo{pages}{63--75}.
\bibitem[{Hulfeld et~al.(2021)Hulfeld, K{\"u}chlin \& Jenny}]{hulfeld2021three}
\bibinfo{author}{Hulfeld, L.}, \bibinfo{author}{K{\"u}chlin, S.}, \&
  \bibinfo{author}{Jenny, P.} (\bibinfo{year}{2021}).
\newblock \bibinfo{title}{Three dimensional atmospheric entry simulation of a
  high altitude cometary dustball meteoroid}.
\newblock {\it \bibinfo{journal}{Astronomy \& Astrophysics}\/},  {\it
  \bibinfo{volume}{650}\/}, \bibinfo{pages}{A101}.
\bibitem[{Jacchia \& Whipple(1961)}]{jacchia1961precision}
\bibinfo{author}{Jacchia, L.~G.}, \& \bibinfo{author}{Whipple, F.~L.}
  (\bibinfo{year}{1961}).
\newblock \bibinfo{title}{Precision orbits of 413 photographic meteors}.
\newblock {\it \bibinfo{journal}{Smithsonian Contributions to Astrophysics}\/},
  .
\bibitem[{Jenniskens et~al.(2021)Jenniskens, Cooper, Baggaley, Heathcote \&
  Lauretta}]{jenniskens2021first}
\bibinfo{author}{Jenniskens, P.}, \bibinfo{author}{Cooper, T.},
  \bibinfo{author}{Baggaley, J.}, \bibinfo{author}{Heathcote, S.}, \&
  \bibinfo{author}{Lauretta, D.} (\bibinfo{year}{2021}).
\newblock \bibinfo{title}{First detection of the arid (ard,\# 1130) meteor
  shower from comet 15p/finlay}.
\newblock {\it \bibinfo{journal}{eMeteorNews}\/},  {\it \bibinfo{volume}{6}\/},
  \bibinfo{pages}{531--533}.
\bibitem[{Jones et~al.(1989)Jones, McIntosh \& Hawkes}]{jones1989age}
\bibinfo{author}{Jones, J.}, \bibinfo{author}{McIntosh, B.}, \&
  \bibinfo{author}{Hawkes, R.} (\bibinfo{year}{1989}).
\newblock \bibinfo{title}{The age of the orionid meteoroid stream}.
\newblock {\it \bibinfo{journal}{Monthly Notices of the Royal Astronomical
  Society}\/},  {\it \bibinfo{volume}{238}\/}, \bibinfo{pages}{179--191}.
\bibitem[{Kambulow et~al.(2022)Kambulow, Vida \& Brown}]{kambulow2022inverting}
\bibinfo{author}{Kambulow, J.}, \bibinfo{author}{Vida, D.}, \&
  \bibinfo{author}{Brown, P.~G.} (\bibinfo{year}{2022}).
\newblock \bibinfo{title}{Inverting physical properties of meteoroids using
  machine learning}.
\newblock \bibinfo{note}{Meteoroids 2022, virtual}.
\bibitem[{Kansal et~al.(2002)Kansal, Torquato \&
  Stillinger}]{kansal2002computer}
\bibinfo{author}{Kansal, A.~R.}, \bibinfo{author}{Torquato, S.}, \&
  \bibinfo{author}{Stillinger, F.~H.} (\bibinfo{year}{2002}).
\newblock \bibinfo{title}{Computer generation of dense polydisperse sphere
  packings}.
\newblock {\it \bibinfo{journal}{The Journal of chemical physics}\/},  {\it
  \bibinfo{volume}{117}\/}, \bibinfo{pages}{8212--8218}.
\bibitem[{Karayiannis et~al.(2009)Karayiannis, Foteinopoulou \&
  Laso}]{karayiannis2009structure}
\bibinfo{author}{Karayiannis, N.~C.}, \bibinfo{author}{Foteinopoulou, K.}, \&
  \bibinfo{author}{Laso, M.} (\bibinfo{year}{2009}).
\newblock \bibinfo{title}{The structure of random packings of freely jointed
  chains of tangent hard spheres}.
\newblock {\it \bibinfo{journal}{The Journal of chemical physics}\/},  {\it
  \bibinfo{volume}{130}\/}, \bibinfo{pages}{164908}.
\bibitem[{Kennedy \& Eberhart(1995)}]{kennedy1995particle}
\bibinfo{author}{Kennedy, J.}, \& \bibinfo{author}{Eberhart, R.}
  (\bibinfo{year}{1995}).
\newblock \bibinfo{title}{Particle swarm optimization}.
\newblock In {\it \bibinfo{booktitle}{Proceedings of ICNN'95-international
  conference on neural networks}\/} (pp. \bibinfo{pages}{1942--1948}).
\newblock \bibinfo{organization}{IEEE} volume~\bibinfo{volume}{4}.
\bibitem[{Kikwaya et~al.(2011)Kikwaya, Campbell-Brown \&
  Brown}]{kikwaya2011bulk}
\bibinfo{author}{Kikwaya, J.-B.}, \bibinfo{author}{Campbell-Brown, M.}, \&
  \bibinfo{author}{Brown, P.} (\bibinfo{year}{2011}).
\newblock \bibinfo{title}{Bulk density of small meteoroids}.
\newblock {\it \bibinfo{journal}{Astronomy \& Astrophysics}\/},  {\it
  \bibinfo{volume}{530}\/}, \bibinfo{pages}{A113}.
\bibitem[{Kohout et~al.(2014)Kohout, Kallonen, Suuronen, Rochette, Hutzler,
  Gattacceca, Badjukov, Skala, B{\"o}hmov{\'a} \&
  {\v{C}}uda}]{kohout2014density}
\bibinfo{author}{Kohout, T.}, \bibinfo{author}{Kallonen, A.},
  \bibinfo{author}{Suuronen, J.-P.}, \bibinfo{author}{Rochette, P.},
  \bibinfo{author}{Hutzler, A.}, \bibinfo{author}{Gattacceca, J.},
  \bibinfo{author}{Badjukov, D.~D.}, \bibinfo{author}{Skala, R.},
  \bibinfo{author}{B{\"o}hmov{\'a}, V.}, \& \bibinfo{author}{{\v{C}}uda, J.}
  (\bibinfo{year}{2014}).
\newblock \bibinfo{title}{Density, porosity, mineralogy, and internal structure
  of cosmic dust and alteration of its properties during high-velocity
  atmospheric entry}.
\newblock {\it \bibinfo{journal}{Meteoritics \& Planetary Science}\/},  {\it
  \bibinfo{volume}{49}\/}, \bibinfo{pages}{1157--1170}.
\bibitem[{Koten et~al.(2019)Koten, Rendtel, Shrben\'{y}, Gural, Borovicka \&
  Kozak}]{koten2019meteors}
\bibinfo{author}{Koten, P.}, \bibinfo{author}{Rendtel, J.},
  \bibinfo{author}{Shrben\'{y}, L.}, \bibinfo{author}{Gural, P.},
  \bibinfo{author}{Borovicka, J.}, \& \bibinfo{author}{Kozak, P.}
  (\bibinfo{year}{2019}).
\newblock \bibinfo{title}{Meteors and meteor showers as observed by optical
  techniques}.
\newblock {\it \bibinfo{journal}{Meteoroids: Sources of Meteors on Earth and
  Beyond}\/},  (pp. \bibinfo{pages}{90--115}).
\bibitem[{Krasnopolsky et~al.(1988)Krasnopolsky, Moroz, Krysko, Tkachuk,
  Moreels, Clairemidi, Parisot, Gogoshev \&
  Gogosheva}]{krasnopolsky1988properties}
\bibinfo{author}{Krasnopolsky, V.}, \bibinfo{author}{Moroz, V.},
  \bibinfo{author}{Krysko, A.}, \bibinfo{author}{Tkachuk, A.~Y.},
  \bibinfo{author}{Moreels, G.}, \bibinfo{author}{Clairemidi, J.},
  \bibinfo{author}{Parisot, J.}, \bibinfo{author}{Gogoshev, M.}, \&
  \bibinfo{author}{Gogosheva, T.} (\bibinfo{year}{1988}).
\newblock \bibinfo{title}{Properties of dust in comet p/halley measured by the
  vega-2 three-channel spectrometer}.
\newblock In {\it \bibinfo{booktitle}{Exploration of Halley’s Comet}\/} (pp.
  \bibinfo{pages}{707--711}).
\newblock \bibinfo{publisher}{Springer}.
\bibitem[{Kres{\'a}k \& Porub\v{c}an(1970)}]{kresak1970dispersion}
\bibinfo{author}{Kres{\'a}k, L.}, \& \bibinfo{author}{Porub\v{c}an, V.}
  (\bibinfo{year}{1970}).
\newblock \bibinfo{title}{The dispersion of meteors in meteor streams. i. the
  size of the radiant areas}.
\newblock {\it \bibinfo{journal}{Bulletin of the Astronomical Institutes of
  Czechoslovakia}\/},  {\it \bibinfo{volume}{21}\/}, \bibinfo{pages}{153}.
\bibitem[{McDonnell et~al.(1987)McDonnell, Evans, Evans, Alexander, Burton,
  Firth, Bussoletti, Grard, Hanner, Sekanina, Stevenson, Turner, Weishaupt,
  Wallis \& Zarnecki}]{McDonnell1987}
\bibinfo{author}{McDonnell, J. A.~M.}, \bibinfo{author}{Evans, G.~C.},
  \bibinfo{author}{Evans, S.~T.}, \bibinfo{author}{Alexander, W.~M.},
  \bibinfo{author}{Burton, W.~M.}, \bibinfo{author}{Firth, J.~G.},
  \bibinfo{author}{Bussoletti, E.}, \bibinfo{author}{Grard, R. J.~L.},
  \bibinfo{author}{Hanner, M.~S.}, \bibinfo{author}{Sekanina, Z.},
  \bibinfo{author}{Stevenson, T.~J.}, \bibinfo{author}{Turner, R.~F.},
  \bibinfo{author}{Weishaupt, U.}, \bibinfo{author}{Wallis, M.~K.}, \&
  \bibinfo{author}{Zarnecki, J.~C.} (\bibinfo{year}{1987}).
\newblock \bibinfo{title}{{The dust distribution within the inner coma of comet
  P/Halley 1982i - Encounter by Giotto's impact detectors}}.
\newblock {\it \bibinfo{journal}{Astronomy and Astrophysics}\/},  {\it
  \bibinfo{volume}{187}\/}, \bibinfo{pages}{719--741}.
\bibitem[{McIntosh \& Hajduk(1983)}]{mcintosh1983comet}
\bibinfo{author}{McIntosh, B.}, \& \bibinfo{author}{Hajduk, A.}
  (\bibinfo{year}{1983}).
\newblock \bibinfo{title}{Comet halley meteor stream: a new model}.
\newblock {\it \bibinfo{journal}{Monthly Notices of the Royal Astronomical
  Society}\/},  {\it \bibinfo{volume}{205}\/}, \bibinfo{pages}{931--943}.
\bibitem[{Moorhead et~al.(2021)Moorhead, Clements \& Vida}]{moorhead2021meteor}
\bibinfo{author}{Moorhead, A.~V.}, \bibinfo{author}{Clements, T.}, \&
  \bibinfo{author}{Vida, D.} (\bibinfo{year}{2021}).
\newblock \bibinfo{title}{Meteor shower radiant dispersions in global meteor
  network data}.
\newblock {\it \bibinfo{journal}{Monthly Notices of the Royal Astronomical
  Society}\/},  {\it \bibinfo{volume}{508}\/}, \bibinfo{pages}{326--339}.
\bibitem[{Moorhead et~al.(2019)Moorhead, Egal, Brown, Moser \&
  Cooke}]{moorhead2019meteor}
\bibinfo{author}{Moorhead, A.~V.}, \bibinfo{author}{Egal, A.},
  \bibinfo{author}{Brown, P.~G.}, \bibinfo{author}{Moser, D.~E.}, \&
  \bibinfo{author}{Cooke, W.~J.} (\bibinfo{year}{2019}).
\newblock \bibinfo{title}{Meteor shower forecasting in near-earth space}.
\newblock {\it \bibinfo{journal}{Journal of Spacecraft and Rockets}\/},  {\it
  \bibinfo{volume}{56}\/}, \bibinfo{pages}{1531--1545}.
\bibitem[{Nelder \& Mead(1965)}]{nelder1965simplex}
\bibinfo{author}{Nelder, J.~A.}, \& \bibinfo{author}{Mead, R.}
  (\bibinfo{year}{1965}).
\newblock \bibinfo{title}{A simplex method for function minimization}.
\newblock {\it \bibinfo{journal}{The computer journal}\/},  {\it
  \bibinfo{volume}{7}\/}, \bibinfo{pages}{308--313}.
\bibitem[{Pecina \& Ceplecha(1983)}]{pecina1983new}
\bibinfo{author}{Pecina, P.}, \& \bibinfo{author}{Ceplecha, Z.}
  (\bibinfo{year}{1983}).
\newblock \bibinfo{title}{New aspects in single-body meteor physics}.
\newblock {\it \bibinfo{journal}{Bulletin of the Astronomical Institutes of
  Czechoslovakia}\/},  {\it \bibinfo{volume}{34}\/}, \bibinfo{pages}{102--121}.
\bibitem[{Picone et~al.(2002)Picone, Hedin, Drob \& Aikin}]{picone2002nrlmsise}
\bibinfo{author}{Picone, J.}, \bibinfo{author}{Hedin, A.},
  \bibinfo{author}{Drob, D.~P.}, \& \bibinfo{author}{Aikin, A.}
  (\bibinfo{year}{2002}).
\newblock \bibinfo{title}{Nrlmsise-00 empirical model of the atmosphere:
  Statistical comparisons and scientific issues}.
\newblock {\it \bibinfo{journal}{Journal of Geophysical Research: Space
  Physics}\/},  {\it \bibinfo{volume}{107}\/}.
\bibitem[{Popova et~al.(2019)Popova, Borovi{\v{c}}ka \&
  Campbell-Brown}]{Popova2019}
\bibinfo{author}{Popova, O.}, \bibinfo{author}{Borovi{\v{c}}ka, J.}, \&
  \bibinfo{author}{Campbell-Brown, M.} (\bibinfo{year}{2019}).
\newblock \bibinfo{title}{{Modelling the entry of meteoroids}}.
\newblock In \bibinfo{editor}{R.~G. O.}, \bibinfo{editor}{A.~D. J.}, \&
  \bibinfo{editor}{C.-B.~M. D} (Eds.), {\it \bibinfo{booktitle}{Meteoroids:
  Sources of Meteors on Earth and Beyond}\/} (p.~\bibinfo{pages}{9}).
\newblock \bibinfo{publisher}{Cambridge University Press}
  volume~\bibinfo{volume}{25}.
\bibitem[{Porub\v{c}an(1973)}]{porubcan1973telescopic}
\bibinfo{author}{Porub\v{c}an, V.} (\bibinfo{year}{1973}).
\newblock \bibinfo{title}{The telescopic radiant areas of the perseids and the
  orionids}.
\newblock {\it \bibinfo{journal}{Bulletin of the Astronomical Institutes of
  Czechoslovakia}\/},  {\it \bibinfo{volume}{24}\/}, \bibinfo{pages}{1}.
\bibitem[{ReVelle \& Ceplecha(2001)}]{ReVelle2001c}
\bibinfo{author}{ReVelle, D.}, \& \bibinfo{author}{Ceplecha, Z.}
  (\bibinfo{year}{2001}).
\newblock \bibinfo{title}{{Bolide physical theory with application to PN and EN
  fireballs}}.
\newblock In {\it \bibinfo{booktitle}{Proceedings of the Meteoroids 2001
  Conference, 6-10 August 2001, Kiruna, Sweden. Ed.: Barbara Warmbein. ESA
  SP-495, Noordwijk: ESA Publications Division, ISBN 92-9092-805-0, 2001, p.
  507-512}\/} (pp. \bibinfo{pages}{507--512}).
\bibitem[{Rudraswami et~al.(2014)Rudraswami, Prasad, Plane, Berg, Feng \&
  Balgar}]{Rudraswami2014}
\bibinfo{author}{Rudraswami, N.~G.}, \bibinfo{author}{Prasad, M.~S.},
  \bibinfo{author}{Plane, J.~M.}, \bibinfo{author}{Berg, T.},
  \bibinfo{author}{Feng, W.}, \& \bibinfo{author}{Balgar, S.}
  (\bibinfo{year}{2014}).
\newblock \bibinfo{title}{{Refractory metal nuggets in different types of
  cosmic spherules}}.
\newblock {\it \bibinfo{journal}{Geochimica et Cosmochimica Acta}\/},  {\it
  \bibinfo{volume}{131}\/}, \bibinfo{pages}{247--266}.
\bibitem[{Ryan et~al.(2008)Ryan, Schaefer, Destefanis \&
  Lambert}]{ryan2008ballistic}
\bibinfo{author}{Ryan, S.}, \bibinfo{author}{Schaefer, F.},
  \bibinfo{author}{Destefanis, R.}, \& \bibinfo{author}{Lambert, M.}
  (\bibinfo{year}{2008}).
\newblock \bibinfo{title}{A ballistic limit equation for hypervelocity impacts
  on composite honeycomb sandwich panel satellite structures}.
\newblock {\it \bibinfo{journal}{Advances in Space Research}\/},  {\it
  \bibinfo{volume}{41}\/}, \bibinfo{pages}{1152--1166}.
\bibitem[{Sato \& Watanabe(2007)}]{sato2007origin}
\bibinfo{author}{Sato, M.}, \& \bibinfo{author}{Watanabe, J.-i.}
  (\bibinfo{year}{2007}).
\newblock \bibinfo{title}{Origin of the 2006 orionid outburst}.
\newblock {\it \bibinfo{journal}{Publications of the Astronomical Society of
  Japan}\/},  {\it \bibinfo{volume}{59}\/}, \bibinfo{pages}{L21--L24}.
\bibitem[{Schult et~al.(2018)Schult, Brown, Pokorn{\'{y}}, Stober \&
  Chau}]{Schult2018}
\bibinfo{author}{Schult, C.}, \bibinfo{author}{Brown, P.~G.},
  \bibinfo{author}{Pokorn{\'{y}}, P.}, \bibinfo{author}{Stober, G.}, \&
  \bibinfo{author}{Chau, J.} (\bibinfo{year}{2018}).
\newblock \bibinfo{title}{{A meteoroid stream survey using meteor head echo
  observations from the Middle Atmosphere ALOMAR Radar System (MAARSY)}}.
\newblock {\it \bibinfo{journal}{Icarus}\/},  {\it \bibinfo{volume}{309}\/},
  \bibinfo{pages}{177--186}.
\bibitem[{Schulze et~al.(1997)Schulze, Kissel \& Jessberger}]{Schulze1997}
\bibinfo{author}{Schulze, H.}, \bibinfo{author}{Kissel, J.}, \&
  \bibinfo{author}{Jessberger, E.~K.} (\bibinfo{year}{1997}).
\newblock \bibinfo{title}{{Chemistry and mineralogy of comet Halley's dust}}.
\newblock In {\it \bibinfo{booktitle}{From stardust to planetesimals}\/} (p.
  \bibinfo{pages}{397}).
\newblock volume \bibinfo{volume}{122}.
\bibitem[{Sekhar \& Asher(2014)}]{sekhar2014resonant}
\bibinfo{author}{Sekhar, A.}, \& \bibinfo{author}{Asher, D.~J.}
  (\bibinfo{year}{2014}).
\newblock \bibinfo{title}{Resonant behavior of comet halley and the orionid
  stream}.
\newblock {\it \bibinfo{journal}{Meteoritics \& Planetary Science}\/},  {\it
  \bibinfo{volume}{49}\/}, \bibinfo{pages}{52--62}.
\bibitem[{Simon et~al.(2008)Simon, Joswiak, Ishii, Bradley, Chi, Grossman,
  Aleon, Brownlee, Fallon, Hutcheon et~al.}]{simon2008refractory}
\bibinfo{author}{Simon, S.~B.}, \bibinfo{author}{Joswiak, D.},
  \bibinfo{author}{Ishii, H.}, \bibinfo{author}{Bradley, J.~P.},
  \bibinfo{author}{Chi, M.}, \bibinfo{author}{Grossman, L.},
  \bibinfo{author}{Aleon, J.}, \bibinfo{author}{Brownlee, D.},
  \bibinfo{author}{Fallon, S.}, \bibinfo{author}{Hutcheon, I.~D.} et~al.
  (\bibinfo{year}{2008}).
\newblock \bibinfo{title}{A refractory inclusion returned by stardust from
  comet 81p/wild 2}.
\newblock {\it \bibinfo{journal}{Meteoritics \& Planetary Science}\/},  {\it
  \bibinfo{volume}{43}\/}, \bibinfo{pages}{1861--1877}.
\bibitem[{Spurn{\`y} \& Shrben{\`y}(2008)}]{spurny2008exceptional}
\bibinfo{author}{Spurn{\`y}, P.}, \& \bibinfo{author}{Shrben{\`y}, L.}
  (\bibinfo{year}{2008}).
\newblock \bibinfo{title}{Exceptional fireball activity of orionids in 2006}.
\newblock {\it \bibinfo{journal}{Earth, Moon, and Planets}\/},  {\it
  \bibinfo{volume}{102}\/}, \bibinfo{pages}{141--150}.
\bibitem[{Subasinghe \& Campbell-Brown(2018)}]{subasinghe2018luminous}
\bibinfo{author}{Subasinghe, D.}, \& \bibinfo{author}{Campbell-Brown, M.}
  (\bibinfo{year}{2018}).
\newblock \bibinfo{title}{Luminous efficiency estimates of meteors. ii.
  application to canadian automated meteor observatory meteor events}.
\newblock {\it \bibinfo{journal}{The Astronomical Journal}\/},  {\it
  \bibinfo{volume}{155}\/}, \bibinfo{pages}{88}.
\bibitem[{Subasinghe et~al.(2017)Subasinghe, Campbell-Brown \&
  Stokan}]{subasinghe2017luminous}
\bibinfo{author}{Subasinghe, D.}, \bibinfo{author}{Campbell-Brown, M.}, \&
  \bibinfo{author}{Stokan, E.} (\bibinfo{year}{2017}).
\newblock \bibinfo{title}{Luminous efficiency estimates of meteors-i.
  uncertainty analysis}.
\newblock {\it \bibinfo{journal}{Planetary and Space Science}\/},  {\it
  \bibinfo{volume}{143}\/}, \bibinfo{pages}{71--77}.
\bibitem[{Subasinghe et~al.(2016)Subasinghe, Campbell-Brown \&
  Stokan}]{subasinghe2016physical}
\bibinfo{author}{Subasinghe, D.}, \bibinfo{author}{Campbell-Brown, M.~D.}, \&
  \bibinfo{author}{Stokan, E.} (\bibinfo{year}{2016}).
\newblock \bibinfo{title}{Physical characteristics of faint meteors by light
  curve and high-resolution observations, and the implications for parent
  bodies}.
\newblock {\it \bibinfo{journal}{Monthly Notices of the Royal Astronomical
  Society}\/},  {\it \bibinfo{volume}{457}\/}, \bibinfo{pages}{1289--1298}.
\bibitem[{T{\'a}rano et~al.(2019)T{\'a}rano, Wheeler, Close \&
  Mathias}]{tarano2019inference}
\bibinfo{author}{T{\'a}rano, A.~M.}, \bibinfo{author}{Wheeler, L.~F.},
  \bibinfo{author}{Close, S.}, \& \bibinfo{author}{Mathias, D.~L.}
  (\bibinfo{year}{2019}).
\newblock \bibinfo{title}{Inference of meteoroid characteristics using a
  genetic algorithm}.
\newblock {\it \bibinfo{journal}{Icarus}\/},  {\it \bibinfo{volume}{329}\/},
  \bibinfo{pages}{270--281}.
\bibitem[{Vaubaillon et~al.(2005)Vaubaillon, Colas \&
  Jorda}]{vaubaillon2005new}
\bibinfo{author}{Vaubaillon, J.}, \bibinfo{author}{Colas, F.}, \&
  \bibinfo{author}{Jorda, L.} (\bibinfo{year}{2005}).
\newblock \bibinfo{title}{A new method to predict meteor showers-i. description
  of the model}.
\newblock {\it \bibinfo{journal}{Astronomy \& Astrophysics}\/},  {\it
  \bibinfo{volume}{439}\/}, \bibinfo{pages}{751--760}.
\bibitem[{Vaubaillon et~al.(2020)Vaubaillon, Egal, Desmars \&
  Baillie}]{vaubaillon2020meteor}
\bibinfo{author}{Vaubaillon, J.}, \bibinfo{author}{Egal, A.},
  \bibinfo{author}{Desmars, J.}, \& \bibinfo{author}{Baillie, K.}
  (\bibinfo{year}{2020}).
\newblock \bibinfo{title}{Meteor shower output caused by comet 15p/finlay}.
\newblock {\it \bibinfo{journal}{WGN, Journal of the International Meteor
  Organization}\/},  {\it \bibinfo{volume}{48}\/}, \bibinfo{pages}{29--35}.
\bibitem[{Verniani(1967)}]{Verniani1967}
\bibinfo{author}{Verniani, F.} (\bibinfo{year}{1967}).
\newblock \bibinfo{title}{{Meteor masses and luminosity}}.
\newblock {\it \bibinfo{journal}{Smithsonian contributions to astrophysics}\/},
  .
\bibitem[{Vida et~al.(2022)Vida, Blaauw~Erskine, Brown, Kambulow,
  Campbell-Brown \& Mazur}]{vida2022flux}
\bibinfo{author}{Vida, D.}, \bibinfo{author}{Blaauw~Erskine, R.~C.},
  \bibinfo{author}{Brown, P.~G.}, \bibinfo{author}{Kambulow, J.},
  \bibinfo{author}{Campbell-Brown, M.}, \& \bibinfo{author}{Mazur, M.~J.}
  (\bibinfo{year}{2022}).
\newblock \bibinfo{title}{Computing optical meteor flux using global meteor
  network data}.
\newblock {\it \bibinfo{journal}{Monthly Notices of the Royal Astronomical
  Society}\/},  {\it \bibinfo{volume}{515}\/}, \bibinfo{pages}{2322--2339}.
\bibitem[{Vida et~al.(2018)Vida, Brown \& Campbell-Brown}]{vida2018modelling}
\bibinfo{author}{Vida, D.}, \bibinfo{author}{Brown, P.~G.}, \&
  \bibinfo{author}{Campbell-Brown, M.} (\bibinfo{year}{2018}).
\newblock \bibinfo{title}{Modelling the measurement accuracy of pre-atmosphere
  velocities of meteoroids}.
\newblock {\it \bibinfo{journal}{Monthly Notices of the Royal Astronomical
  Society}\/},  {\it \bibinfo{volume}{479}\/}, \bibinfo{pages}{4307--4319}.
\bibitem[{Vida et~al.(2021{\natexlab{a}})Vida, Brown, Campbell-Brown, Weryk,
  Stober \& McCormack}]{vida2021high}
\bibinfo{author}{Vida, D.}, \bibinfo{author}{Brown, P.~G.},
  \bibinfo{author}{Campbell-Brown, M.}, \bibinfo{author}{Weryk, R.~J.},
  \bibinfo{author}{Stober, G.}, \& \bibinfo{author}{McCormack, J.~P.}
  (\bibinfo{year}{2021}{\natexlab{a}}).
\newblock \bibinfo{title}{High precision meteor observations with the canadian
  automated meteor observatory: Data reduction pipeline and application to
  meteoroid mechanical strength measurements}.
\newblock {\it \bibinfo{journal}{Icarus}\/},  {\it \bibinfo{volume}{354}\/},
  \bibinfo{pages}{114097}.
\bibitem[{Vida et~al.(2020{\natexlab{a}})Vida, Brown, Campbell-Brown, Wiegert
  \& Gural}]{vida2019meteorresults}
\bibinfo{author}{Vida, D.}, \bibinfo{author}{Brown, P.~G.},
  \bibinfo{author}{Campbell-Brown, M.}, \bibinfo{author}{Wiegert, P.}, \&
  \bibinfo{author}{Gural, P.~S.} (\bibinfo{year}{2020}{\natexlab{a}}).
\newblock \bibinfo{title}{Estimating trajectories of meteors: an observational
  monte carlo approach--ii. results}.
\newblock {\it \bibinfo{journal}{Monthly Notices of the Royal Astronomical
  Society}\/},  {\it \bibinfo{volume}{491}\/}, \bibinfo{pages}{3996--4011}.
\bibitem[{Vida et~al.(2023)Vida, Brown, Devillepoix, Wiegert, Moser,
  Matlovi{\v{c}}, Herd, Hill, Sansom, Towner et~al.}]{vida2023direct}
\bibinfo{author}{Vida, D.}, \bibinfo{author}{Brown, P.~G.},
  \bibinfo{author}{Devillepoix, H.~A.}, \bibinfo{author}{Wiegert, P.},
  \bibinfo{author}{Moser, D.~E.}, \bibinfo{author}{Matlovi{\v{c}}, P.},
  \bibinfo{author}{Herd, C.~D.}, \bibinfo{author}{Hill, P.~J.},
  \bibinfo{author}{Sansom, E.~K.}, \bibinfo{author}{Towner, M.~C.} et~al.
  (\bibinfo{year}{2023}).
\newblock \bibinfo{title}{Direct measurement of decimetre-sized rocky material
  in the oort cloud}.
\newblock {\it \bibinfo{journal}{Nature Astronomy}\/},  {\it
  \bibinfo{volume}{7}\/}, \bibinfo{pages}{318--329}.
\bibitem[{Vida et~al.(2020{\natexlab{b}})Vida, Campbell-Brown, Brown, Egal \&
  Mazur}]{vida2020new}
\bibinfo{author}{Vida, D.}, \bibinfo{author}{Campbell-Brown, M.},
  \bibinfo{author}{Brown, P.~G.}, \bibinfo{author}{Egal, A.}, \&
  \bibinfo{author}{Mazur, M.~J.} (\bibinfo{year}{2020}{\natexlab{b}}).
\newblock \bibinfo{title}{A new method for measuring the meteor mass index:
  application to the 2018 draconid meteor shower outburst}.
\newblock {\it \bibinfo{journal}{Astronomy \& Astrophysics}\/},  {\it
  \bibinfo{volume}{635}\/}, \bibinfo{pages}{A153}.
\bibitem[{Vida et~al.(2020{\natexlab{c}})Vida, Gural, Brown, Campbell-Brown \&
  Wiegert}]{vida2019meteortheory}
\bibinfo{author}{Vida, D.}, \bibinfo{author}{Gural, P.~S.},
  \bibinfo{author}{Brown, P.~G.}, \bibinfo{author}{Campbell-Brown, M.}, \&
  \bibinfo{author}{Wiegert, P.} (\bibinfo{year}{2020}{\natexlab{c}}).
\newblock \bibinfo{title}{Estimating trajectories of meteors: an observational
  monte carlo approach--i. theory}.
\newblock {\it \bibinfo{journal}{Monthly Notices of the Royal Astronomical
  Society}\/},  {\it \bibinfo{volume}{491}\/}, \bibinfo{pages}{2688--2705}.
\bibitem[{Vida et~al.(2021{\natexlab{b}})Vida, {\v{S}}egon, Gural, Brown,
  McIntyre, Dijkema, Pavleti{\'c}, Kuki{\'c}, Mazur, Eschman
  et~al.}]{vida2021global}
\bibinfo{author}{Vida, D.}, \bibinfo{author}{{\v{S}}egon, D.},
  \bibinfo{author}{Gural, P.~S.}, \bibinfo{author}{Brown, P.~G.},
  \bibinfo{author}{McIntyre, M.~J.}, \bibinfo{author}{Dijkema, T.~J.},
  \bibinfo{author}{Pavleti{\'c}, L.}, \bibinfo{author}{Kuki{\'c}, P.},
  \bibinfo{author}{Mazur, M.~J.}, \bibinfo{author}{Eschman, P.} et~al.
  (\bibinfo{year}{2021}{\natexlab{b}}).
\newblock \bibinfo{title}{The global meteor network--methodology and first
  results}.
\newblock {\it \bibinfo{journal}{Monthly Notices of the Royal Astronomical
  Society}\/},  {\it \bibinfo{volume}{506}\/}, \bibinfo{pages}{5046--5074}.
\bibitem[{Voj{\'a}{\v{c}}ek et~al.(2019)Voj{\'a}{\v{c}}ek, Borovi{\v{c}}ka,
  Koten, Spurn{\`y} \& {\v{S}}tork}]{vojavcek2019properties}
\bibinfo{author}{Voj{\'a}{\v{c}}ek, V.}, \bibinfo{author}{Borovi{\v{c}}ka, J.},
  \bibinfo{author}{Koten, P.}, \bibinfo{author}{Spurn{\`y}, P.}, \&
  \bibinfo{author}{{\v{S}}tork, R.} (\bibinfo{year}{2019}).
\newblock \bibinfo{title}{Properties of small meteoroids studied by meteor
  video observations}.
\newblock {\it \bibinfo{journal}{Astronomy \& Astrophysics}\/},  {\it
  \bibinfo{volume}{621}\/}, \bibinfo{pages}{A68}.
\bibitem[{Weryk \& Brown(2013)}]{weryk2013simultaneous}
\bibinfo{author}{Weryk, R.~J.}, \& \bibinfo{author}{Brown, P.~G.}
  (\bibinfo{year}{2013}).
\newblock \bibinfo{title}{Simultaneous radar and video meteors—ii: Photometry
  and ionisation}.
\newblock {\it \bibinfo{journal}{Planetary and Space Science}\/},  {\it
  \bibinfo{volume}{81}\/}, \bibinfo{pages}{32--47}.
\bibitem[{Ye et~al.(2014)Ye, Wiegert, Brown, Campbell-Brown \&
  Weryk}]{ye2014unexpected}
\bibinfo{author}{Ye, Q.}, \bibinfo{author}{Wiegert, P.~A.},
  \bibinfo{author}{Brown, P.~G.}, \bibinfo{author}{Campbell-Brown, M.~D.}, \&
  \bibinfo{author}{Weryk, R.~J.} (\bibinfo{year}{2014}).
\newblock \bibinfo{title}{The unexpected 2012 draconid meteor storm}.
\newblock {\it \bibinfo{journal}{Monthly Notices of the Royal Astronomical
  Society}\/},  {\it \bibinfo{volume}{437}\/}, \bibinfo{pages}{3812--3823}.
\bibitem[{Ye et~al.(2015)Ye, Brown, Bell, Gao, Ma{\v{s}}ek \&
  Hui}]{ye2015bangs}
\bibinfo{author}{Ye, Q.-Z.}, \bibinfo{author}{Brown, P.~G.},
  \bibinfo{author}{Bell, C.}, \bibinfo{author}{Gao, X.},
  \bibinfo{author}{Ma{\v{s}}ek, M.}, \& \bibinfo{author}{Hui, M.-T.}
  (\bibinfo{year}{2015}).
\newblock \bibinfo{title}{Bangs and meteors from the quiet comet 15p/finlay}.
\newblock {\it \bibinfo{journal}{The Astrophysical Journal}\/},  {\it
  \bibinfo{volume}{814}\/}, \bibinfo{pages}{79}.
\bibitem[{Yeomans \& Kiang(1981)}]{yeomans1981long}
\bibinfo{author}{Yeomans, D.~K.}, \& \bibinfo{author}{Kiang, T.}
  (\bibinfo{year}{1981}).
\newblock \bibinfo{title}{The long-term motion of comet halley}.
\newblock {\it \bibinfo{journal}{Monthly Notices of the Royal Astronomical
  Society}\/},  {\it \bibinfo{volume}{197}\/}, \bibinfo{pages}{633--646}.
\bibitem[{Zangmeister et~al.(2014)Zangmeister, Radney, Dockery, Young, Ma, You
  \& Zachariah}]{zangmeister2014packing}
\bibinfo{author}{Zangmeister, C.~D.}, \bibinfo{author}{Radney, J.~G.},
  \bibinfo{author}{Dockery, L.~T.}, \bibinfo{author}{Young, J.~T.},
  \bibinfo{author}{Ma, X.}, \bibinfo{author}{You, R.}, \&
  \bibinfo{author}{Zachariah, M.~R.} (\bibinfo{year}{2014}).
\newblock \bibinfo{title}{Packing density of rigid aggregates is independent of
  scale}.
\newblock {\it \bibinfo{journal}{Proceedings of the National Academy of
  Sciences}\/},  {\it \bibinfo{volume}{111}\/}, \bibinfo{pages}{9037--9041}.
\bibitem[{Znojil(1968)}]{znojil1968frequency}
\bibinfo{author}{Znojil, V.} (\bibinfo{year}{1968}).
\newblock \bibinfo{title}{Frequency occurrence of small particles in meteor
  showers. ii. orionids, $\varepsilon$ geminids}.
\newblock {\it \bibinfo{journal}{Bulletin of the Astronomical Institutes of
  Czechoslovakia}\/},  {\it \bibinfo{volume}{19}\/}, \bibinfo{pages}{306}.

\end{thebibliography}

\newpage

\appendix
\section{Comparison between observations and model fits} \label{app:obs_sim_comp}

\begin{figure*}
    \includegraphics[width=\linewidth]{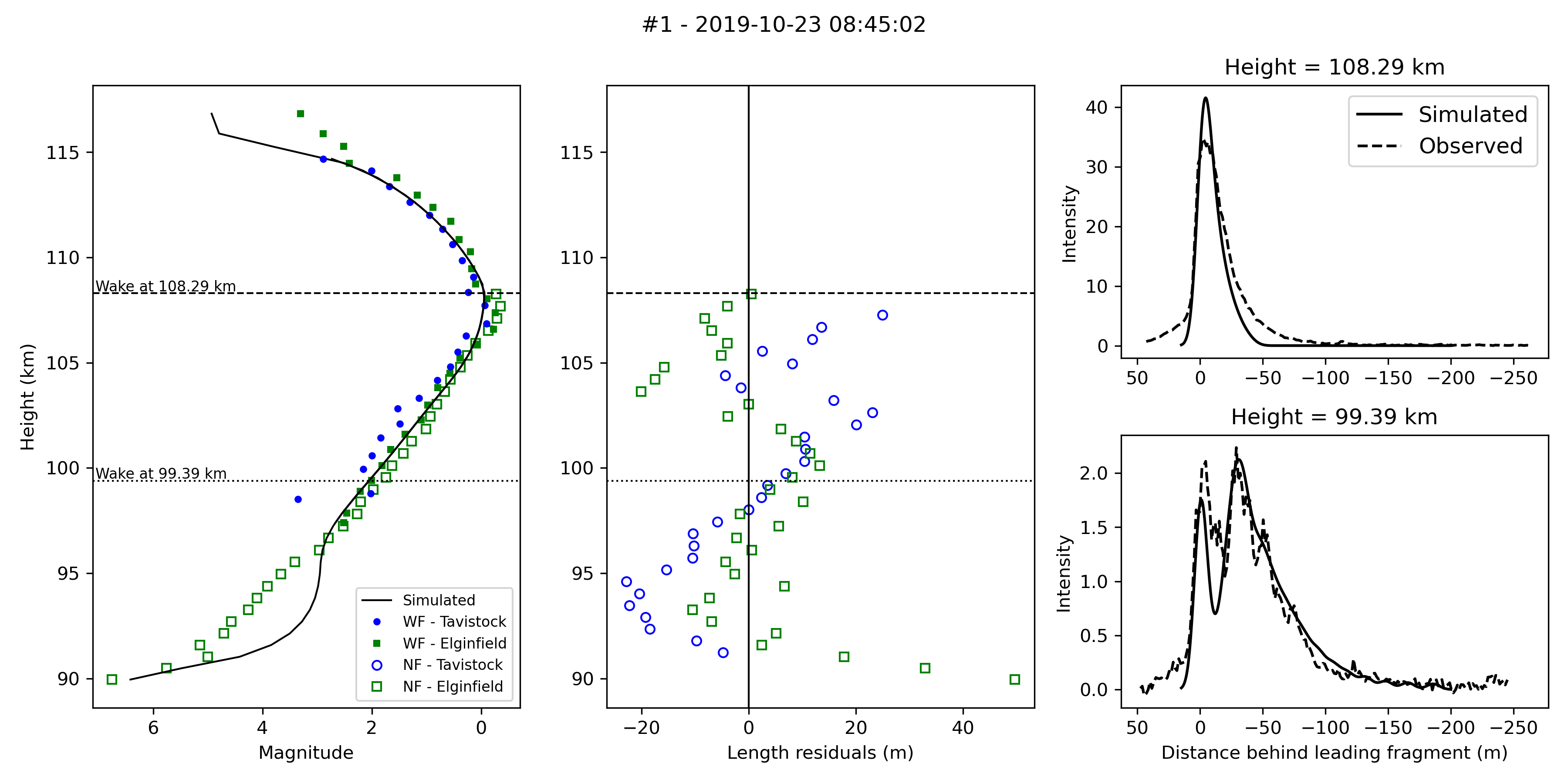}
    \includegraphics[width=\linewidth]{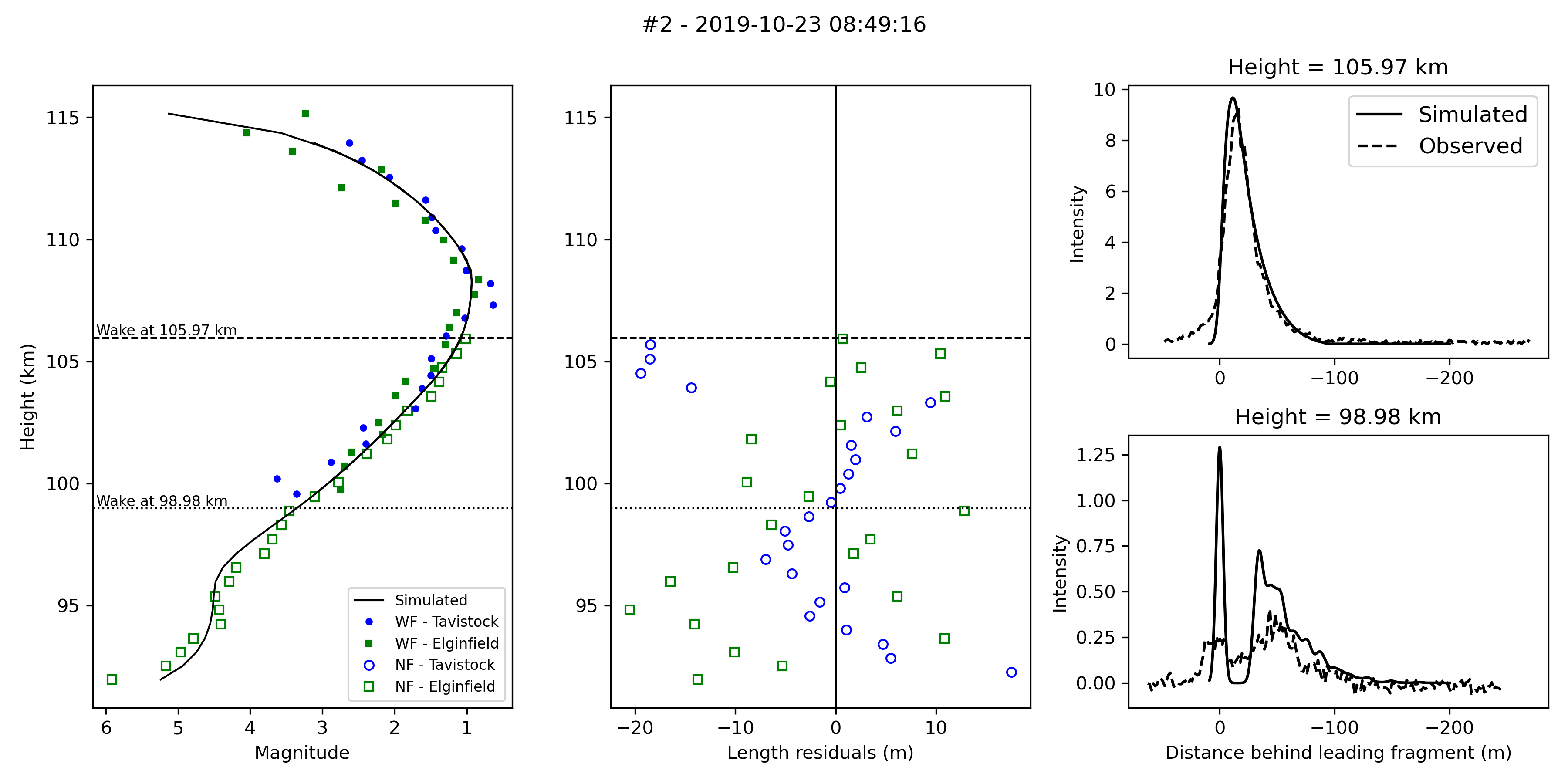}
    \includegraphics[width=\linewidth]{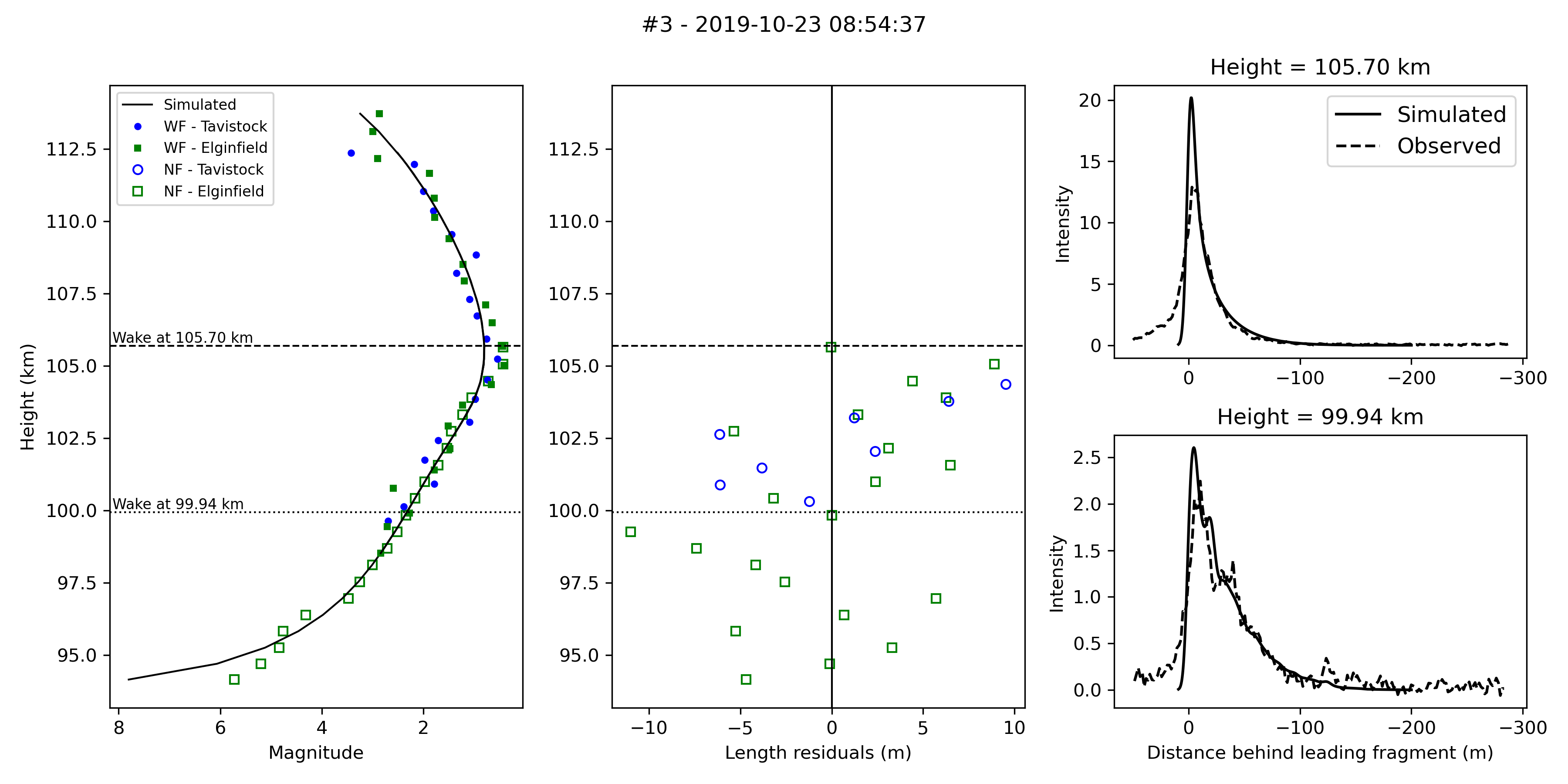}
    \caption[Comparison between observations and simulations]{Comparison between observations and simulations. WF are wide-field and NF are narrow-field observations. Narrow-field observations are delayed until the mirrors lock onto the meteor. Left: Magnitude comparison. Middle: Residuals in length vs. time. Right: Wake at two sampled height, at the beginning of tracking and midway until the end of tracking.}
  \label{fig:obs_sim_comp1}
\end{figure*}

\begin{figure*}
    \includegraphics[width=\linewidth]{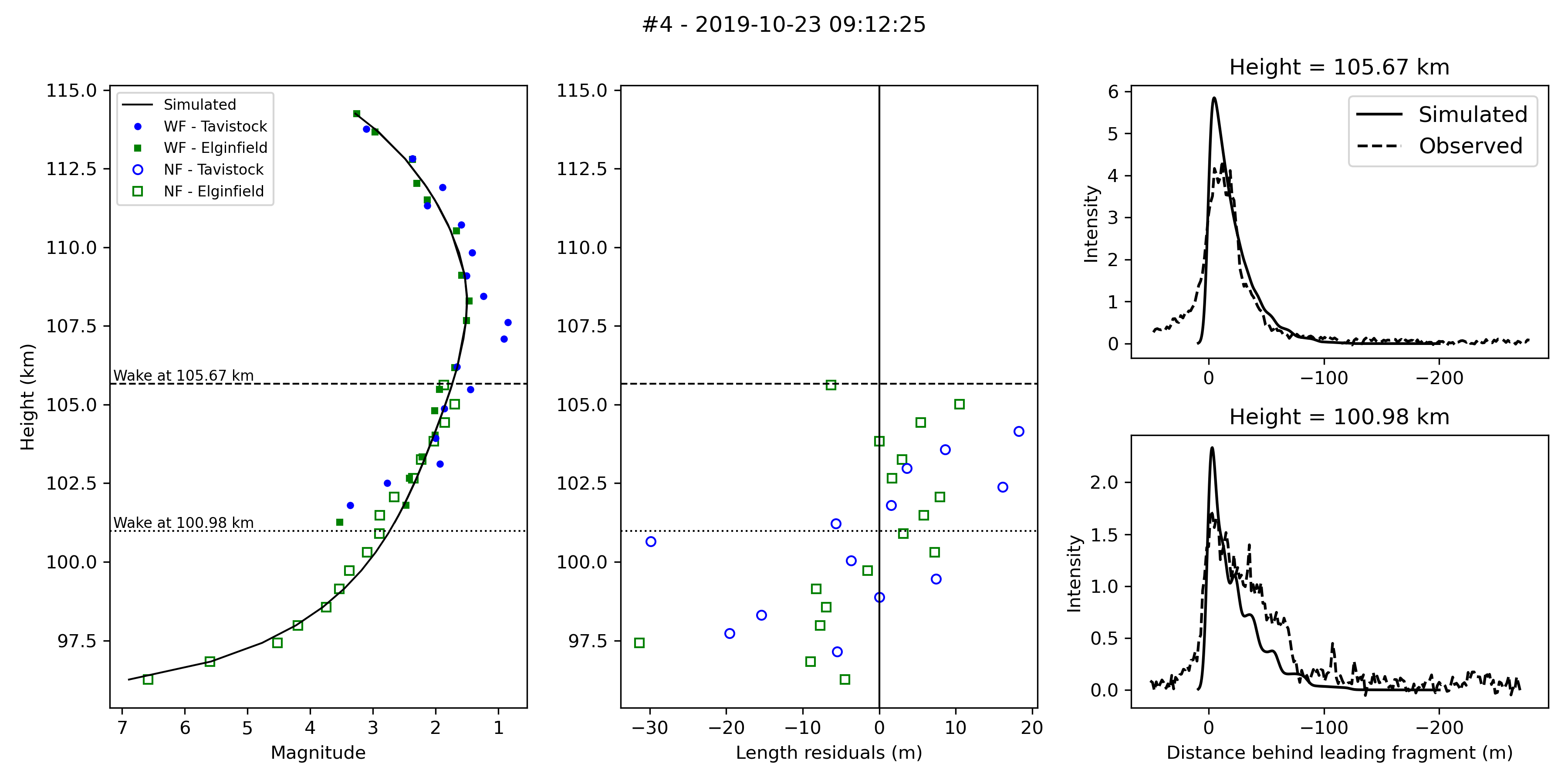}
    \includegraphics[width=\linewidth]{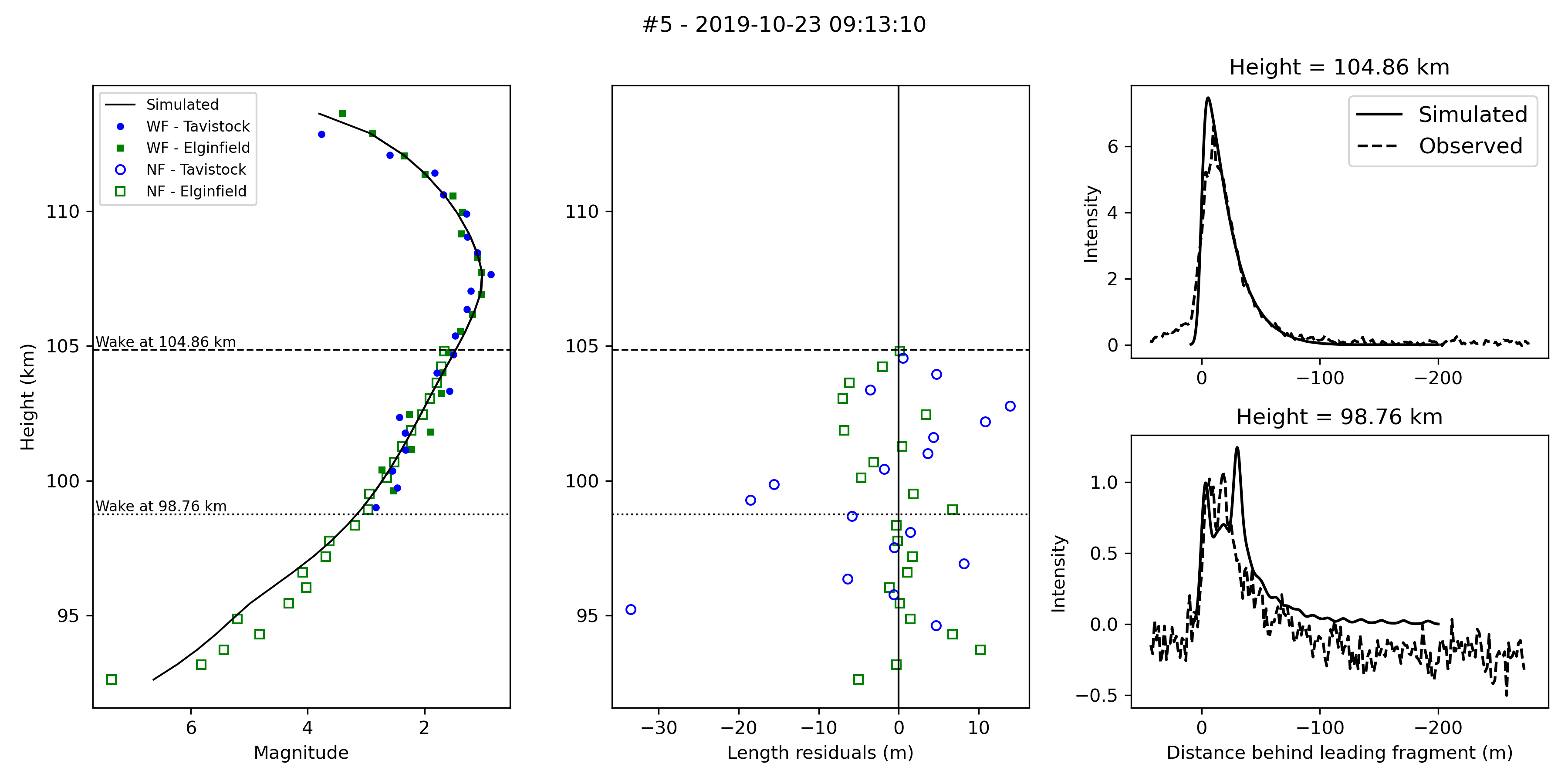}
    \includegraphics[width=\linewidth]{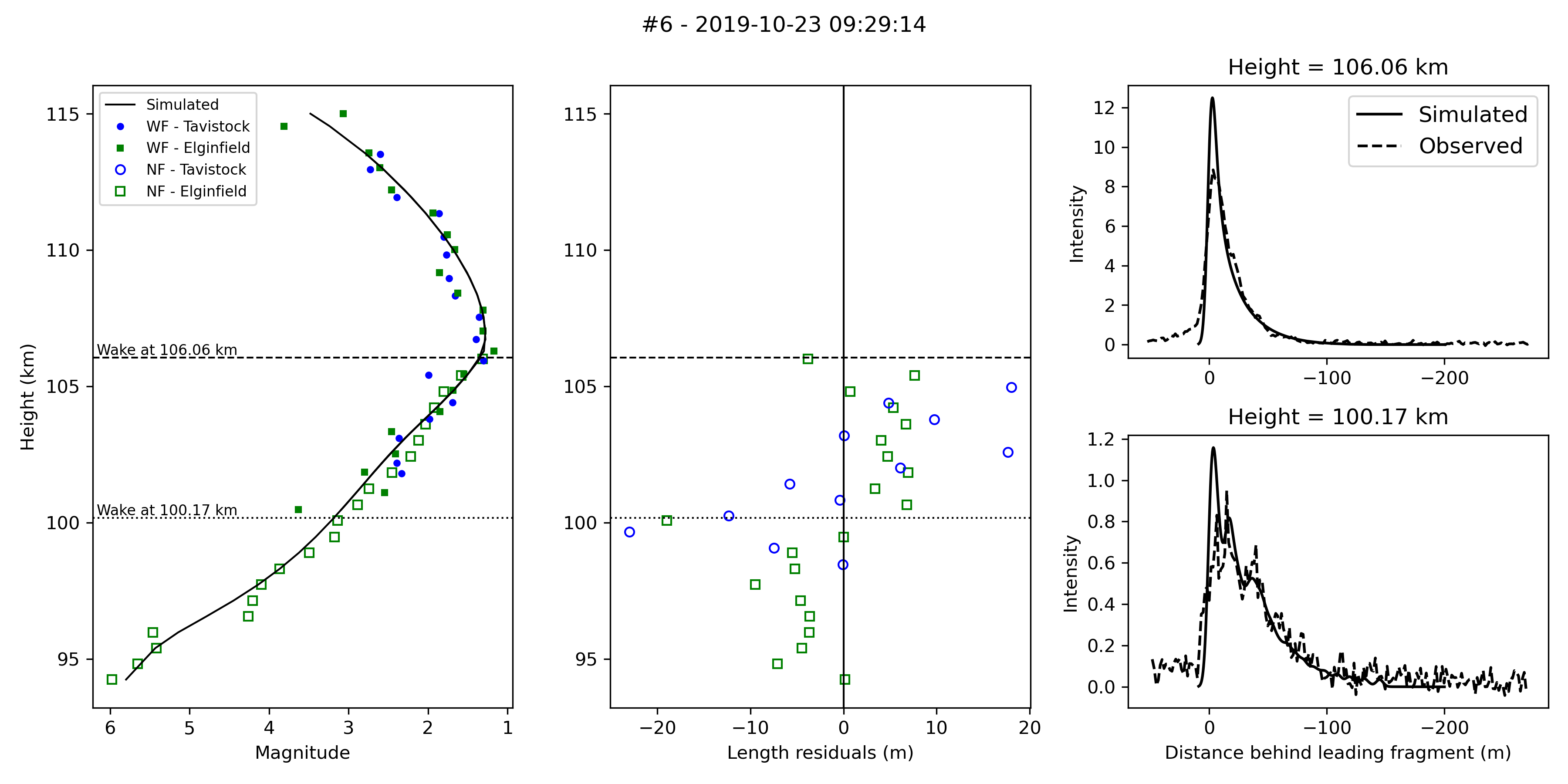}
    \caption[Comparison between observations and simulations]{Comparison between observations and simulations.}
  \label{fig:obs_sim_comp2}
\end{figure*}

\begin{figure*}
    \includegraphics[width=\linewidth]{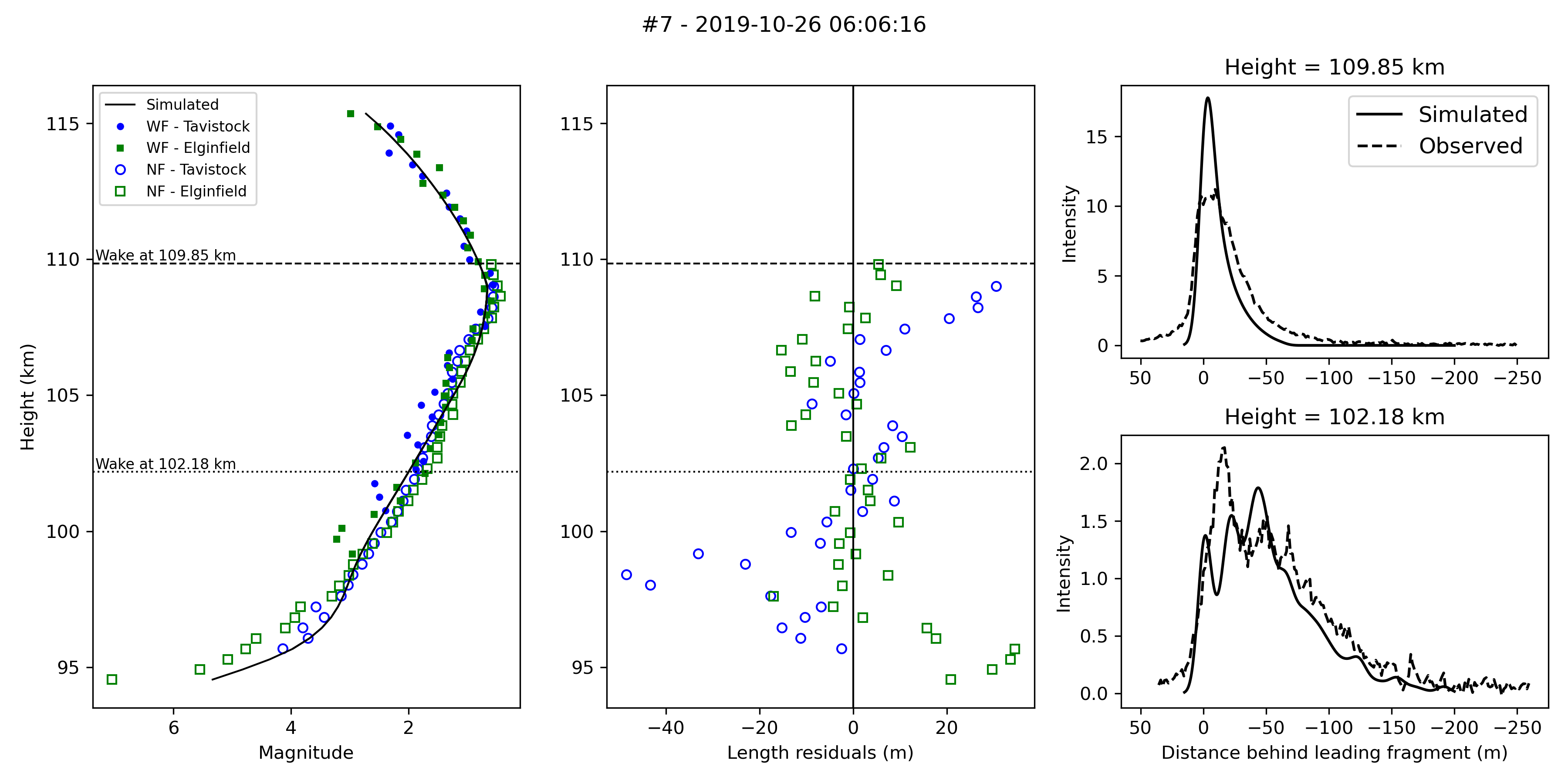}
    \includegraphics[width=\linewidth]{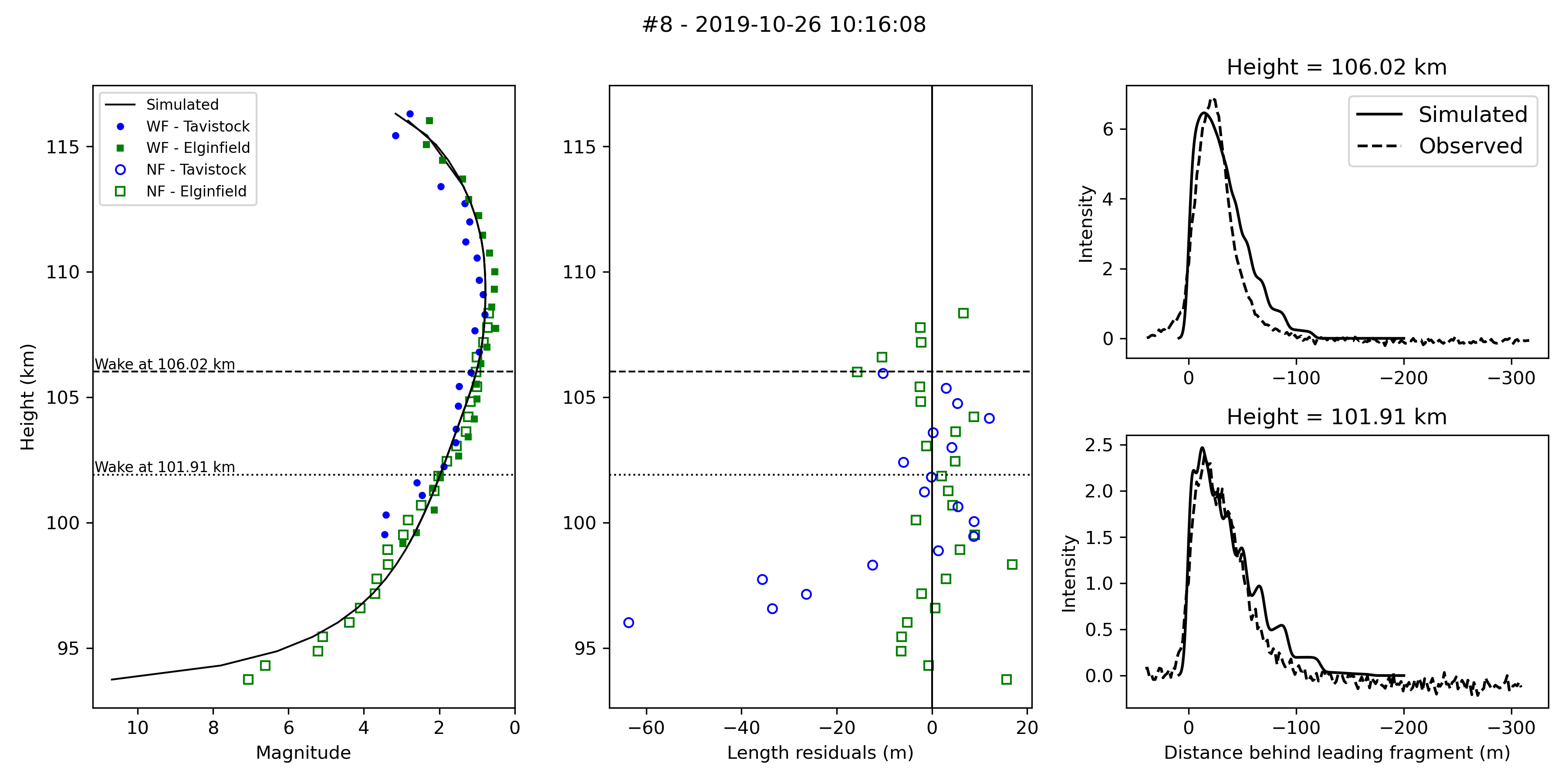}
    \includegraphics[width=\linewidth]{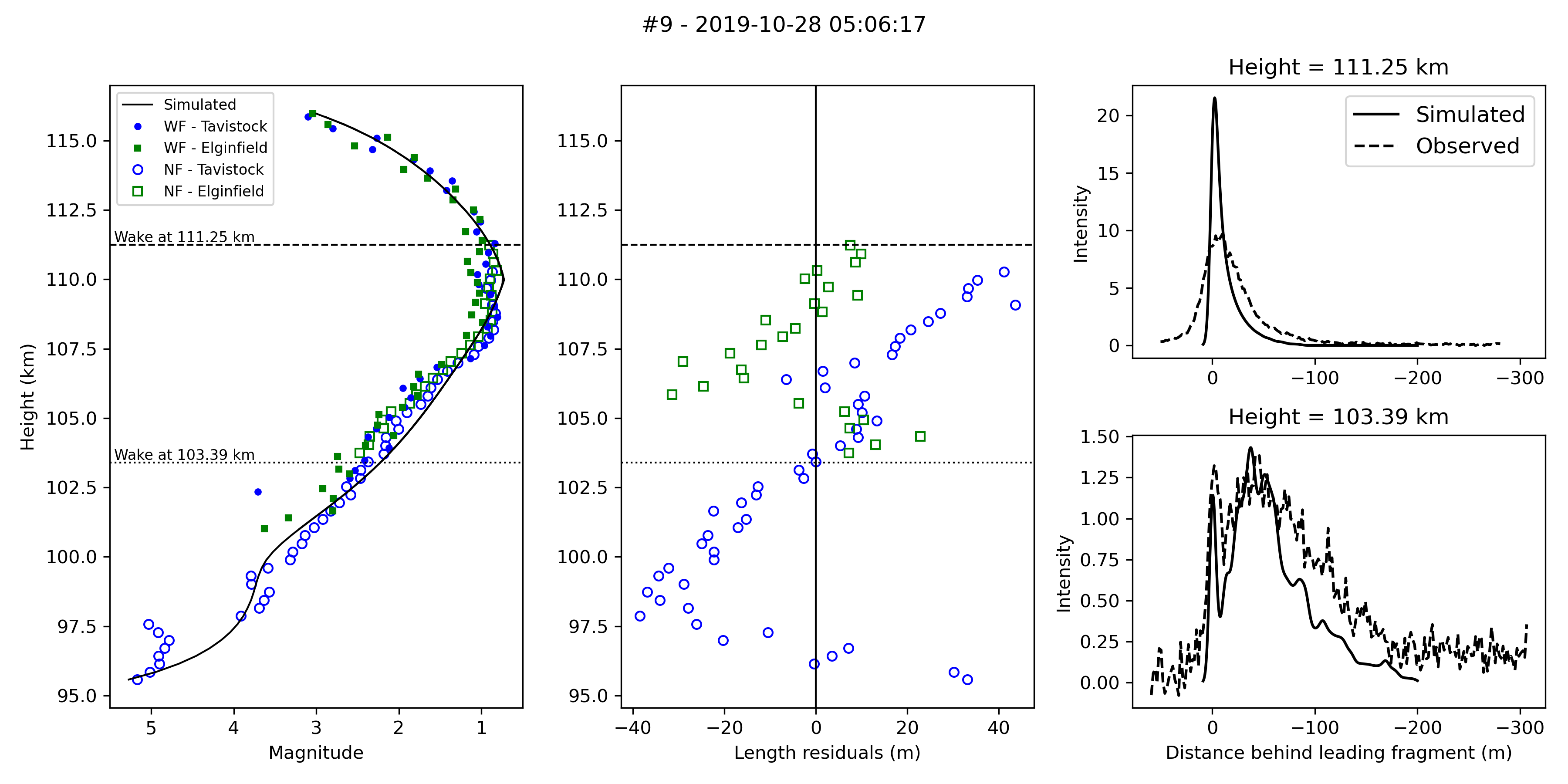}
    \caption[Comparison between observations and simulations]{Comparison between observations and simulations.}
  \label{fig:obs_sim_comp3}
\end{figure*}

\begin{figure*}
    \includegraphics[width=\linewidth]{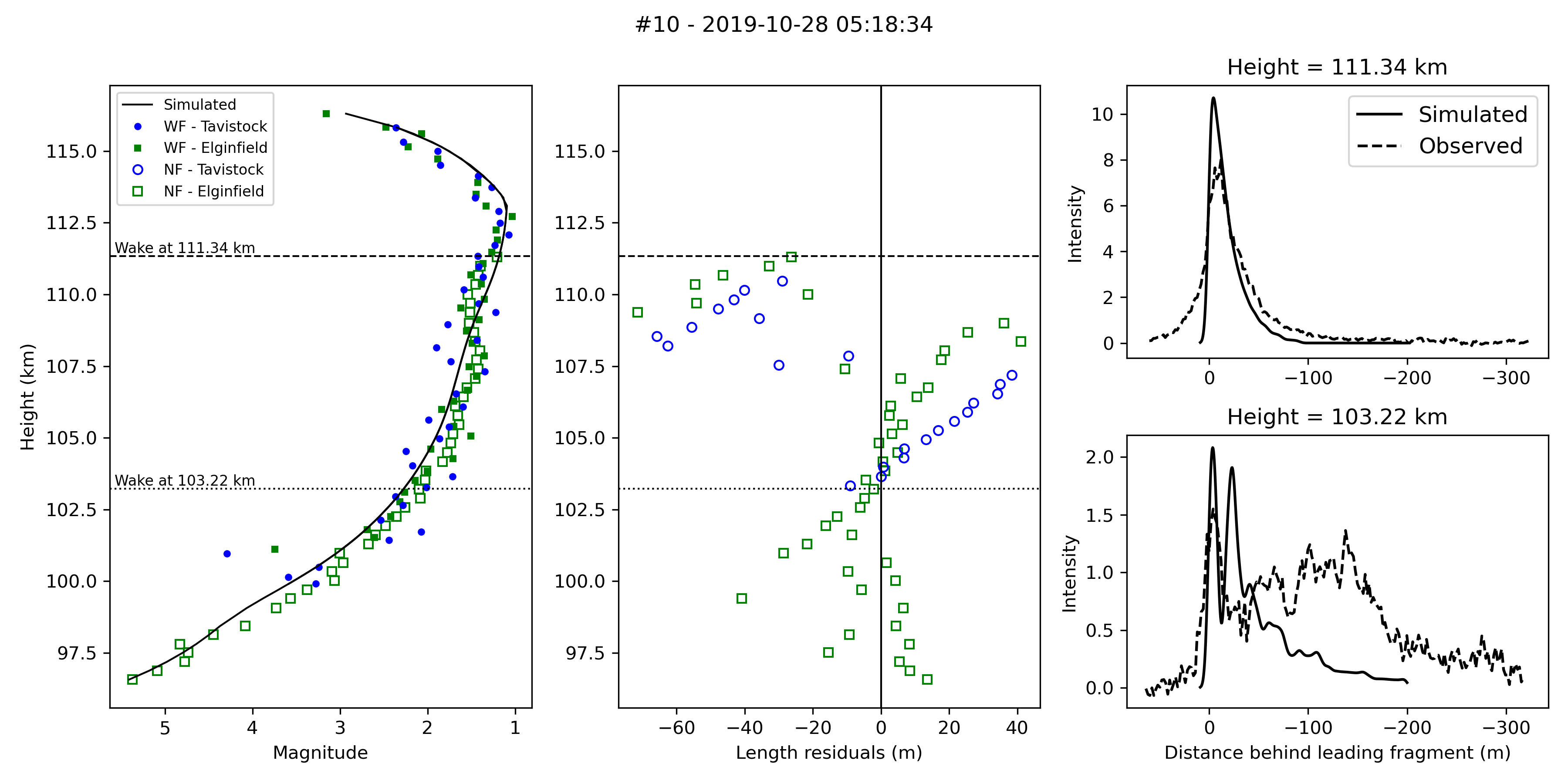}
    \includegraphics[width=\linewidth]{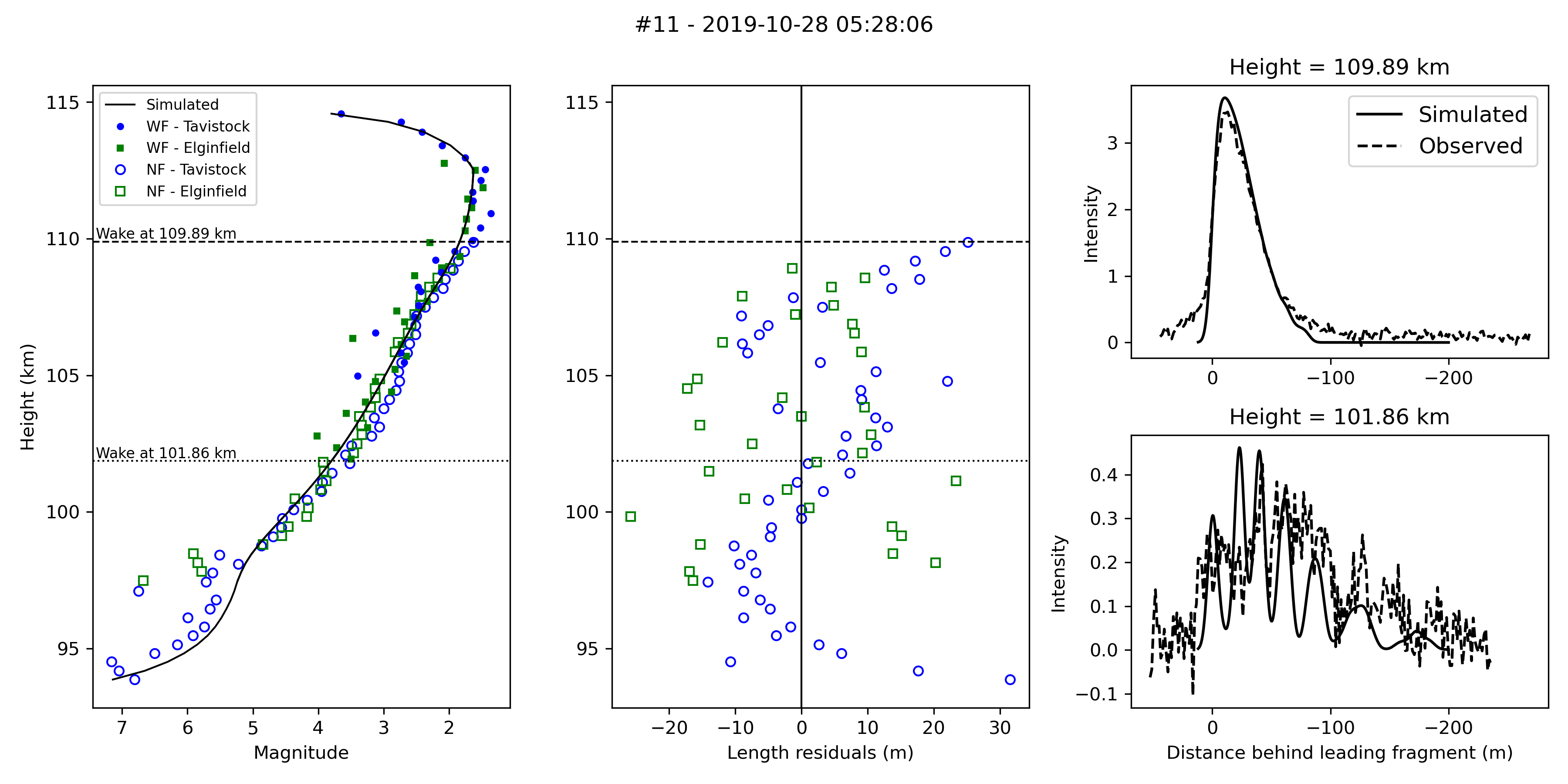}
    \includegraphics[width=\linewidth]{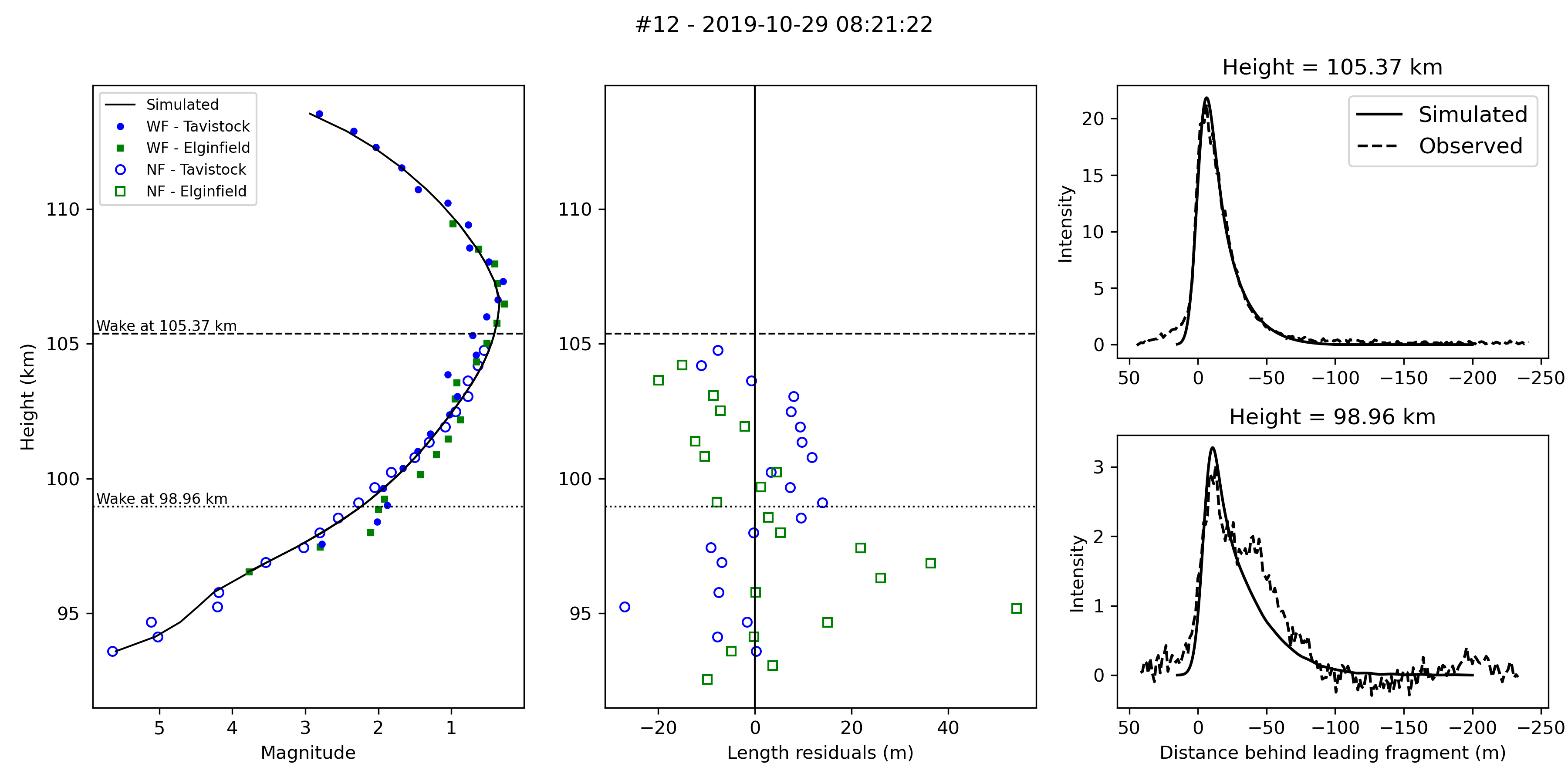}
    \caption[Comparison between observations and simulations]{Comparison between observations and simulations.}
  \label{fig:obs_sim_comp4}
\end{figure*}

\begin{figure*}
    \includegraphics[width=\linewidth]{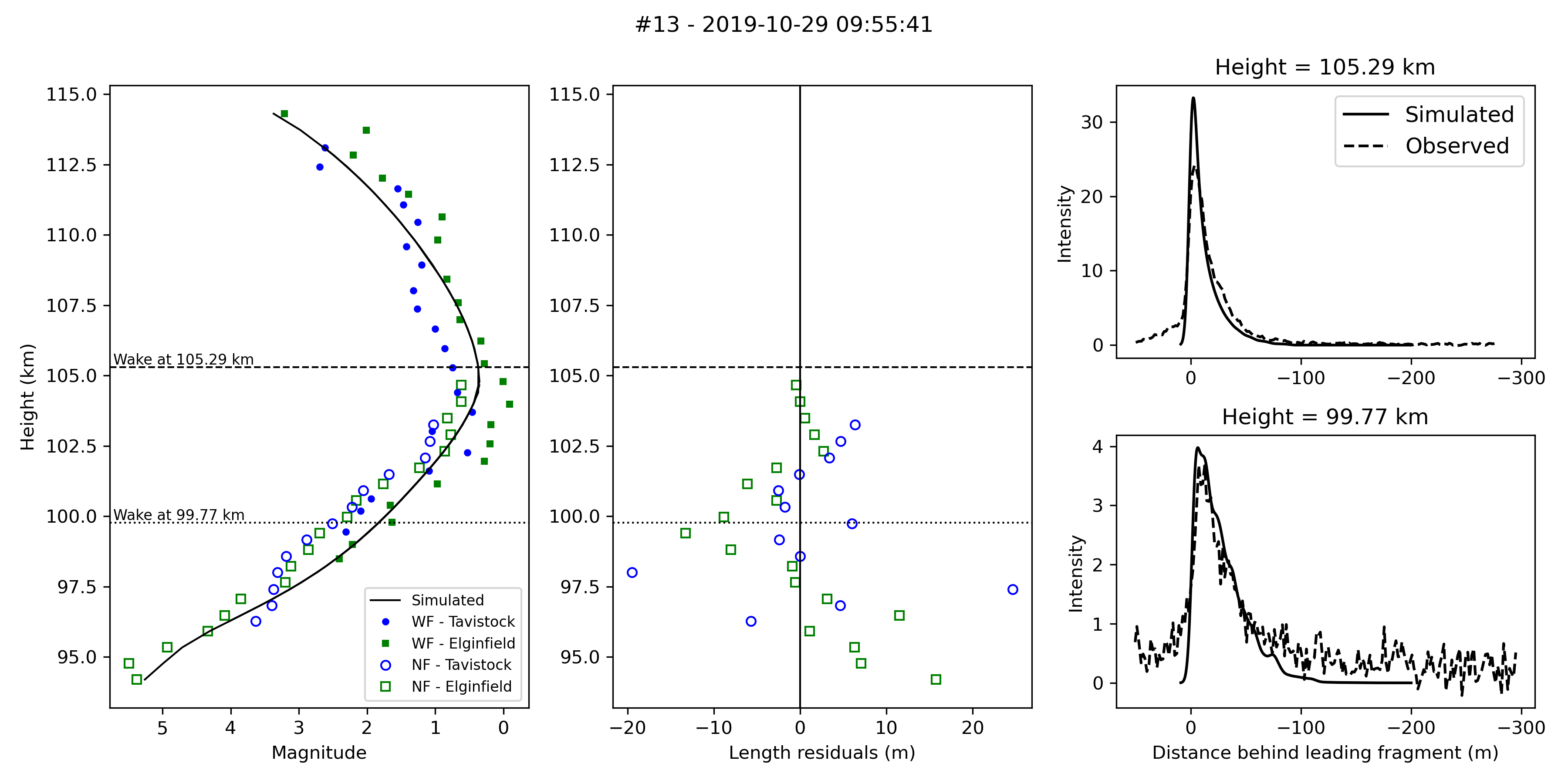}
    \includegraphics[width=\linewidth]{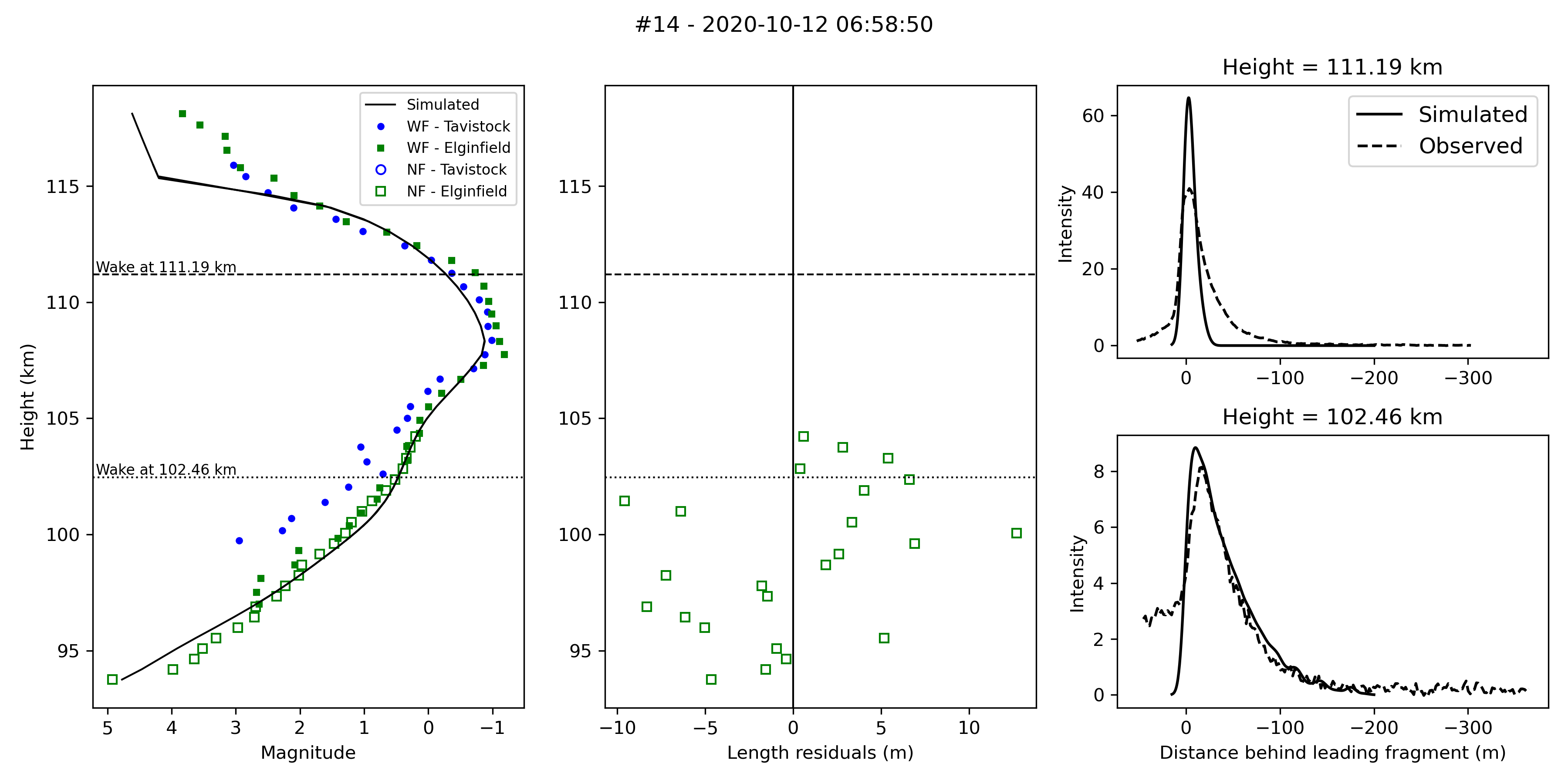}
    \includegraphics[width=\linewidth]{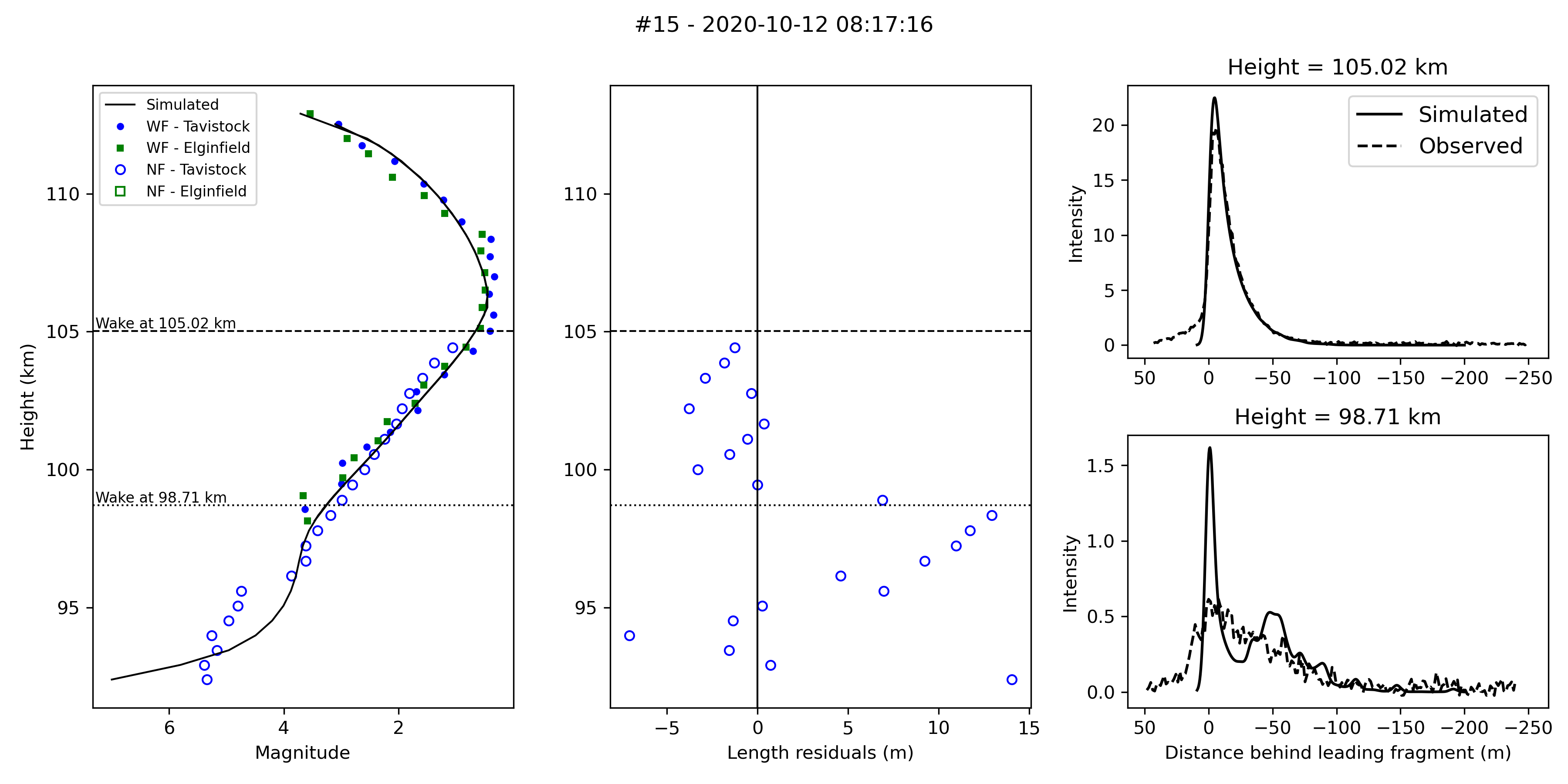}
    \caption[Comparison between observations and simulations]{Comparison between observations and simulations.}
  \label{fig:obs_sim_comp5}
\end{figure*}

\begin{figure*}
    \includegraphics[width=\linewidth]{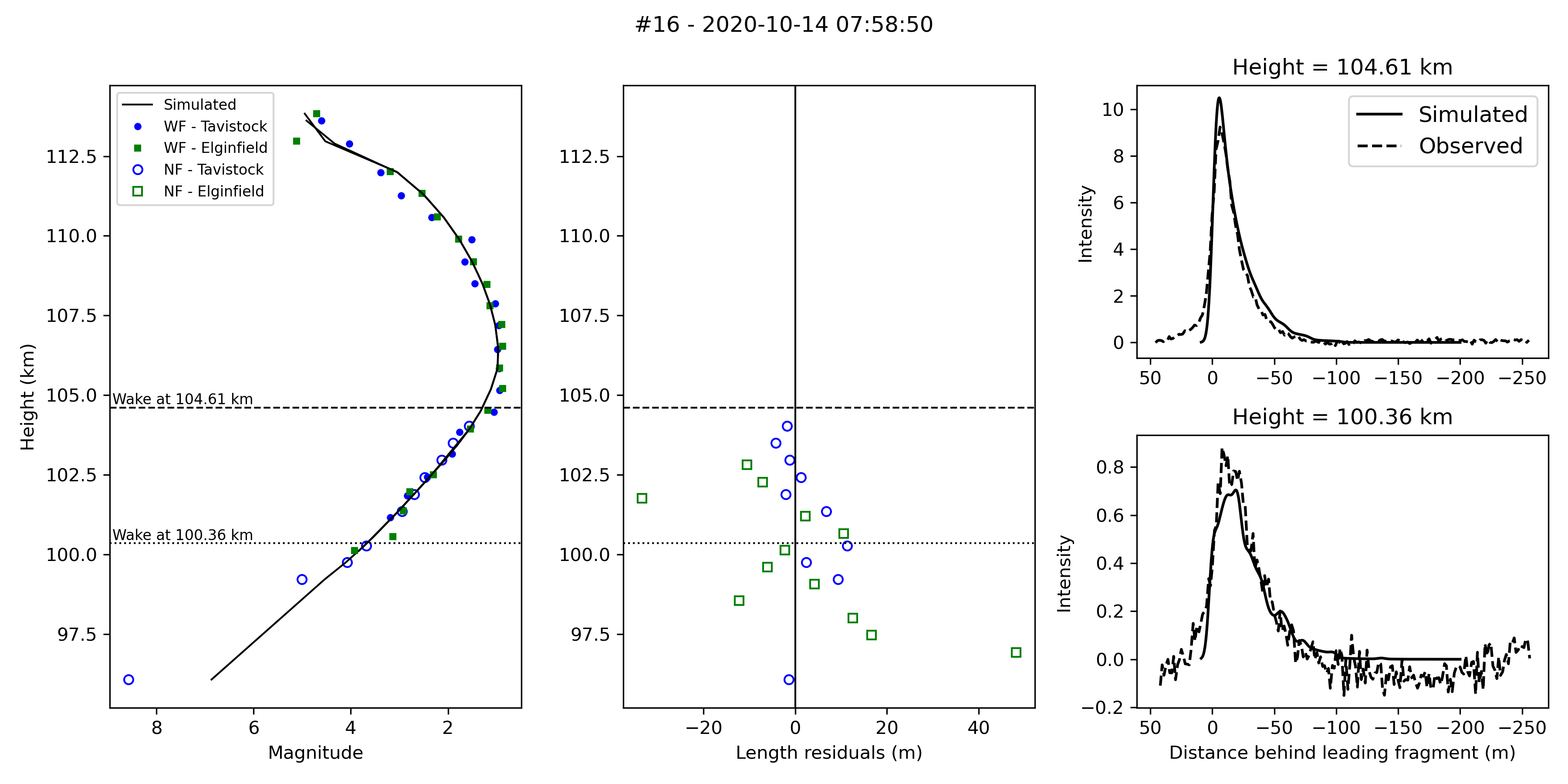}
    \includegraphics[width=\linewidth]{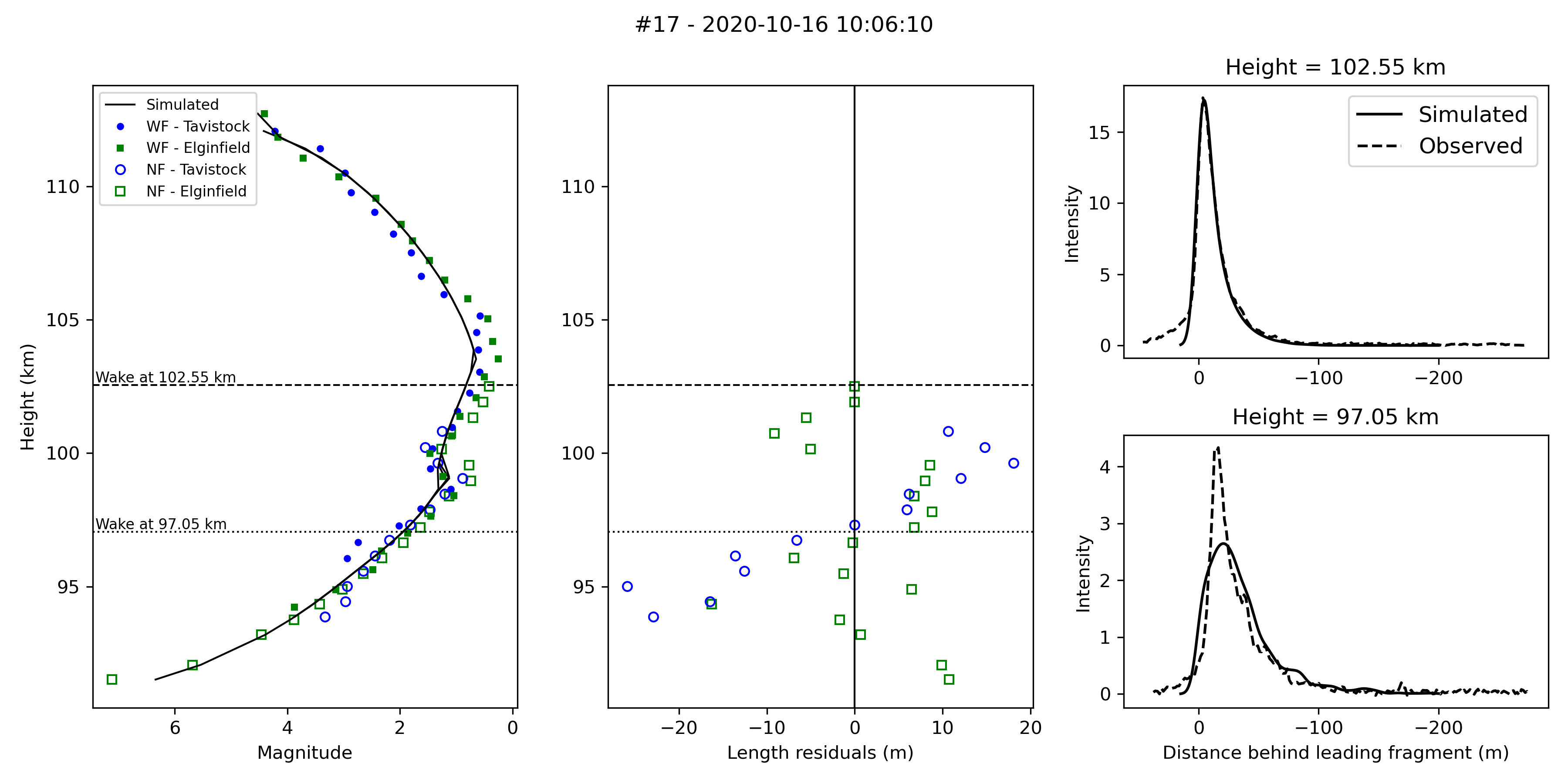}
    \includegraphics[width=\linewidth]{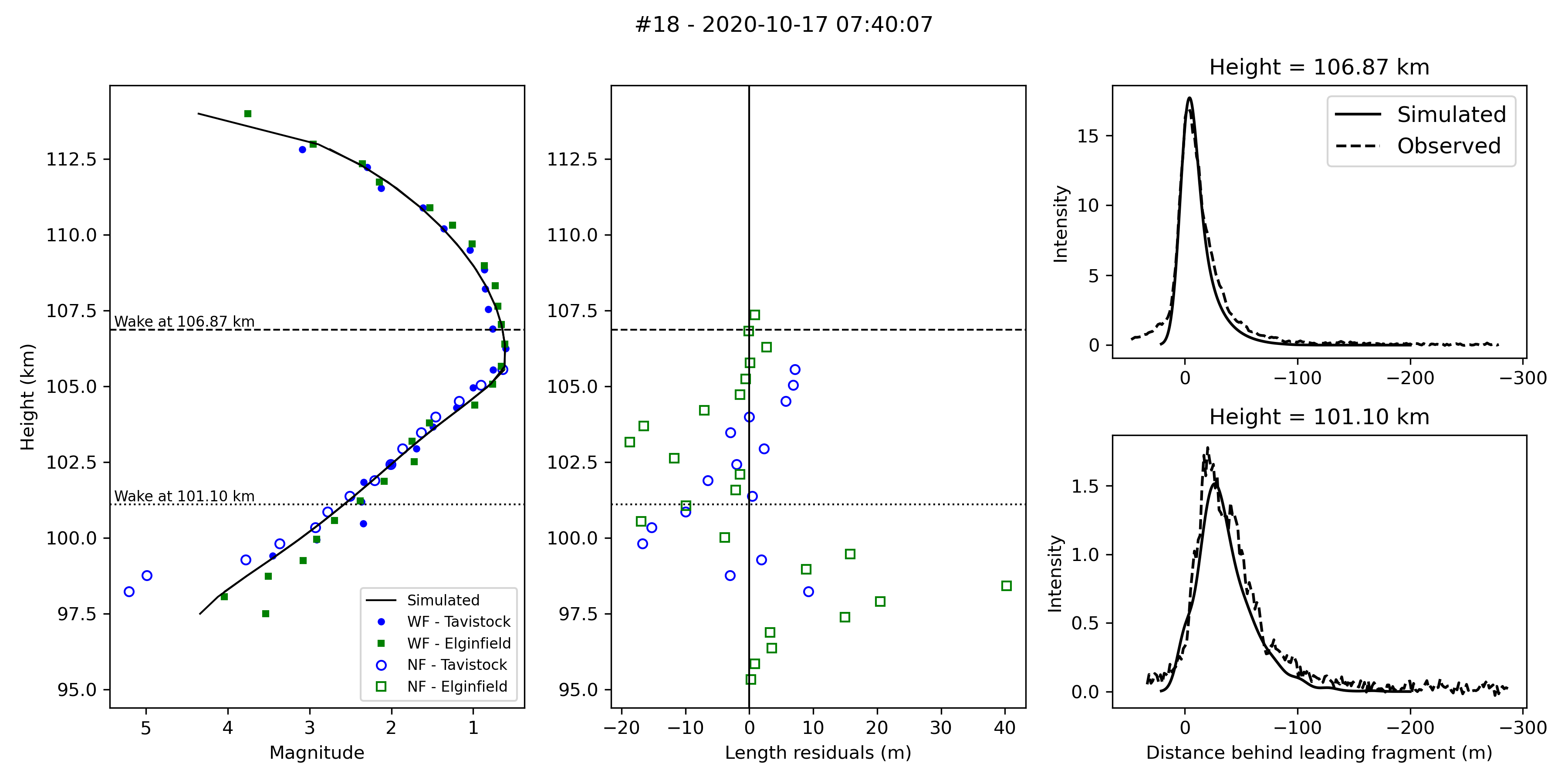}
    \caption[Comparison between observations and simulations]{Comparison between observations and simulations.}
  \label{fig:obs_sim_comp6}
\end{figure*}

\end{document}